# Chandra Early-Type Galaxy Atlas

Dong-Woo Kim, Craig Anderson, Douglas Burke, Raffaele D'Abrusco, Giuseppina Fabbiano, Antonella Fruscione, Jennifer Lauer, Michael McCollough, Douglas Morgan, Amy Mossman, Ewan O'Sullivan, Alessandro Paggi, Saeqa Vrtilek

Center for Astrophysics | Harvard & Smithsonian
60 Garden Street, Cambridge, MA 02138, USA

Ginevra Trinchieri

INAF-Osservatorio Astronomico di Brera,
Via Brera 28, 20121 Milan, Italy

**abstract**

The hot ISM in early type galaxies (ETGs) plays a crucial role in understanding their formation and evolution. The structural features of the hot gas identified by Chandra observations point to key evolutionary mechanisms, (e.g., AGN and stellar feedback, merging history). In our Chandra Galaxy Atlas (CGA) project, taking full advantage of the Chandra capabilities, we systematically analyzed the archival Chandra data of 70 ETGs and produced uniform data products for the hot gas properties. The primary data products are spatially resolved 2D spectral maps of the hot gas from individual galaxies. We emphasize that new features can be identified in the spectral maps which are not readily visible in the surface brightness maps. The high-level images can be viewed at the dedicated CGA website, and the CGA data products can be downloaded to compare with data at other wavelengths and to perform further analyses. Utilizing our data products, we address a few focused science topics.

## 1. INTRODUCTION

For the last two decades, the Chandra X-ray Observatory has revolutionized many essential science subjects in astronomy and astrophysics. The main driving force is its unprecedented high spatial resolution[1] - the capability to resolve fine structures in a sub-arcsec spatial scale. The study of the hot interstellar medium (ISM) in early type galaxies (ETGs) can be taken to a new level thanks to this resolution. The hot ISM, which is the dominant phase in ETGs (e.g., see Kim & Pellegrini 2012), plays a crucial role in understanding the formation and evolution of the host galaxies. Various structural features of the hot ISM, which was previously considered as smooth and featureless, have been identified by Chandra. They include cavities, cold fronts, filaments and tails which are closely related to critical astrophysical mechanisms for the galaxy formation and evolution, e.g., AGN feedback, merging history, accretion, stripping and star formation (SF) and its quenching (e.g., see Kim & Pellegrini 2012 and references therein).

Using the first four years of Chandra observations, Diehl & Statler (2007, 2008a, 2008b) assembled a sample of 54 nearby (D $\lesssim$ 100 Mpc) ETGs (or 36 with temperature profiles). These authors produced a systematic characterization of the hot gas morphology and compared this morphology with the optical stellar distribution, radio emission, and AGN properties. Using the first 15 years of Chandra observations, we have expanded the nearby ETG sample to 70 galaxies and produced spatially resolved homogeneous data products with additional spectral information. Previous archival studies have often focused on the global properties or 1D radial profiles (e.g., Kim & Fabbiano 2015; Lakhchaura et al. 2018). To explore the 2D distribution of spectral properties of the hot gas, we applied 4 different spatial binning techniques. In particular, the 2D *spectral* maps (of the hot gas temperature, normalized emission measure, projected pseudo-pressure and projected pseudo-entropy) can reveal unique features, which may not be visible in 1D radial profiles or the 2D surface brightness maps alone.

Although many galaxies were previously investigated, those studies were mostly done individually (see Appendix B). We provide uniformly reduced data products so that the entire sample of early type galaxies can be examined consistently. We also note that some Chandra data are extensively investigated for the first time as part of this project (e.g., NGC 1132 by Kim et al. 2018). We make use of new and improved data analysis (and statistical) techniques (e.g., CIAO[2] s/w package v4.9), instrument calibration data (in Chandra CALDB[3] v4.7), and atomic data (in ATOMDB[4] v3).

This paper is organized as follows. In section 2, we describe the sample selection and observational information of each Chandra archival data. In Section 3, we describe the Chandra data analysis techniques and related issues. In Section 4, we present the data products of Chandra Galaxy Atlas, our guidelines for general users on how to use the CGA data and data caveats in the current version. In section 5, we explain current and future focused science goals. In Appendix A, we list the CGA data products (downloadable) and their descriptions in detail. In Appendix B, we present the notes on individual galaxies. Throughout this paper, we quote errors at the 1$\sigma$ significance level.

---

[1] http://cxc.harvard.edu/cdo/about_chandra/overview_cxo.html

[2] http://cxc.harvard.edu/ciao/

[3] http://cxc.harvard.edu/caldb/

[4] http://www.atomdb.org

## 2. SAMPLE SELECTION AND CHANDRA OBSERVATIONS

Our sample contains 70 E and S0 galaxies (type < 0, based on RC3[5]) for which observations are available from the public Chandra archive (up to AO15). As many giant galaxies are in galaxy groups/clusters in the RC3 catalog, our sample includes several examples of brightest group/cluster galaxies (BCGs). However, we exclude large groups/clusters by limiting $T_{GAS}$ below ~1.5 keV, because $T_{GAS}$ is a good measure of the total mass of the system. About 80% of the sample galaxies have $T_{GAS}$ = 0.3 - 1.0 keV.

We use both ACIS-I and ACIS-S Chandra observations[6], but exclude observations with grating. We limit the minimum exposure time to 10 ksec to exclude snapshot observations for AGN studies. In Table 1, we list our sample with the basic galaxy information (e.g., RA, DEC, distance, size, K-band luminosity, etc.) and the total Chandra exposure time (see the footnotes below).

As multiple Chandra observations are often merged for a given galaxy, we assign a unique merge id, or mid in short (column 12 in Table 1; column 3 in Table 2) to each galaxy data set. If only one Chandra observation is used, the mid is the same as the Chandra observation ID (obsid). For multiple observations, the mid consists of 5 digits, like an obsid. The first digit is 9, followed by two digits indicating the number of obsids used. The last two digits (usually "01") are unique serial numbers to separate different combinations of observation parameters (e.g., set of obsids, set of chips, etc.). The exposure time of this merged dataset (column 13 in Table 1) is determined by summing the LIVETIME of the chip where the galaxy center lies, after removing background flares which were determined for each chip of each obsid. See section 3 for details on how we removed background flares.

**Table 1 CGA sample**

| name | RA | DEC | D | type | r_maj | r_min | PA | Re | log(LK) | $N_H$ | mid | exp |
|---|---|---|---|---|---|---|---|---|---|---|---|---|
| | h m s | d m s | Mpc | | arcmin | arcmin | deg | arcmin | log(Lo) | $10^{20}$ $cm^{-2}$ | | ksec |
| (1) | (2) | (3) | (4) | (5) | (6) | (7) | (8) | (9) | (10) | (11) | (12) | (13) |
| I1262 | 17 33 2.0 | +43 45 34.6 | 130.0 | -5.0 | 0.60 | 0.32 | 80.0 | 0.20 | 11.42 | 2.43 | 90401 | 138.8 |
| I1459 | 22 57 10.6 | -36 27 44.0 | 29.2 | -5.0 | 2.62 | 1.90 | 42.5 | 0.62 | 11.54 | 1.17 | 02196 | 52.5 |
| I1860 | 02 49 33.7 | -31 11 21.0 | 93.8 | -5.0 | 0.87 | 0.60 | 6.4 | 0.31 | 11.57 | 2.05 | 10537 | 36.8 |
| I4296 | 13 36 39.0 | -33 57 57.2 | 50.8 | -5.0 | 1.69 | 1.62 | 45.0 | 0.80 | 11.74 | 4.09 | 03394 | 24.3 |
| N0193 | 00 39 18.6 | +03 19 52.0 | 47.0 | -2.5 | 0.72 | 0.60 | 70.0 | 0.32 | 10.99 | 2.79 | 90201 | 106.7 |
| N0315 | 00 57 48.9 | +30 21 8.8 | 69.8 | -4.0 | 1.62 | 1.02 | 45.0 | 0.62 | 11.84 | 5.92 | 90201 | 55.9 |
| N0383 | 01 07 24.9 | +32 24 45.0 | 63.4 | -3.0 | 0.79 | 0.71 | 25.0 | 0.34 | 11.54 | 5.41 | 02147 | 42.9 |
| N0499 | 01 23 11.5 | +33 27 38.0 | 54.5 | -2.5 | 0.81 | 0.64 | 70.0 | 0.28 | 11.31 | 5.21 | 90401 | 37.4 |
| N0507 | 01 23 40.0 | +33 15 20.0 | 63.8 | -2.0 | 1.55 | 1.55 | 60.0 | 0.69 | 11.62 | 5.23 | 90201 | 60.3 |
| N0533 | 01 25 31.4 | +01 45 32.8 | 76.9 | -5.0 | 1.90 | 1.17 | 47.5 | 0.72 | 11.73 | 3.07 | 02880 | 35.3 |
| N0720 | 01 53 0.5 | -13 44 19.2 | 27.7 | -5.0 | 2.34 | 1.20 | 140.0 | 0.60 | 11.31 | 1.58 | 90401 | 92.4 |
| N0741 | 01 56 21.0 | +05 37 44.0 | 70.9 | -5.0 | 1.48 | 1.44 | 90.0 | 0.64 | 11.72 | 4.44 | 02223 | 29.1 |
| N1052 | 02 41 4.8 | -08 15 20.8 | 19.4 | -5.0 | 1.51 | 1.04 | 120.0 | 0.56 | 10.93 | 3.07 | 05910 | 57.2 |
| N1132 | 02 52 51.8 | -01 16 28.8 | 95.0 | -4.5 | 1.26 | 0.67 | 150.0 | 0.56 | 11.58 | 5.19 | 90201 | 38.3 |
| N1316 | 03 22 41.7 | -37 12 29.6 | 21.5 | -2.0 | 6.01 | 4.26 | 47.5 | 1.22 | 11.76 | 2.13 | 02022 | 23.6 |
| N1332 | 03 26 17.3 | -21 20 7.3 | 22.9 | -3.0 | 2.34 | 0.72 | 112.5 | 0.46 | 11.23 | 2.30 | 90201 | 53.7 |
| N1380 | 03 36 27.9 | -34 58 32.9 | 17.6 | -2.0 | 2.39 | 1.15 | 7.0 | 0.63 | 11.08 | 1.42 | 09526 | 37.0 |
| N1387 | 03 36 57.1 | -35 30 23.9 | 20.3 | -3.0 | 1.41 | 1.41 | 110.0 | 0.59 | 10.98 | 14.50 | 04168 | 45.4 |
| N1395 | 03 38 29.8 | -23 01 39.7 | 24.1 | -5.0 | 2.94 | 2.23 | 92.5 | 0.78 | 11.34 | 1.94 | 00799 | 16.4 |
| N1399 | 03 38 29.1 | -35 27 2.7 | 19.9 | -5.0 | 3.46 | 3.23 | 150.0 | 0.81 | 11.41 | 1.49 | 90602 | 206.8 |
| N1400 | 03 39 30.8 | -18 41 17.0 | 26.4 | -3.0 | 1.15 | 1.00 | 94.0 | 0.38 | 11.05 | 5.17 | 90202 | 54.6 |
| N1404 | 03 38 51.9 | -35 35 39.8 | 21.0 | -5.0 | 1.66 | 1.48 | 162.5 | 0.45 | 11.25 | 1.51 | 91101 | 486.3 |
| N1407 | 03 40 11.9 | -18 34 48.4 | 28.8 | -5.0 | 2.29 | 2.13 | 60.0 | 1.06 | 11.57 | 5.42 | 00791 | 41.9 |
| N1550 | 04 19 37.9 | +02 24 35.7 | 51.1 | -3.2 | 1.12 | 0.97 | 30.0 | 0.43 | 11.24 | 11.25 | 90401 | 106.8 |

[5] Third Reference Catalogue of Bright Galaxies (RC3) de Vaucouleurs G., et al. 1991

[6] http://cxc.harvard.edu/proposer/POG/

| | | | | | | | | | | | | |
|---|---|---|---|---|---|---|---|---|---|---|---|---|
| N1553 | 04 16 10.5 | -55 46 48.5 | 18.5 | -2.0 | 2.23 | 1.41 | 150.0 | 0.95 | 11.36 | 1.49 | 00783 | 16.6 |
| N1600 | 04 31 39.9 | -05 05 10.0 | 57.4 | -5.0 | 1.23 | 0.83 | 5.0 | 0.81 | 11.63 | 4.86 | 90201 | 48.7 |
| N1700 | 04 56 56.3 | -04 51 56.7 | 44.3 | -5.0 | 1.66 | 1.04 | 85.0 | 0.30 | 11.39 | 4.76 | 02069 | 29.5 |
| N2300 | 07 32 20.0 | +85 42 34.2 | 30.4 | -2.0 | 1.41 | 1.02 | 108.0 | 0.55 | 11.25 | 5.49 | 90201 | 65.7 |
| N2563 | 08 20 35.7 | +21 04 4.0 | 67.8 | -2.0 | 1.04 | 0.76 | 70.0 | 0.32 | 11.39 | 4.25 | 07925 | 48.1 |
| N3115 | 10 05 14.0 | -07 43 6.9 | 9.7 | -3.0 | 3.62 | 1.23 | 45.0 | 0.57 | 10.95 | 4.61 | 91101 | 1088.5 |
| N3379 | 10 47 49.6 | +12 34 53.9 | 10.6 | -5.0 | 2.69 | 2.39 | 67.5 | 0.78 | 10.87 | 2.78 | 90501 | 325.2 |
| N3402 | 10 50 26.1 | -12 50 42.3 | 64.9 | -4.0 | 1.04 | 1.04 | 170.0 | 0.47 | 11.39 | 4.50 | 03243 | 28.0 |
| N3607 | 11 16 54.6 | +18 03 7.0 | 22.8 | -2.0 | 2.45 | 1.23 | 125.0 | 0.76 | 11.25 | 1.48 | 02073 | 38.5 |
| N3608 | 11 16 59.0 | +18 08 55.3 | 22.9 | -5.0 | 1.58 | 1.29 | 80.0 | 0.49 | 10.81 | 1.48 | 02073 | 37.7 |
| N3842 | 11 44 2.1 | +19 56 59.0 | 97.0 | -5.0 | 0.71 | 0.51 | 175.0 | 0.63 | 11.67 | 2.27 | 04189 | 42.4 |
| N3923 | 11 51 1.8 | -28 48 22.0 | 22.9 | -5.0 | 2.94 | 1.95 | 47.5 | 0.88 | 11.45 | 6.30 | 90201 | 94.5 |
| N4104 | 12 06 39.0 | +28 10 27.1 | 120.0 | -2.0 | 1.29 | 0.77 | 35.0 | 0.57 | 11.89 | 1.67 | 06939 | 33.8 |
| N4125 | 12 08 6.0 | +65 10 26.9 | 23.9 | -5.0 | 2.88 | 1.58 | 82.5 | 0.85 | 11.35 | 1.82 | 02071 | 60.6 |
| N4261 | 12 19 23.2 | +05 49 30.8 | 31.6 | -5.0 | 2.04 | 1.82 | 172.5 | 0.75 | 11.43 | 1.58 | 90201 | 130.3 |
| N4278 | 12 20 6.8 | +29 16 50.7 | 16.1 | -5.0 | 2.04 | 1.90 | 27.5 | 0.56 | 10.87 | 1.76 | 90901 | 557.6 |
| N4291 | 12 20 17.8 | +75 22 14.8 | 26.2 | -5.0 | 0.95 | 0.79 | 110.0 | 0.27 | 10.80 | 2.88 | 11778 | 28.2 |
| N4325 | 12 23 6.7 | +10 37 16.0 | 110.0 | 0.0 | 0.48 | 0.32 | 175.0 | 0.33 | 11.29 | 2.14 | 03232 | 28.0 |
| N4342 | 12 23 39.0 | +07 03 14.4 | 16.5 | -3.0 | 0.64 | 0.30 | 165.0 | 0.10 | 10.16 | 1.60 | 90201 | 74.4 |
| N4374 | 12 25 3.7 | +12 53 13.1 | 18.4 | -5.0 | 3.23 | 2.81 | 122.5 | 1.02 | 11.37 | 2.78 | 90301 | 112.4 |
| N4382 | 12 25 24.1 | +18 11 27.9 | 18.4 | -1.0 | 3.54 | 2.75 | 12.5 | 1.38 | 11.41 | 2.50 | 02016 | 38.2 |
| N4406 | 12 26 11.7 | +12 56 46.0 | 17.1 | -5.0 | 4.46 | 2.88 | 125.0 | 2.07 | 11.36 | 2.69 | 90202 | 22.0 |
| N4438 | 12 27 45.6 | +13 00 31.8 | 18.0 | 0.0 | 4.26 | 1.58 | 20.5 | 0.95 | 10.94 | 2.60 | 90201 | 25.0 |
| N4472 | 12 29 46.8 | +08 00 1.7 | 16.3 | -5.0 | 5.12 | 4.16 | 162.5 | 1.74 | 11.60 | 1.62 | 90501 | 362.5 |
| N4477 | 12 30 2.2 | +13 38 11.8 | 16.5 | -2.0 | 1.90 | 1.73 | 40.0 | 0.73 | 10.83 | 2.65 | 90401 | 117.4 |
| N4526 | 12 34 3.0 | +07 41 56.9 | 16.9 | -2.0 | 3.62 | 1.20 | 113.0 | 0.68 | 11.20 | 1.63 | 03925 | 36.1 |
| N4552 | 12 35 39.8 | +12 33 22.8 | 15.3 | -5.0 | 2.56 | 2.34 | 150.0 | 0.68 | 11.01 | 2.56 | 90401 | 197.0 |
| N4555 | 12 35 41.2 | +26 31 23.0 | 91.5 | -5.0 | 0.95 | 0.81 | 120.0 | 0.50 | 11.59 | 1.33 | 02884 | 26.4 |
| N4594 | 12 39 59.4 | -11 37 23.0 | 9.8 | 1.0 | 4.35 | 1.77 | 87.5 | 1.19 | 11.33 | 3.67 | 90301 | 186.7 |
| N4636 | 12 42 49.9 | +02 41 16.0 | 14.7 | -5.0 | 3.01 | 2.34 | 142.5 | 1.56 | 11.10 | 1.82 | 90401 | 191.1 |
| N4649 | 12 43 40.0 | +11 33 9.7 | 16.8 | -5.0 | 3.71 | 3.01 | 107.5 | 1.28 | 11.49 | 2.13 | 90601 | 284.2 |
| N4782 | 12 54 35.7 | -12 34 7.1 | 60.0 | -5.0 | 0.89 | 0.85 | 5.0 | 0.25 | 11.79 | 3.58 | 03220 | 48.3 |
| N5044 | 13 15 24.0 | -16 23 7.9 | 31.2 | -5.0 | 1.48 | 1.48 | 10.0 | 0.42 | 11.24 | 4.94 | 90201 | 102.1 |
| N5129 | 13 24 10.0 | +13 58 36.0 | 103.0 | -5.0 | 0.85 | 0.71 | 5.0 | 0.48 | 11.66 | 1.76 | 90201 | 43.2 |
| N5171 | 13 29 21.5 | +11 44 6.0 | 100.0 | -3.0 | 0.55 | 0.41 | 0.0 | 0.43 | 11.32 | 1.94 | 03216 | 34.4 |
| N5813 | 15 01 11.3 | +01 42 7.1 | 32.2 | -5.0 | 2.08 | 1.51 | 130.0 | 0.89 | 11.38 | 4.25 | 90901 | 616.8 |
| N5846 | 15 06 29.3 | +01 36 20.2 | 24.9 | -5.0 | 2.04 | 1.90 | 27.5 | 0.99 | 11.34 | 4.24 | 90201 | 109.4 |
| N5866 | 15 06 29.5 | +55 45 47.6 | 15.3 | -1.0 | 2.34 | 0.97 | 123.0 | 0.64 | 10.96 | 1.47 | 02879 | 30.7 |
| N6107 | 16 17 20.1 | +34 54 5.0 | 127.9 | -5.0 | 0.43 | 0.33 | 27.5 | 0.44 | 11.79 | 1.49 | 08180 | 18.7 |
| N6338 | 17 15 23.0 | +57 24 40.0 | 123.0 | -2.0 | 0.76 | 0.51 | 15.0 | 0.48 | 11.75 | 2.60 | 04194 | 46.6 |
| N6482 | 17 51 48.8 | +23 04 19.0 | 58.4 | -5.0 | 1.00 | 0.85 | 65.0 | 0.37 | 11.52 | 7.77 | 03218 | 16.4 |
| N6861 | 20 07 19.4 | -48 22 11.5 | 28.1 | -3.0 | 1.41 | 0.91 | 140.0 | 0.38 | 11.14 | 5.01 | 90201 | 110.1 |
| N6868 | 20 09 54.1 | -48 22 46.0 | 26.8 | -5.0 | 1.77 | 1.41 | 80.0 | 0.50 | 11.26 | 4.96 | 90201 | 94.5 |
| N7618 | 23 19 47.2 | +42 51 9.5 | 74.0 | -5.0 | 0.60 | 0.50 | 10.0 | 0.36 | 11.46 | 11.93 | 90301 | 75.2 |
| N7619 | 23 20 14.5 | +08 12 22.5 | 53.0 | -5.0 | 1.26 | 1.15 | 40.0 | 0.57 | 11.57 | 5.04 | 90201 | 55.0 |
| N7626 | 23 20 42.5 | +08 13 1.0 | 56.0 | -5.0 | 1.32 | 1.17 | 10.0 | 0.74 | 11.62 | 5.05 | 02074 | 26.2 |

Column 1. Galaxy name (NGC or IC name)
Column 2-3. RA and DEC (J2000) from 2MASS via NED[7]
Column 4. Distance in Mpc, primarily taken from Tonry et al. (2001), Cappellari et al. (2011) and Tully et al. (2013). If not listed in the above references, we take a mean value from NED.
Column 5. Type taken from RC3
Column 6-7. Semi-major and semi-minor axis of the $D_{25}$ ellipse in arcmin taken from RC3
Column 8. Position angle of the $D_{25}$ ellipse from 2MASS via NED, measured eastward from the north
Column 9. Effective radius in arcmin taken from RC3
Column 10. K-band luminosity from 2MASS (K_tot mag) via NED (assuming $M_K$(sun) = 3.28 mag and D in col 4)
Column 11. Galactic line of sight column Hydrogen density in unit of $10^{20}$ $cm^{-2}$ by colden[8]
Column 12. The Chandra merge id (mid in short – see Table 2 for individual obsids).

[7] http://ned.ipac.caltech.edu

[8] http://asc.harvard.edu/toolkit/colden.jsp

Column 13. The total effective exposure in ksec (see Table 2 for the exposures for individual obsids).

In Table 2, we list the Chandra observational information for each obsid, grouped by galaxy. They include the observation date (column 4), off-axis angle (in arcmin) of the galaxy center from the telescope aim point (column 5), detectors indicating ACIS-I or ACIS-S (column 6) and chips (ccdid) used in this study (column 7), ccdid where the galaxy center lies (column 8), effective exposure after removing background flares (see section 3.1) and observation-specific notes on individual obsid (column 10).

**Table 2 Chandra Observations**

| obsid (1) | name (2) | mid (3) | obs_date year-month-day (4) | OAA arcmin (5) | detector (6) | ccdid (7) | target (8) | eff_exp ksec (9) | notes (10) |
|---|---|---|---|---|---|---|---|---|---|
| 02018 | I1262 | 90401 | 2001-08-23 | 0.2 | ACIS-S | 6,7 | 7 | 28.7 | |
| 06949 | I1262 | 90401 | 2006-04-17 | 0.0 | ACIS-I | 0,1,2,3 | 3 | 38.6 | |
| 07321 | I1262 | 90401 | 2006-04-19 | 0.0 | ACIS-I | 0,1,2,3 | 3 | 35.0 | |
| 07322 | I1262 | 90401 | 2006-04-22 | 0.0 | ACIS-I | 0,1,2,3 | 3 | 36.5 | |
| 02196 (skip) | I1459 | 02196 | 2001-08-12 | 0.0 | ACIS-S | 6,7 | 7 | 52.5 | a c |

(A full version is available at the end.)

Column 1. Unique Chandra observation identification number
Column 2. Galaxy name (NGC or IC name)
Column 3. The merge id (mid in short) is an identification number when data from multiple obsids are used.
Column 4. observation date
Column 5. OAA (off-axis-angle) of the galaxy center in arcmin
Column 6. Detector (ACIS-I or ACIS-S)
Column 7. Chips used in CGA
Column 8. The chip where the center of the galaxy lies
Column 9. Effective exposure time (in ksec) of the target chip, after removing background flares
Column 10. Notes on individual obsids: a subarray (512 rows), b subarray (256 array), c CCD readout streaks, d no fid light

## 3. DATA ANALYSIS

Using the CIAO science threads[9] as a guide, we, we have developed our own analysis pipelines to apply robust data reduction methods in three main steps: (1) merging multiple observations and imaging diffuse emission after excluding point sources, (2) adaptively binning to determine optimal spectral extraction regions, (3) extracting spectra from each spatial bin, fitting and mapping spectral parameters.

### 3. 1 Merging Multiple Observations

Since 45% of galaxies were observed multiple times with a range of observing configurations (e.g., different CCDs, pointing, roll angle, field-of-view, etc.), it is critical to properly combine all observations without losing coherency.

[9] http://cxc.harvard.edu/ciao/threads/index.html

For ACIS-I observations, we use the four front-illuminated chips I0-I3 (CCDID=0-3), while for ACIS-S, we use the back-illuminated chip S3 (CCDID=7), where the target lies, and the front-illuminated chip S2 (CCDID=6), where the extended diffuse emission is often visible. Because the point spread function (PSF)[10] becomes large at large off-axis angles (OAAs), the other chips are generally not useful for our purposes. However, for a few special cases, we use a non-standard set of chips (see Table 2). Similarly, we do not use those observations where the target galaxy is at a large OAA (typically OAA > 4 arcmin) because of the large PSF.

Figure 1a illustrates an example of merging six (two ACIS-I and four ACIS-S) Chandra observations of NGC 1399. Figure 1b shows the DSS[11] optical image of the same region of the sky. In all observations, the center of NGC 1399 is within 1.5 arcmin from the aim point.

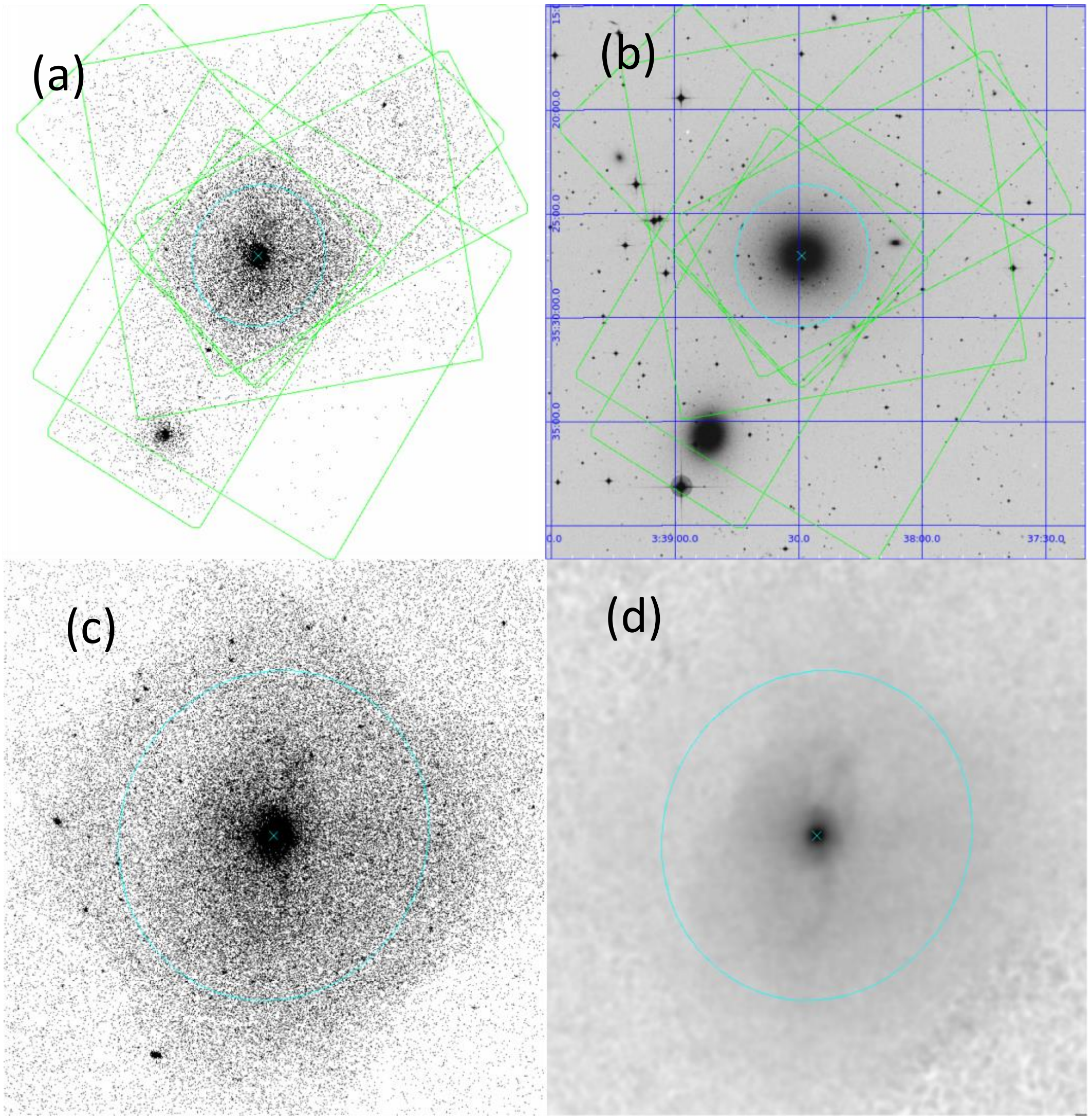

[10] http://cxc.harvard.edu/ciao/PSFs/psf_central.html

[11] https://archive.stsci.edu/cgi-bin/dss_form

**Figure 1.** (a) The B-band (0.3-8 keV) image of NGC 1399. Six observations are merged. The green rectangles are the detector frames of ACIS-I (I0-I3 4 chips with an approximate size of 16'x16') and ACIS-S (S2-S3 2 chips with an approximate size of 8' x 16'). The cyan ellipse is the optical $D_{25}$ ellipse. A companion galaxy, NGC 1404, is also seen at the lower left corner. (b) The optical DSS image in the same field-of-view. (c) A zoomed in view of the C-band (0.5-5 keV) image of NGC 1399. A few hundreds of point sources are embedded in the extended diffuse emission. (d) The G-band (0.5–2 keV) image after the detected point sources are excluded, filled with photons in the surrounding regions. This image was also exposure-corrected and smoothed with a Gaussian kernel of 35 pixels (17") to best illustrate the diffuse gas emission of NGC 1399. The bright central region and the northern and southern filaments are clearly visible inside the $D_{25}$ ellipse. In all figures, north is up and east is to the left.

Figure 2 shows the entire data processing workflow. Once ACIS data are downloaded from the Chandra archive, we run the ACIS level 2 processing using the CIAO tool ***chandra_repro***[12]. This step makes sure that the up-to-date calibration data (e.g., CTI correction, time-dependent ACIS gain, bad pixels) are applied. To correct the relative positional error among multiple observations, we detect point sources from individual observations using the CIAO tool ***wavdetect*** (with scale=1,2,4,8 and sigthresh = $10^{-6}$) on a C-band image (0.5-5 keV, see Table 3 for the definitions of energy bands used in CGA) made with the CIAO tool ***fluximage***. Note that we run ***fluximage*** again after removing point sources detected on the merged data in section 3.2. Then, we register bright (net counts > 30) point sources commonly detected in a pair of observations, using a CIAO tool ***reproject_aspect*** (with radius=2 and residlim=1). This positional correction is usually small ($\Delta$RA and $\Delta$Dec < 0.5") but necessary for the Chandra observations with a high spatial resolution. We rerun ***chandra_repro*** for the 2nd time to apply the newly determined aspect solutions.

**Table 3. Energy band**

| Band name | short name | Minimum Energy | Maximum Energy | Effective Energy* |
|---|---|---|---|---|
| broad | B | 0.3 | 8.0 | 2.3 |
| hard | H | 2.0 | 7.0 | 3.8 |
| center | C | 0.5 | 5.0 | 2.0 |
| gas | G | 0.5 | 2.0 | 1.0 |

Notes.
B - used to overview the entire observation
C and H - used in detecting point sources (section 3.2)
C - used in adaptive spatial binning (section 3.3)
G - used to show diffuse images (section 3.2)
* Effective energy: effective area weighted mean energy within a given energy band.

To remove the time intervals of background flares[13], we generate light curves from the point-source removed event files, determine the mean rate and its standard deviation, then apply a CIAO tool ***deflare*** with 2$\sigma$ clipping (see Markevitch's note[14] for more details about the ACIS background). While this method works for most observations, it fails in extreme cases when the flare occurs during a significant fraction of a given observation or the background rate changes gradually throughout the observation. We examine each light curve and manually remove the time interval with flares before determining the mean rate and standard deviation and rerun ***deflare***. Background flares have varying effects on different chips, hence we apply this step to each CCD for a given observation.

[12] http://cxc.harvard.edu/ciao/threads/createL2/
[13] http://cxc.harvard.edu/ciao/threads/flare/
[14] http://cxc.harvard.edu/contrib/maxim/bg/

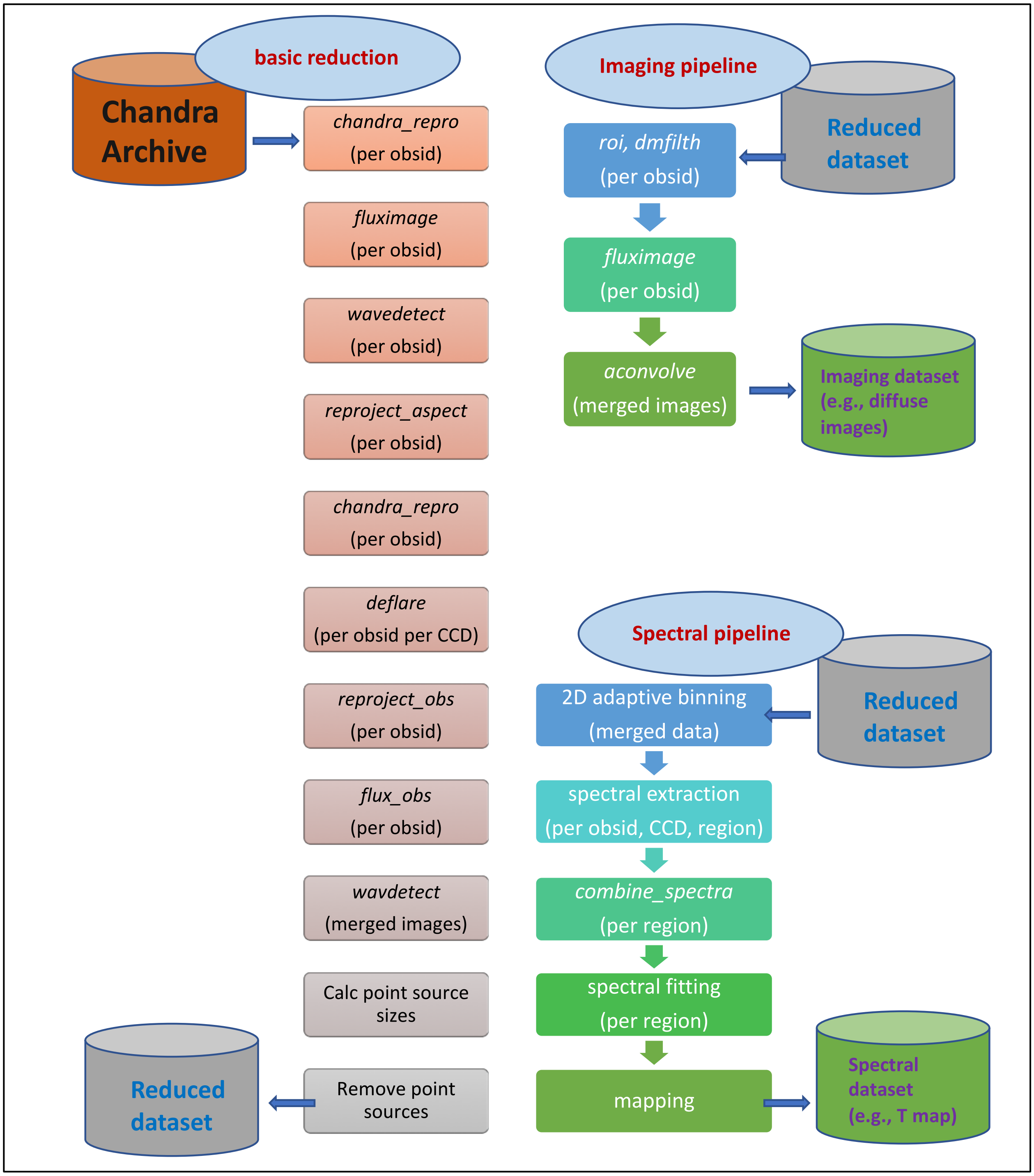

**Figure 2**. Overview of the data analysis workflow. The individual processing steps are in rectangles and associated datasets in cylinders. Those in italics indicate the CIAO tools used in each step.

To reproject individual observations onto a single tangent plane, we use a CIAO tool ***reproject_obs***. To make merged images in multiple energy bands (before and after exposure-correction), we use a CIAO tool ***flux_obs***. Figure 1c shows an example of the merged Chandra observations of NGC 1399.

## 3.2 Point Source Detection and Removal

We run ***wavdetect*** for the 2nd time to detect point sources from the merged images in multiple energy bands. To remove point sources, we primarily use the point sources detected in the C-band (0.5-5 keV). We have experimented with various energy bands and conclude that adding lower or higher energies (below 0.5 keV and above 5 keV) does not result in more secure or more detections, since we include more background than source signal, thus actually reducing the signal-to-noise ratio (S/N).

Since our goal is to investigate the diffuse emission, we do not remove the faint (possibly false) sources and remove only real sources by applying net counts > 10 and S/N > 3. We will discuss the point source properties including faint sources in a separate paper.

To determine the size of each point source, we simulate the PSF for each source for each observation, using MARX[15]. We have compared MARX-PSFs with those made by more accurate ray tracing (SAO-Trace) and found no significant difference for our purpose. We re-project the simulated PSF onto the same tangent plane as the observation and merge them to produce an exposure-weighted mean PSF image as in the merged real observations. We apply variable encircled energy (EE) fractions from 90% to 98%, depending on the net counts to optimally remove photons from point sources (see Table 4), i.e., higher fractions for brighter sources so that we can effectively remove the wing of a bright source, but lower fractions for faint sources to remove only the PSF core.

**Table 4. point source size**

| net counts | EE fraction (%) |
|---|---|
| < 100 | 90 |
| 100 - 200 | 94 |
| 200 - 500 | 95 |
| 500 - 1000 | 96 |
| 1000 - 2000 | 97 |
| > 2000 | 98 |

Near the galaxy center where the hot gas emission peaks, ***wavdetect*** may detect false sources or miss real sources, and the source positions may be uncertain. These problems are particularly severe in systems with narrow, high surface brightness gas substructures (e.g., in NGC 4374, NGC 4636). Instead of the C-band, we use the wavdetect sources detected in the H-band (2-7 keV), as the point sources (mostly LMXBs and background AGNs) are expected to have a harder spectrum than the hot gas (e.g., Boroson et al. 2011). We empirically determined the optimal radius inside which we use the H-band (rather than the C-band) by examining several galaxies with complex hot gas structures. We found this radius where the count in one ACIS pixel in the C-band is 1. In Figure 3, we compare the point sources detected in the C-band and H-band in the central region of NGC 4374. The two sources detected only in C-band (but not in H-band) are marked by the red arrows. As they are likely hot gas blobs, we do not exclude them as X-ray binaries. Another two sources with uncertain positions and sizes (in C-band) are marked by the black arrows. We use the H-band positions and PSF-based source sizes for them.

[15] http://space.mit.edu/cxc/marx/

To make diffuse gas images, we refill the point source elliptical regions with values interpolated from surrounding pixels[16] by using a series of CIAO tools, ***roi***, ***splitroi*** and ***dmfilth***. Then, we generate point-source-excluded, refilled, exposure-corrected images with a CIAO tool, ***fluximage*** and smoothed images by a CIAO tool, ***aconvolve***. Figure 1d shows an example of the diffuse emission of NGC 1399.

We note that while we use the merged data to detect point sources and to perform spatial binning (see section 3.3), we analyze the individual observations for extracting the X-ray spectra and producing the corresponding calibration files, rmf/arf (see section 3.4), because the CCD responses and PSFs vary as a function of detector location, photon energy and observing time.

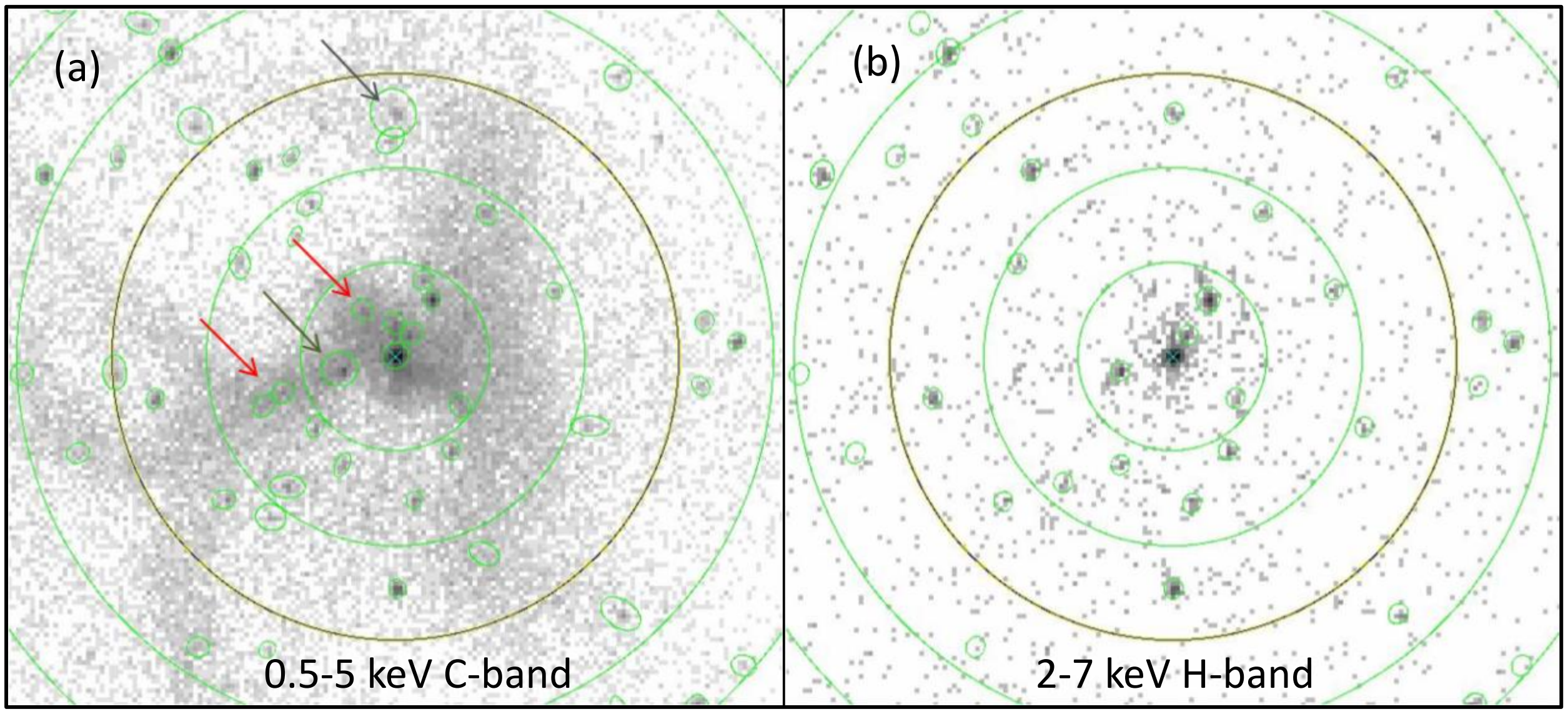

**Figure 3**. Point sources in the central region of NGC 4374 (M84) detected in two energy bands (a) in the 0.5-5 keV C-band and (b) in the 2-7 keV H-band. The red arrow indicates sources detected only in the C band and the black arrows indicates sources with uncertain positions and sizes. The small green ellipses are (a) the source sizes determined by wavdetect and (b) our new psf-based source sizes. The large green circles are at r = 10, 20, 30, 40". The yellow circle at 30" indicates the radius inside which the H-band sources are used.

When a central AGN is very bright (count rate ≳ 0.1 per second), its ACIS observation suffers from a readout streak problem[17]. We determine the streak region by ***acis_streak_map*** for each observation and remove the region from individual event files for the adaptive binning and spectral analysis (section 3.3 – 3.5), as well as removing the point sources (see the related issue in section 4.3).

---

[16] http://cxc.harvard.edu/ciao/threads/diffuse_emission/index.html#bkgreg

[17] http://cxc.harvard.edu/ciao/threads/acisreadcorr/

### 3.3 Adaptive Binning

The most innovative step in our CGA project is to apply four adaptive binning methods to characterize the two-dimensional (2D) spatial/spectral properties of the hot gas.

1) Annulus Binning (AB). Use circular annuli spanning an entire 360 deg or partial annuli with a set of specific sectors. The latter is particularly useful when the gas distribution is not spherically symmetric or at the edge of the detector to avoid unwanted sources. The inner and outer radii of each annulus are adaptively determined, based on a given signal-to-noise ratio (S/N). We apply S/N=20, 30 and 50.
2) WVT Binning (WB). Use the weighted Voronoi tessellation (WVT) adaptive binning. This method was originally developed to analyze optical integral field spectroscopic data by Cappellari and Copin (2003) and later applied to X-ray data by Diehl & Statler (2006).
3) Contour Binning (CB). This method is similar to WB, but additionally takes into account the fact that the areas with similar surface brightness have similar spectral properties (Sanders 2006). As the regions within an iso-intensity contour are grouped together into a single bin, each spatial bin often has a partial annulus-like shape.
4) Hybrid Binning (HB). This is described in detail in O'Sullivan et al. (2014). A fixed grid of a chosen spatial resolution is laid out covering the region of interest. For each grid square, the algorithm adaptively determines the radius of a circular spectral extraction region, centered at that position, necessary to achieve the desired S/N or number of counts, up to some maximum size. Spectra are extracted and fitted, and the fit values used to populate the map. Since the extraction regions may be larger than the grid squares they may overlap, and may therefore not be statistically independent. The resulting maps are therefore analogous to adaptively smoothed images, with high spatial resolution and low smoothing in regions of high surface brightness, and heavier smoothing of fainter emission.

The first method is the most straightforward, and has been widely used to measure the one-dimensional (1D) radial profiles and global properties of ETGs. The next two methods provide the 2D distributions of gas properties in a statistically rigorous manner. The fourth method uses neighboring bins which are not independent; hence statistics are not straightforward. However, it provides complementary information at higher spatial resolution in regions of lower surface brightness, which may be lost in the first three methods. All binning methods are applied to the C-band (0.5-5 keV). The bin size is controlled by S/N or net counts per bin. We typically use 3 sets of S/N values (20, 30, 50) to optimally balance resolution and statistics. We note that in calculating S/N, the background count was properly treated in AB and HB where S/N = src_count / (src_counts + bg_counts)$^{1/2}$, but not in WB and CB where S/N = (src_counts + bg_count)$^{1/2}$. While this makes a negligible effect near the center, the s/n (in WB and CB) is actually lower than specified when the background is important in the outskirts. We will implement the correct S/N calculation in a future release. We note that we apply a kT error cut to the resulting maps, which should remove any low S/N regions where the spectra were of poor quality (see section 3.5).

In Figure 4, we show an example of the four G-band (0.5- 2 keV) images of NGC 1399 (the same galaxy as in Figure 1) binned by the four adaptive binning methods. As described above, each binning method has pros and cons. While AB provides only the 1D characterization, WB provides additional 2D azimuthal variations, CB further shows similar surface brightness regions, and finally, HB retains the highest resolution in low surface brightness regions. In analyzing a galaxy, it is recommended that all four binning results should be considered since they convey different and complementary information,

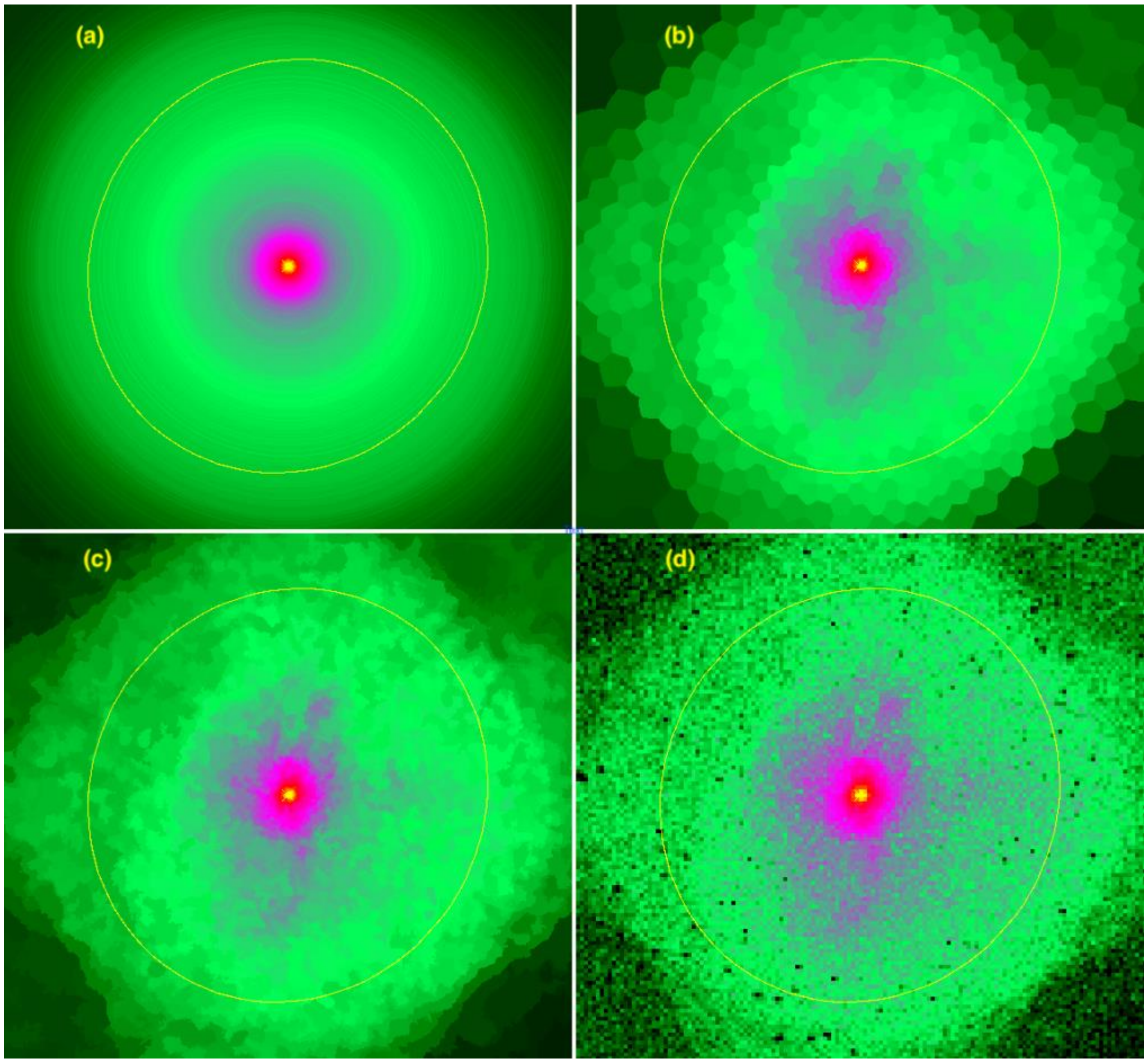

**Figure 4.** Adaptively binned G-band (0.5-2 keV) images of NGC 1399 (in Fig 1). (a) AB, (b) WB, (c) CB, and (d) HB. In all cases, we set S/N=20.

### 3.4 Spectral Extraction

Once the spatial adaptive binning is done, the X-ray spectra are extracted from each spatial bin. The spectral extraction is done on each individual observation. If the spatial bin lies on more than one chip, the spectra are extracted separately per each chip. To take into account time-dependent and position-dependent ACIS responses. the corresponding arf and rmf files are also extracted per observation per chip.

To properly subtract the background emission, we download the blank sky data from the Chandra archive[18]. We match the sky background event file per obsid per chip to the real observation by re-projecting them to the same tangent plane as each observation (see Section 3.1) and excluding the same point source regions as done in each observation (see Section 3.2). We then rescale them to match the rate at higher energies (9-12 keV) where the photons are primarily from the background (Markevitch 2003). To confirm the validity of the sky background and to check temporal and spatial variations of the soft X-ray background, we used off-axis, source-free regions from the same observation in a few test cases and did not find any significant difference.

[18] http://cxc.harvard.edu/ciao/threads/acisbackground/index.html#choosefile.lookup

Once source spectrum, background spectrum, arf and rmf per obsid per chip are generated, we use a CIAO tool, ***combine_spectra***, to combine them to make a single data set per bin. This way the spectral fitting is more straightforward and quicker. We also performed a joint fit by simultaneously fitting individual spectra and found no significant difference.

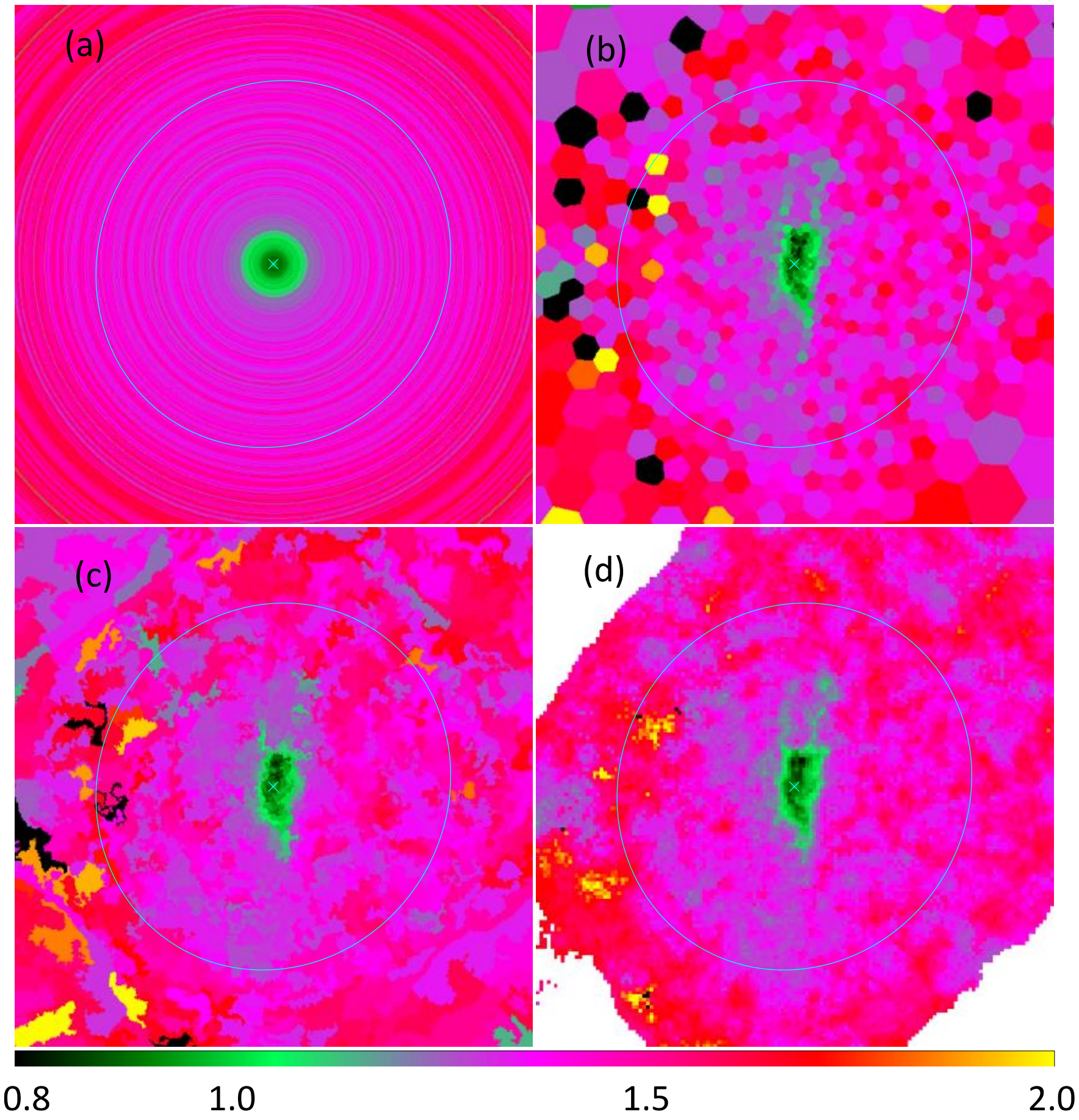

**Figure 5**. Temperature maps of NGC 1399, determined by four binning methods as in Fig 3. The color bar indicates the temperature ranging from 0.8 keV to 2 keV.

### 3.5 Spectral Fitting

We primarily use a two-component emission model, VAPEC[19] for hot gas (collisionally-ionized diffuse gas) and power-law for undetected LMXBs. Taking advantage of the adaptive binning to separate different regions with different spectral properties, this 2-component model is sufficient in most cases. We fix the power-law index to be 1.7, which is appropriate for the hard spectra of LMXBs (e.g., Boroson et al. 2011) and AGNs (see AGN-related issues in section 4.3) and $N_H$ to be the Galactic HI column density (Dickey and Lockman 1990). We also fix the metal abundance to be solar at GRSA (Grevesse, N. & Sauval, A.J. 1998). Although the abundance is known to vary from a few tenth to a few times solar inside the hot ISM, the hot gas temperatures do not significantly depend on the abundance (e.g. see Kim & Pellegrini 2012). The proper determination of the metal abundance requires high S/N data, and we will further investigate the 2D abundance distribution in a future paper.

[19] https://heasarc.gsfc.nasa.gov/xanadu/xspec/manual/XSmodelApec.html

We then produce spectral maps with various spectral parameters, e.g., the gas temperature, the normalization parameter of the VAPEC model divided by the area of each spatial bin (we call it normalized emission measure[20]) and the lower and upper limits of each parameter and corresponding reduced $\chi^2$ values. We apply a mask to remove the bins where the spectral parameters are not well constrained with a significant error in spectral fitting. We select the mask such that only gas temperatures with an error of less than 30% are shown. In this way, the resulting map shows only bins with reliable measurements. Additionally, we also produce the projected pseudo pressure maps ($P_P = S_X^{1/2}$ T) and projected pseudo entropy maps ($K_P = S_X^{-1/3}$ T) in arbitrary units, where Sx is the normalized emission measure. We note that although they are not the 3D pressure and entropy, they still provide useful information, particularly for discontinuities, e.g., cold front and shock front, etc. (see e.g., Werner et al. 2012, Kim et al. 2018). We also note that the APEC normalization (pressure and entropy as well) depend on the abundance variation. Unless the abundance is constant, the abundance gradient (likely negative) would make the normalization and the pseudo pressure flatter, and the pseudo entropy steeper (e.g., see Kim & Pellegrini 2012).

In Figure 5, we show the temperature maps of NGC 1399 made by four adaptive binning methods as in Figure 4. The temperature is 0.8 keV in the central region and gradually increases with increasing radius to ~1.5 keV at the $D_{25}$ ellipse. In contrast to AB, the other three binning methods clearly show the 2D thermal structure, the cooler gas extending along the filaments seen in the surface brightness maps (see more in Section 5.1).

### 3.6 Multi-Processing

Because the number of spatial bins is large, of the order of 10,000, for some deep observations of hot gas rich galaxies, processing can be time consuming and becomes impractical for a single processor. To optimally process a large number of repeated tasks, particularly in the spectral extraction, generating arf/rmf and spectral fitting, we make use of the Smithsonian Institution High Performance Cluster (SI/HPC)[21], a Beowulf cluster consisting of nearly 4,000 CPU cores, distributed over 108 compute nodes and over 24TB of total RAM.

## 4. CHANDRA GALAXY ATLAS

### 4.1 Chandra Galaxy Atlas Data Products

We provide the Chandra Galaxy Atlas data products in a dedicated CGA website[22]. The 1D radial profiles and the 2D images can be directly viewed on the browser, and the necessary data products can be downloaded for further analysis. For those who want to take a quick look at the hot gas distribution and the thermal structure, the figures posted on the CGA website should be useful. Examples are shown in Figure 6, 7 and 8.

---

[20] The APEC normalization (for its definition see the above footnote 19) depends on the bin size. The normalization divided by the bin area is proportional to the integral of $n_e n_H$ along the line of sight. The unit of the area is in pixels. An ACIS pixel size is 0.492 arcsec on each side.

[21] https://confluence.si.edu/pages/viewpage.action?pageId=9995361

[22] http://cxc.cfa.harvard.edu/GalaxyAtlas/v1

In Figure 6, the radial profiles of the surface brightness ($L_{X,GAS}$ / area) and the gas temperature of NGC 4649 are presented. We show the profiles measured in two spatial binning methods, AB and WB. While AB well represents the hot gas structure when the gas is symmetric, WB further shows the degree of asymmetry as in the outer region of NGC 4649. Also shown are the total $L_{X,GAS}$ and log(L)-weighted mean T (see section 5.2) within a given r. The vertical lines indicate r = 5 arcsec (red), r = 1 and 5 x $R_e$ (cyan) and r = $R_{MAX}$ (blue) which is the maximum radius where the hot gas is reliably detected with an azimuthal coverage larger than 95% (see section 5.2).

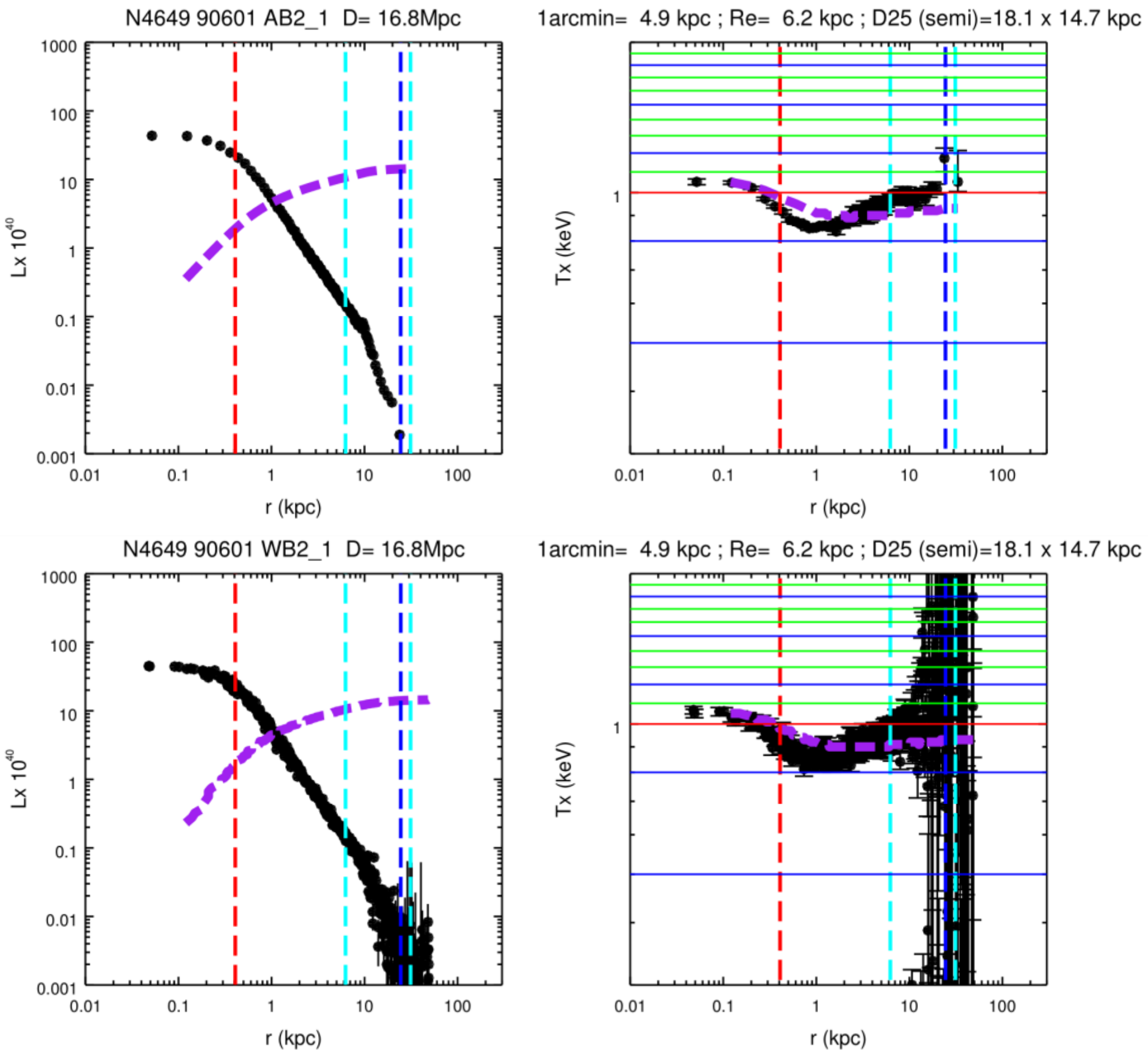

**Figure 6**. Radial profiles of (left) the surface brightness ($L_{X,GAS}$ from a given bin in $10^{40}$ erg $s^{-1}$ / area of the bin in $arcmin^2$) and (right) the gas temperature in two spatial binning methods, (top) AB and (bottom) WB. The thick dashed purple lines indicate (left) $L_{X,GAS}$ (< r) in $10^{40}$ erg $s^{-1}$ and (right) $L_X$-weighted mean T (< r). The vertical lines indicate r = 5 arcsec (red), r = 1 and 5 x $R_e$ (cyan) and r = $R_{MAX}$ (blue) which is the maximum radius where the hot gas is reliably detected with an azimuthal coverage larger than 95% (see section 5.2).

Figure 7 shows a part of the CGA image gallery with postage stamp 2D images of diffuse gas of individual galaxies and Figure 8 shows the 2D spectra maps of an example galaxy, NGC 4649 ($3^{th}$-row $3^{nd}$-column in the left panel) which is displayed by simply clicking the galaxy image on the left panel. The spectral maps include the binned images, T maps, normalized emission measure maps (= APEC normalization parameter / bin area), projected pressure maps, and projected entropy maps measured in 4 binning methods (AB, WB, CB, and HB).

We produce downloadable data products and make them available in two packages. The main package (package 1) consists of the high-level data products: point source removed and filled, exposure corrected images (jpg/png and FITS format) in multiple energy bands to best illustrate the diffuse hot gas distribution, derived T-maps (jpg/png and FITS format) in various adaptive binning methods to show the hot gas thermal structure. They can be used directly to get an overview, to obtain the necessary hot gas quantities and to compare them with data at other wavelengths with no additional X-ray data reduction.

For those who intend to perform their own analyses, e.g., to extract and fit X-ray spectra from user-specified regions, we provide all the necessary data in the supplementary package (package 2) which includes all the event files per obsid and per ccd. The entire sets of data products in the main and supplementary packages are described in detail in Appendix A.

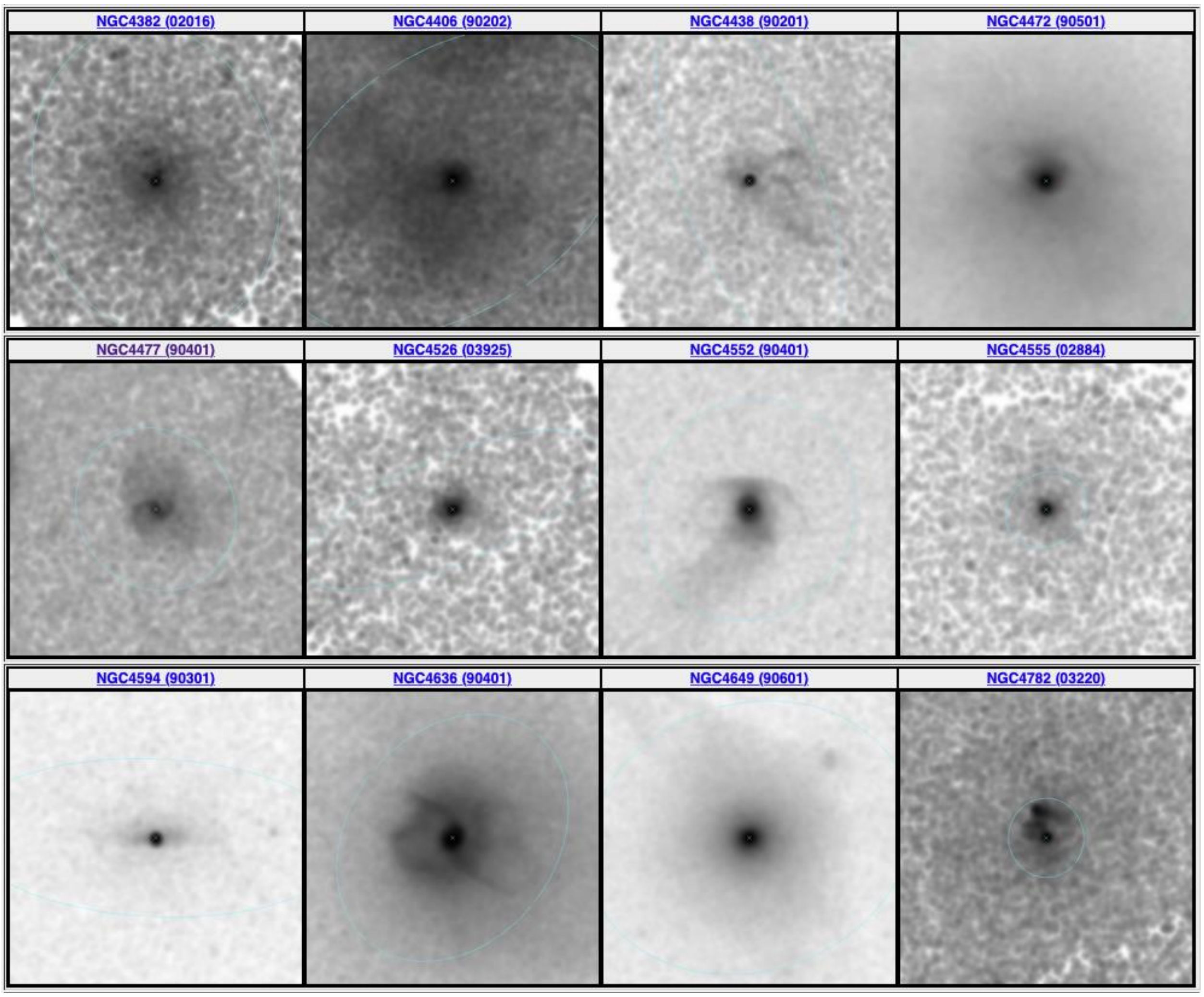

**Figure 7**. The image gallery of diffuse gas from individual galaxies available in the CGA web site.

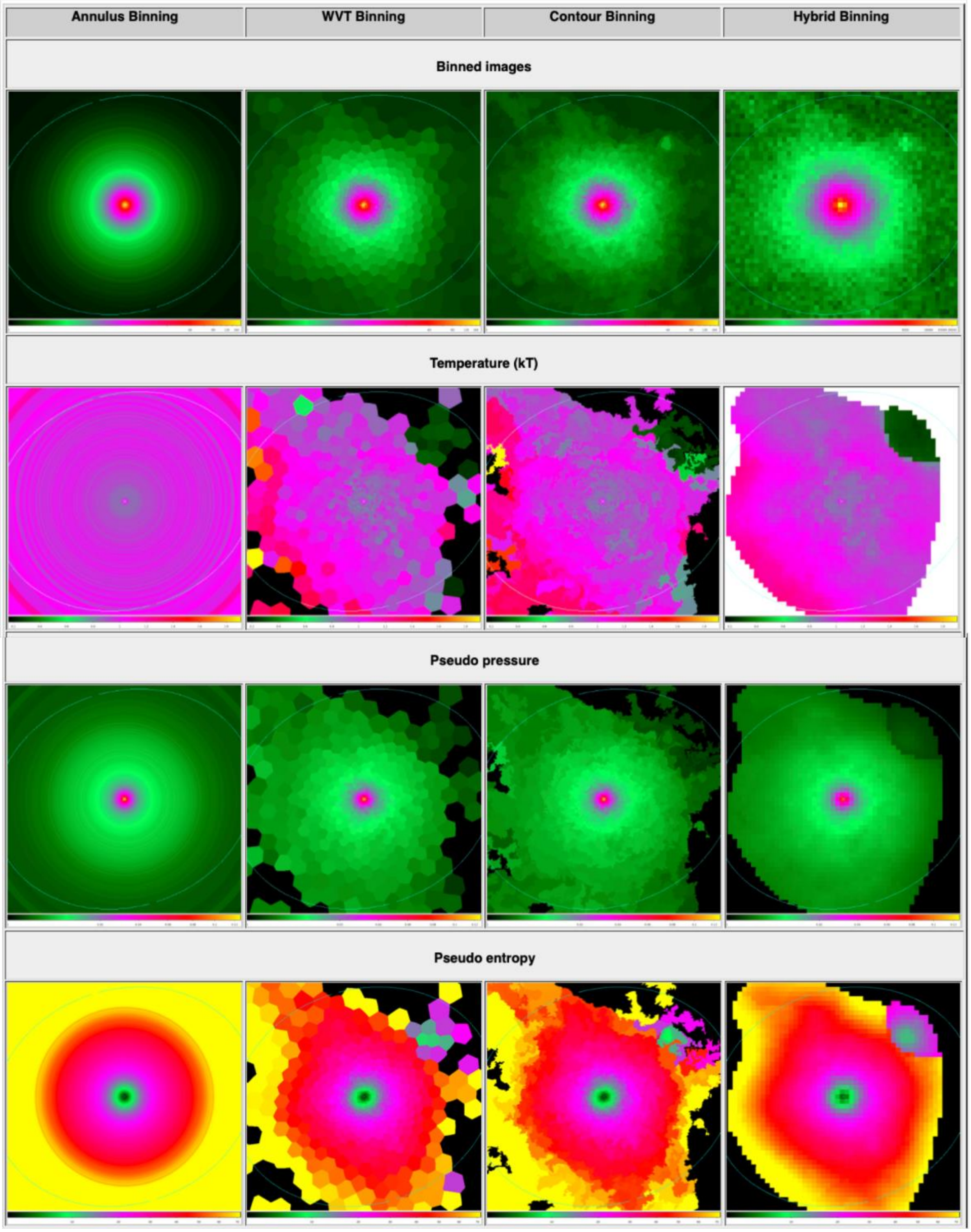

**Figure 8**. An example of a set of spectral maps (binned image, temperature map, pressure map and entropy map from top to bottom) of NGC 4649.

## 4.2 Example Usages of Chandra Galaxy Atlas Data Products

While the high-level products (e.g., diffuse images and temperature maps) are shown in a pre-determined format in the CGA website, users can manipulate the CGA data products for their own use for various purposes. Here we provide a few possible applications.

#### 4.2.1 How to view the X-ray image as observed

The main package includes the merged images, exposure maps and diffuse images (point source excluded and exposure corrected) for multiple energy bands (in Table 3). These fits files can be viewed, e.g. using SAOImage ds9[23]. Similarly, the temperature maps (in fits) can be downloaded for review in ds9 and to compare with images at other wavelengths.

#### 4.2.2 How to review spectral fitting results of individual spatial bins

The spectral fitting results for all spatial bins are available in the data package. The 'binno.fits' file in the main package can be used to identify the bin number of the region of interest (e.g., using ds9). An ascii table (sum.dat in the main package) contains a summary of fitting results including T, norm, reduced $\chi^2$, distance from the galaxy center and area for each bin.

#### 4.2.3 How to rerun spectral fitting in individual spatial bins

Because of a large number of spatial bins, the source and background spectra, arf and rmf are not included in the data package. Instead, we provide a full set of source and background event files (for all obsids and for all CCDs). A user can use these data files and perform their own spectral fitting with their own models and parameters as described in section 3.4-3.5 (see also CIAO science threads for spectral fitting[24]). Users can further use their own region files.

#### 4.2.4 How to make a radial profile in a specified pie sector

Given that a full 2D spectral maps are provided, one can easily generate 1D projected radial profiles with any spectral parameter as a function of r (from sum.dat in the main package). They include temperature, norm/area, luminosity/area, projected entropy and projected pressure. If the hot gas morphology is not azimuthally symmetric, one can select the proper region in a given pie sector, for example away from or across the discontinuity or cavity. WB (section 3.3) is particularly useful for this purpose.

### 4.3 Data Caveats

1) CCD readout streak. When the central AGN is very bright, the CCD readout may cause streaks along the CCD column (see column 10 in Table 2). In this case, we determine the streak region by ***acis_streak_map*** for each obsid and remove this region from the image and event files for adaptive binning (section 3.3) and spectral analysis (section 3.4-3.5). Effectively we treat the streak the same way as the point sources removed in section 3.2. However, unlike the point sources, we do not refill the excluded region for the diffuse hot gas image, because the streaks are often in the middle of complex hot gas structures.

---

[23] http://ds9.si.edu/site/Home.html

[24] http://cxc.harvard.edu/ciao/threads/ispec.html

2) CCD gaps and boundaries. The gaps between chips and node boundaries are often less exposed and are most evident in raw images before the exposure correction, but sometimes still visible even after the exposure corrections. They are least visible in the spectral maps (e.g., in a T map), because the different exposures are appropriately treated bin by bin. One good example is NGC 1550 which was observed four times and the significant exposure variation is clearly visible near the southern edge of the $D_{25}$ ellipse. Although the raw data and binned image (counts/area) show a sharp discontinuity, the T map and EM (normalization/area) map are generally smooth with no clear artifact at that location. We note that the gap may appear more significant in CB than other binning methods because the CB spatial bin is determined in a region with similar counts/area. Users should be cautious about the gaps when considering any physical changes across the gap.
3) Strong nuclear sources. When an AGN at the center dominates the entire X-ray emission with a strong nuclear X-ray source and little amount of hot gas which is confined near the central region, the hot gas properties are difficult to measure. In this release, we adopt a power-law with a spectral index of 1.7 for the nuclear source. If the AGN X-ray spectrum is significantly different, the measured hot gas properties are subject to systematic uncertainties. We will carefully investigate the circum-nuclear hot gas as one of our focused studies in the near future.

## 5. DISCUSSIONS

The CGA data products provide a basis for the investigation of a number of important scientific questions. We describe some examples below, including results from previous studies and investigations we plan to publish in later papers.

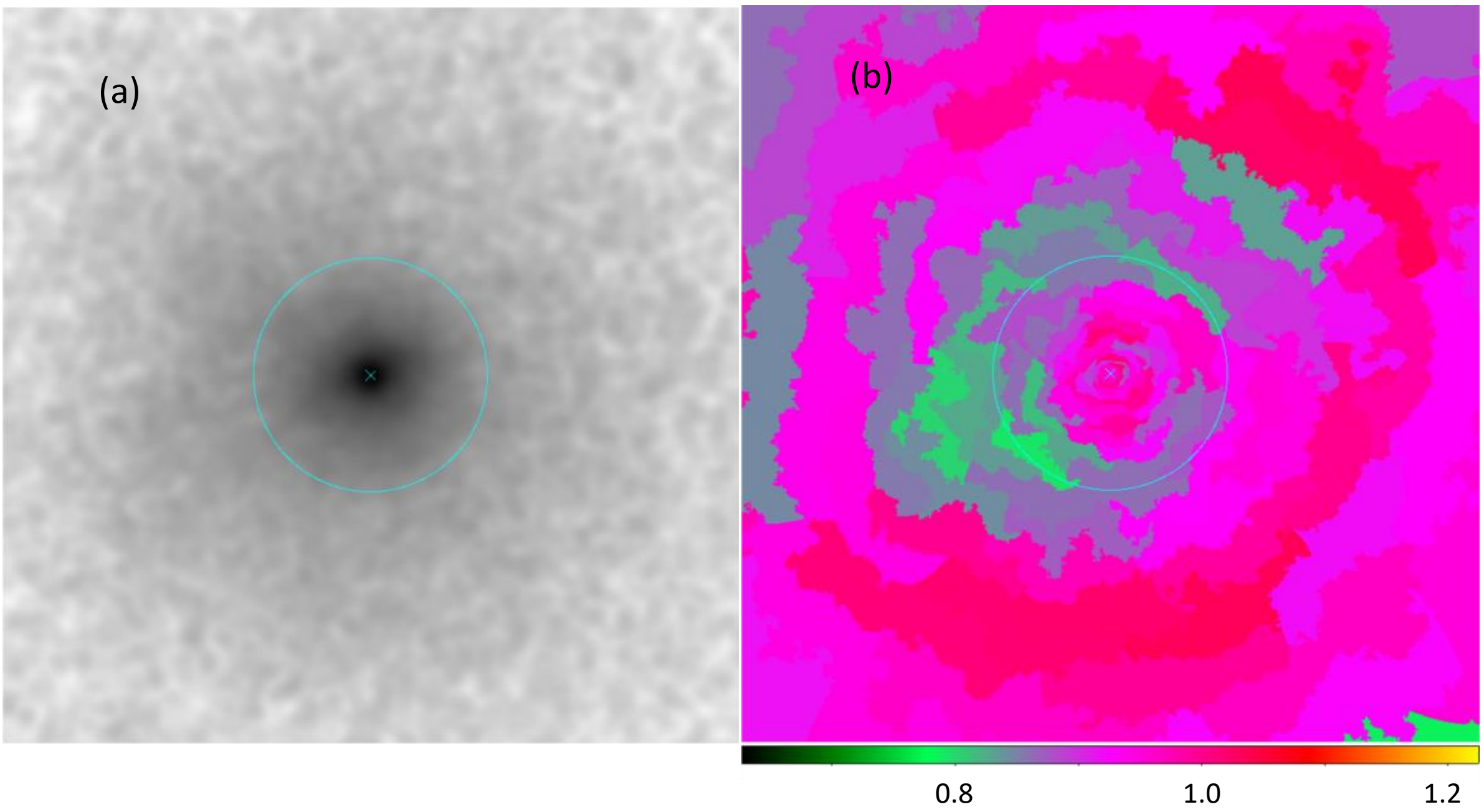

**Figure 9**. (a) The diffuse image (0,5-2 keV) and (b) temperature map (CB) of the hot gas in NGC 3402. The cyan ellipse indicates the optical $D_{25}$ ellipse.

## 5.1 Hot Gas Morphology

The high spatial resolution Chandra data show various spatial features in the hot ISM which were previously considered as smooth and relaxed. Those features seen in the X-ray surface brightness maps include X-ray jets (e.g., NGC 315), cavities (e.g., NGC 5813), cold fronts (e.g., NGC 1404), filaments (e.g., NGC 1399), and tails (e.g., NGC 7619). Furthermore, new features can be identified in the spectral maps which are not readily visible in the surface brightness maps. An excellent example is NGC 3402 (as reported by O'Sullivan et al. 2007), where the X-ray surface brightness of the diffuse gas appears to be relaxed, but the temperature map clearly indicates a cooler shell structure at 20-40 kpc (see Figure 9 and additional notes in Appendix B).

Another good example is NGC 4649 (see Figure 8). The top row shows the asymmetric hot gas distribution with the extended tail toward the north-east and the south-west (upper-left and lower-right, see also Wood et al. 2017). The first (AB) T map shows a rather complex profile. T decreases with increasing radius in the inner region and then increases in the outer region (see also Humphrey et al. 2013 and Paggi et al. 2014). The negative gradient in the central region is likely related to the AGN while the positive gradient in the outskirts is likely related to the hotter gas associated with the galaxy cluster in which NGC 4649 resides. The next three 2D T-maps further show more complex temperature structure; the cooler region is more extended toward the north-east and the south-west, indicating the asymmetrical features are extended from the center to the outskirts. In spite of this galaxy being extensively investigated, this 2D T structure was discovered for the first time in our program.

Another excellent example of new features discovered in our program is the disturbed fossil group galaxy NGC 1132. It is a well-known fossil group, i.e., an old system expected to have formed several Gyr ago, and to be completely relaxed by now. Instead, the hot gas morphology is disturbed and asymmetrical, with a cold front following a possible bow shock (Kim et al. 2018). We have found a few more interesting cases and will present them in separate papers.

## 5.2 Global Properties of Hot Gas

The global quantities of hot gaseous halos in early type galaxies have been extensively used in the literature to compare with data at other wavelengths and theoretical predictions to better understand the physical properties of the hot ISM and the evolution of the ETGs (e.g., Kim & Fabbiano 2013, 2015; Goulding et al. 2016; Negri et al. 2014, Choi, et al. 2017; Ciotti et al. 2017; Forbes et al. 2017). In Table 5, we list the hot gas X-ray luminosity $L_{X,GAS}$ and temperature $T_{GAS}$, measured within three radii, one effective radius (Re), five Re and $R_{MAX}$. The hot gas within one Re is closely related to the galaxy, as opposed to the large scale surrounding medium (e.g., Goulding et al. 2016; see also Voit et al. 2017 for the usage of $L_X$(Re) to compare with model predictions.) The second radius (5 x Re) is where the total mass can be reliably measured with optical mass tracers, for example, GCs and PNe (see Deason et al. 2012; Alabi et al. 2016) and often used in scaling relation to correlate with the total gas luminosity and temperature (Kim & Fabbiano 2013; Forbes et al. 2017). We provide $L_{X,GAS}$ within five Re so that the total mass and the global X-ray properties from the same region can be compared. Finally, $R_{MAX}$ is the maximum radius where the hot gas emission is reliably detected with an azimuthal coverage larger than 95% (calculated by the exposure map). In Table 5, we present the results from WB (WVT binning – see section 3.3) which can reflect the 2D spatial variation more accurately than AB (annulus binning). However, the results from AB are mostly consistent.

**Table 5. Global X-ray properties of hot ISM**

| name (1) | mid (2) | d (3) | Re_kpc (4) | Re5_kpc (5) | Rmax_kpc (6) | Lx1 (7) | eLx1 (8) | Lx5 (9) | eLx5 (10) | Lxm (11) | eLxm (12) | Tx1 (13) | eTx1 (14) | Tx5 (15) | eTx5 (16) | Txm (17) | eTxm (18) |
|---|---|---|---|---|---|---|---|---|---|---|---|---|---|---|---|---|---|
| I1262 | 90401 | 130.0 | 7.7 | 38.6 | 271.1 | 34.268 | 2.568 | 356.478 | 11.499 | 1194.051 | 25.122 | 1.48 | 0.06 | 1.56 | 0.02 | 1.68 | 0.02 |
| I1459 | 02196 | 29.2 | 5.2 | 26.2 | 19.9 | 1.329 | 0.068 | --- | --- | 2.537 | 0.100 | 0.51 | 0.02 | --- | --- | 0.49 | 0.01 |
| I1860 | 10537 | 93.8 | 8.4 | 42.0 | 118.2 | 48.288 | 2.101 | 197.020 | 6.315 | 360.236 | 10.236 | 1.07 | 0.02 | 1.28 | 0.01 | 1.37 | 0.01 |
| I4296 | 03394 | 50.8 | 11.9 | 59.4 | 32.0 | 14.834 | 0.539 | --- | --- | 22.614 | 1.018 | 0.79 | 0.01 | --- | --- | 0.81 | 0.01 |
| N0193 | 90201 | 47.0 | 4.4 | 22.1 | 61.5 | 0.437 | 0.054 | 10.611 | 0.237 | 16.006 | 0.341 | 0.67 | 0.05 | 0.84 | 0.01 | 0.86 | 0.01 |
| N0315 | 90201 | 69.8 | 12.5 | 62.5 | 47.4 | 19.553 | 0.609 | --- | --- | 24.168 | 0.753 | 0.64 | 0.01 | --- | --- | 0.67 | 0.01 |
| N0383 | 02147 | 63.4 | 6.3 | 31.5 | 83.0 | 5.479 | 0.325 | 18.359 | 1.846 | --- | --- | 0.76 | 0.03 | 1.35 | 0.06 | --- | --- |
| N0499 | 90401 | 54.5 | 4.4 | 21.8 | 71.3 | 13.034 | 0.420 | 68.994 | 0.886 | 125.578 | 1.480 | 0.82 | 0.01 | 0.74 | 0.01 | 0.80 | 0.01 |
| N0507 | 90201 | 63.8 | 12.9 | 64.4 | 133.0 | 39.935 | 1.268 | 254.337 | 4.528 | 391.513 | 6.278 | 1.14 | 0.01 | 1.27 | 0.01 | 1.30 | 0.01 |
| N0533 | 02880 | 76.9 | 16.2 | 81.1 | 96.9 | 81.607 | 1.702 | 154.586 | 3.651 | 167.047 | 4.073 | 0.96 | 0.01 | 1.12 | 0.01 | 1.13 | 0.01 |
| N0720 | 90401 | 27.7 | 4.8 | 24.1 | 45.6 | 2.557 | 0.057 | 7.264 | 0.107 | 8.547 | 0.138 | 0.61 | 0.01 | 0.56 | 0.01 | 0.54 | 0.01 |
| N0741 | 02223 | 70.9 | 13.2 | 65.9 | 89.4 | 21.661 | 1.010 | 45.690 | 3.434 | 51.319 | 3.682 | 0.87 | 0.02 | 1.19 | 0.04 | 1.16 | 0.03 |
| N1052 | 05910 | 19.4 | 3.1 | 15.7 | 24.5 | 0.236 | 0.024 | 0.418 | 0.042 | --- | --- | 0.49 | 0.03 | 0.41 | 0.02 | --- | --- |
| N1132 | 90201 | 95.0 | 15.5 | 77.5 | 129.0 | 39.365 | 1.944 | 170.013 | 4.270 | 239.697 | 5.408 | 1.03 | 0.02 | 1.09 | 0.01 | 1.09 | 0.01 |
| N1316 | 02022 | 21.5 | 7.6 | 38.1 | 14.6 | 4.039 | 0.107 | --- | --- | 5.137 | 0.131 | 0.64 | 0.01 | --- | --- | 0.62 | 0.01 |
| N1332 | 90201 | 22.9 | 3.0 | 15.3 | 28.9 | 1.592 | 0.052 | 2.408 | 0.071 | 2.754 | 0.084 | 0.65 | 0.02 | 0.62 | 0.01 | 0.63 | 0.01 |
| N1380 | 09526 | 17.6 | 3.2 | 16.1 | 22.2 | 0.463 | 0.027 | 0.847 | 0.062 | 1.003 | 0.078 | 0.33 | 0.01 | 0.48 | 0.02 | 0.55 | 0.02 |
| N1387 | 04168 | 20.3 | 3.5 | 17.5 | 42.4 | 2.858 | 0.361 | 5.750 | 0.425 | 14.045 | 0.589 | 0.45 | 0.05 | 0.54 | 0.03 | 0.94 | 0.02 |
| N1395 | 00799 | 24.1 | 5.4 | 27.2 | 39.7 | 1.772 | 0.144 | 5.052 | 0.291 | 7.053 | 0.377 | 0.82 | 0.04 | 0.87 | 0.03 | 0.89 | 0.02 |
| N1399 | 90602 | 19.9 | 4.7 | 23.5 | 47.4 | 12.327 | 0.094 | 40.498 | 0.213 | 67.533 | 0.523 | 1.03 | 0.00 | 1.29 | 0.00 | 1.38 | 0.00 |
| N1400 | 90202 | 26.4 | 2.9 | 14.7 | 34.6 | 0.361 | 0.051 | 0.898 | 0.079 | --- | --- | 0.49 | 0.06 | 0.55 | 0.04 | --- | --- |
| N1404 | 91101 | 21.0 | 2.7 | 13.7 | 60.0 | 7.820 | 0.046 | 14.845 | 0.069 | --- | --- | 0.65 | 0.00 | 0.63 | 0.00 | --- | --- |
| N1407 | 00791 | 28.8 | 8.9 | 44.3 | 36.4 | 8.605 | 0.194 | --- | --- | 14.060 | 0.414 | 0.86 | 0.01 | --- | --- | 1.01 | 0.01 |
| N1550 | 90401 | 51.1 | 6.3 | 31.7 | 104.0 | 62.400 | 0.977 | 248.324 | 2.885 | 506.931 | 5.534 | 1.11 | 0.01 | 1.29 | 0.00 | 1.34 | 0.01 |
| N1553 | 00783 | 18.5 | 5.1 | 25.7 | 24.3 | 0.448 | 0.040 | --- | --- | 2.222 | 0.095 | 0.42 | 0.04 | --- | --- | 0.35 | 0.01 |
| N1600 | 90201 | 57.4 | 13.5 | 67.5 | 72.3 | 12.356 | 0.774 | 32.174 | 1.871 | 33.050 | 1.953 | 1.13 | 0.03 | 1.43 | 0.04 | 1.43 | 0.04 |
| N1700 | 02069 | 44.3 | 3.9 | 19.3 | 57.9 | 4.994 | 0.227 | 9.709 | 0.386 | 9.709 | 0.386 | 0.50 | 0.02 | 0.43 | 0.01 | 0.43 | 0.01 |
| N2300 | 90201 | 30.4 | 4.8 | 24.2 | 39.8 | 4.173 | 0.154 | 8.251 | 0.301 | 10.474 | 0.377 | 0.63 | 0.02 | 0.75 | 0.02 | 0.81 | 0.02 |
| N2563 | 07925 | 67.8 | 6.4 | 31.9 | 124.9 | 10.125 | 0.708 | 20.832 | 1.962 | 70.908 | 3.934 | 0.95 | 0.03 | 1.24 | 0.05 | 1.48 | 0.04 |
| N3115 | 91101 | 9.7 | 1.6 | 8.1 | 14.6 | 0.028 | 0.001 | 0.049 | 0.003 | 0.064 | 0.007 | 0.46 | 0.02 | 0.39 | 0.01 | 0.34 | 0.01 |
| N3379 | 90501 | 10.6 | 2.4 | 12.1 | 14.3 | 0.023 | 0.002 | 0.059 | 0.005 | --- | --- | 0.44 | 0.02 | 0.34 | 0.01 | --- | --- |
| N3402 | 03243 | 64.9 | 8.8 | 44.1 | 78.7 | 106.600 | 2.099 | 270.035 | 3.075 | 342.002 | 3.525 | 0.95 | 0.01 | 0.92 | 0.00 | 0.92 | 0.00 |
| N3607 | 02073 | 22.8 | 5.0 | 25.1 | 35.4 | 0.310 | 0.049 | 2.559 | 0.159 | --- | --- | 0.63 | 0.10 | 0.49 | 0.03 | --- | --- |
| N3608 | 02073 | 22.9 | 3.3 | 16.5 | 38.9 | 0.000 | 0.000 | 0.837 | 0.088 | --- | --- | 1.00 | 0.00 | 0.47 | 0.04 | --- | --- |
| N3842 | 04189 | 97.0 | 17.8 | 88.9 | 80.0 | 7.977 | 1.099 | --- | --- | --- | --- | 1.04 | 0.04 | --- | --- | --- | --- |
| N3923 | 90201 | 22.9 | 5.8 | 29.2 | 35.5 | 3.646 | 0.069 | 5.313 | 0.115 | 5.422 | 0.136 | 0.51 | 0.01 | 0.49 | 0.01 | 0.49 | 0.01 |
| N4104 | 06939 | 120.0 | 20.0 | 100.2 | 157.1 | 79.862 | 5.307 | 175.693 | 12.780 | 236.625 | 15.603 | 1.29 | 0.04 | 1.49 | 0.05 | 1.52 | 0.05 |
| N4125 | 02071 | 23.9 | 5.9 | 29.5 | 30.1 | 1.834 | 0.051 | 4.211 | 0.090 | 4.211 | 0.090 | 0.46 | 0.01 | 0.40 | 0.01 | 0.40 | 0.01 |
| N4261 | 90201 | 31.6 | 6.9 | 34.3 | 39.9 | 6.624 | 0.157 | 11.791 | 0.331 | 12.507 | 0.358 | 0.70 | 0.01 | 0.96 | 0.01 | 0.97 | 0.01 |
| N4278 | 90901 | 16.1 | 2.6 | 13.0 | 24.9 | 0.307 | 0.008 | 0.502 | 0.012 | 0.706 | 0.018 | 0.41 | 0.01 | 0.42 | 0.01 | 0.41 | 0.01 |
| N4291 | 11778 | 26.2 | 2.0 | 10.2 | 33.0 | 0.307 | 0.038 | 4.323 | 0.168 | 6.832 | 0.252 | 0.57 | 0.09 | 0.48 | 0.02 | 0.56 | 0.01 |
| N4325 | 03232 | 110.0 | 10.5 | 52.3 | 138.7 | 149.721 | 3.476 | 562.512 | 7.153 | 743.875 | 8.733 | 0.82 | 0.01 | 0.93 | 0.00 | 0.95 | 0.00 |
| N4342 | 90201 | 16.5 | 0.5 | 2.3 | 21.6 | 0.003 | 0.003 | 0.185 | 0.008 | 1.343 | 0.029 | 0.86 | 0.38 | 0.61 | 0.02 | 0.66 | 0.01 |
| N4374 | 90301 | 18.4 | 5.5 | 27.3 | 24.9 | 4.752 | 0.050 | --- | --- | 8.153 | 0.157 | 0.68 | 0.00 | --- | --- | 0.92 | 0.01 |
| N4382 | 02016 | 18.4 | 7.4 | 37.0 | 24.1 | 0.851 | 0.037 | --- | --- | 1.999 | 0.070 | 0.41 | 0.01 | --- | --- | 0.37 | 0.01 |
| N4406 | 90202 | 17.1 | 10.3 | 51.7 | 30.8 | 11.955 | 0.168 | --- | --- | 46.383 | 0.487 | 0.86 | 0.01 | --- | --- | 0.93 | 0.00 |
| N4438 | 90201 | 18.0 | 5.0 | 24.8 | 18.3 | 0.986 | 0.053 | --- | --- | 2.807 | 0.242 | 0.61 | 0.02 | --- | --- | 0.83 | 0.03 |
| N4472 | 90501 | 16.3 | 8.2 | 41.2 | 38.7 | 14.562 | 0.054 | --- | --- | 34.672 | 0.221 | 0.99 | 0.00 | --- | --- | 1.14 | 0.00 |
| N4477 | 90401 | 16.5 | 3.5 | 17.6 | 21.6 | 0.643 | 0.025 | 1.331 | 0.058 | --- | --- | 0.36 | 0.01 | 0.64 | 0.02 | --- | --- |
| N4526 | 03925 | 16.9 | 3.3 | 16.7 | 21.3 | 0.267 | 0.021 | 0.446 | 0.038 | --- | --- | 0.38 | 0.02 | 0.42 | 0.02 | --- | --- |
| N4552 | 90401 | 15.3 | 3.0 | 15.3 | 22.3 | 1.927 | 0.022 | 2.523 | 0.032 | 2.774 | 0.041 | 0.60 | 0.01 | 0.60 | 0.01 | 0.65 | 0.01 |
| N4555 | 02884 | 91.5 | 13.2 | 66.1 | 115.3 | 11.911 | 0.822 | 39.345 | 2.110 | 51.000 | 2.488 | 0.85 | 0.03 | 1.03 | 0.02 | 1.00 | 0.02 |
| N4594 | 90301 | 9.8 | 3.4 | 16.9 | 26.1 | 0.201 | 0.006 | 0.636 | 0.025 | 0.757 | 0.047 | 0.55 | 0.01 | 0.49 | 0.01 | 0.46 | 0.01 |
| N4636 | 90401 | 14.7 | 6.7 | 33.3 | 34.8 | 14.122 | 0.055 | 28.106 | 0.090 | 28.381 | 0.091 | 0.69 | 0.00 | 0.79 | 0.00 | 0.79 | 0.00 |
| N4649 | 90601 | 16.8 | 6.2 | 31.2 | 24.5 | 10.397 | 0.049 | --- | --- | 13.807 | 0.065 | 0.90 | 0.00 | --- | --- | 0.91 | 0.00 |
| N4782 | 03220 | 60.0 | 4.4 | 21.8 | 75.6 | 1.470 | 0.168 | 8.614 | 0.795 | 24.721 | 2.138 | 0.57 | 0.05 | 0.97 | 0.04 | 1.15 | 0.04 |
| N5044 | 90201 | 31.2 | 3.9 | 19.3 | 40.8 | 24.859 | 0.221 | 186.834 | 0.683 | 289.017 | 1.007 | 0.84 | 0.00 | 0.94 | 0.00 | 1.03 | 0.00 |
| N5129 | 90201 | 103.0 | 14.3 | 71.6 | 144.8 | 29.011 | 1.220 | 88.676 | 2.581 | 147.756 | 3.629 | 0.81 | 0.02 | 0.90 | 0.01 | 0.78 | 0.01 |
| N5171 | 03216 | 100.0 | 12.4 | 62.0 | 111.5 | 2.735 | 0.550 | 25.561 | 2.226 | 59.950 | 3.138 | 0.96 | 0.07 | 0.98 | 0.04 | 0.86 | 0.02 |
| N5813 | 90901 | 32.2 | 8.3 | 41.6 | 43.7 | 37.910 | 0.102 | 108.779 | 0.170 | 110.017 | 0.173 | 0.65 | 0.00 | 0.71 | 0.00 | 0.71 | 0.00 |
| N5846 | 90201 | 24.9 | 7.2 | 35.8 | 50.7 | 15.893 | 0.140 | 40.604 | 0.291 | 44.409 | 0.339 | 0.70 | 0.00 | 0.82 | 0.00 | 0.82 | 0.00 |
| N5866 | 02879 | 15.3 | 2.8 | 14.2 | 19.4 | 0.125 | 0.014 | 0.262 | 0.023 | --- | --- | 0.36 | 0.03 | 0.35 | 0.02 | --- | --- |
| N6107 | 08180 | 127.9 | 16.3 | 81.5 | 155.0 | 22.905 | 3.944 | 68.813 | 12.564 | 136.801 | 17.111 | 1.22 | 0.06 | 1.39 | 0.09 | 1.86 | 0.17 |
| N6338 | 04194 | 123.0 | 17.1 | 85.3 | 286.3 | 333.765 | 15.807 | 650.347 | 35.560 | 1076.741 | 46.423 | 1.49 | 0.03 | 1.74 | 0.04 | 1.79 | 0.04 |
| N6482 | 03218 | 58.4 | 6.3 | 31.7 | 76.5 | 42.411 | 1.435 | 80.526 | 1.818 | 104.898 | 2.333 | 0.83 | 0.01 | 0.72 | 0.01 | 0.66 | 0.01 |
| N6861 | 90201 | 28.1 | 3.1 | 15.5 | 62.6 | 2.098 | 0.095 | 5.331 | 0.235 | 19.839 | 0.424 | 0.87 | 0.02 | 1.10 | 0.02 | 1.08 | 0.01 |
| N6868 | 90201 | 26.8 | 3.9 | 19.6 | 62.3 | 1.161 | 0.057 | 7.196 | 0.140 | 20.430 | 0.366 | 0.72 | 0.03 | 0.70 | 0.01 | 0.69 | 0.01 |
| N7618 | 90301 | 74.0 | 7.8 | 38.8 | 122.0 | 28.334 | 0.862 | 104.457 | 1.507 | 207.757 | 3.241 | 0.94 | 0.01 | 0.87 | 0.01 | 0.89 | 0.01 |
| N7619 | 90201 | 53.0 | 8.8 | 44.1 | 105.3 | 16.562 | 0.466 | 40.008 | 0.881 | 69.881 | 1.928 | 0.83 | 0.01 | 0.92 | 0.01 | 0.99 | 0.01 |
| N7626 | 02074 | 56.0 | 12.0 | 60.1 | 103.2 | 6.595 | 1.241 | 14.096 | 1.577 | 29.785 | 2.427 | 0.56 | 0.08 | 0.79 | 0.05 | 0.95 | 0.04 |

Column 1. Galaxy name (NGC or IC name)
Column 2. Merge id (or mid in short)
Column 3. Distance in Mpc (same as in Table 2)
Column 4. Re (effective radius) in kpc

```
Column 5. 5 x Re in kpc
Column 6. Rmax in kpc (the maximum radius to which the hot gas can be measured)
Column 7, 9, 11  LX,GAS within three radii given in Column 4, 5, 6.
Column 8, 10, 12  error in LX,GAS within three radii given in Column 4, 5, 6.
Column 13, 15, 17  TGAS within three radii given in Column 4, 5, 6.
Column 14, 16, 18  error in TGAS within three radii given in Column 4, 5, 6.
```

We note that if the hot gas is extended beyond $R_{MAX}$, particularly for hot gas rich ETGs, the luminosity within $R_{MAX}$ is less than the total gas luminosity (see below). We also note that $L_X$ within a given radius is a projected quantity (i.e., 2D rather than 3D), such that $L_{X,GAS}(< R)$ is the gas luminosity from a cylindrical volume, rather than a spherical volume. As we are working on de-projecting the hot gas properties, we will present the 3D $L_{X,GAS}$ within a given 3D radius in the near future.

In some gas poor galaxies where the X-ray emission is dominated by the central region and the hot gas is not extended, we stop at five Re and do not list the quantities at $R_{MAX}$. In some nearby large galaxies where 5 Re > $R_{MAX}$, we do not list the quantities at five Re.

Since we excluded the regions of the detected point sources (see section 3), we correct $L_{X,GAS}$ in each spatial bin by the ratio of areas with and without the point sources before summing them up within a given radius. This correction is 20% for NGC 3115 where a large number of point sources are detected in the ultra-deep (1 Msec) Chandra observations, and the amount of hot ISM is small. In all the other cases, the correction is less than 10%. We have also excluded the contribution from nearby galaxies if they are within a given radius. The X-ray luminosity of the nearby galaxy is determined mostly by CB (contour binning – see section 3.3) which makes a separate region around the nearby galaxy.

The gas temperature is also measured within the same three radii - one Re, five Re, and $R_{MAX}$. Since the region of interest consists of many spatial bins, we take an $L_X$-weighted average temperature, but in a logarithmic scale (i.e., geometric average instead of arithmetic average). For line-dominated spectra (for T ≤ 2 keV), the temperature is mainly determined by the energy of the peak intensity and this energy is linearly proportional to the plasma temperature in a log scale (see Figure 2 in Vikhlinin 2006). In this case, the geometric average (rather than the arithmetic average) is close (within a few %) to what one would measure by a single spectral fit with a single spectrum extracted from the region of interest (e.g., r < 1 Re, or r < 5Re), often called $T_{SPEC}$ (see Figure 3 in Vikhlinin 2006). The temperature in Table 5 are calculated by $L_X$-weighted averaging log(T), where T is the gas temperature of each bin. We compare the geometric and arithmetic average values and find that the difference is usually less than 6%.

For completeness for hot gas rich galaxies, in Table 6 we also list the total gas luminosity and temperature, compiled from the literature. We plan to include XMM-Newton data to supplement our Chandra results so that we can obtain complete gas properties by adding the outskirts of the extended hot halos. We will present the XMM results in a separate paper.

**Table 6. Total luminosity and temperature of hot ISM**

| name | d | $L_{X,GAS}$ | Lxgas_lo | Lxgas_up | $T_{GAS}$ | Tgas_lo | Tgas_up | ref | note |
|---|---|---|---|---|---|---|---|---|---|
| I1262 | 130.20 | 1639.585 | 245.937 | 3033.234 | 1.30 | 1.29 | 1.31 | 3 | |
| I1459 | 29.24 | 4.311 | 1.103 | 7.520 | 0.48 | 0.46 | 0.50 | 3 | |
| I1860 | 93.76 | 499.164 | 498.177 | 500.150 | 1.37 | 1.36 | 1.38 | 8 | |
| I4296 | 50.82 | 10.669 | 6.789 | 14.549 | 0.88 | 0.86 | 0.90 | 3 | |
| N0193 | 47.00 | 14.939 | 10.158 | 19.719 | 0.77 | 0.76 | 0.78 | 3 | |
| N0315 | 69.80 | 8.858 | 5.654 | 12.063 | 0.64 | 0.63 | 0.65 | 3 | |
| N0383 | 63.39 | 18.353 | 16.507 | 20.198 | 1.35 | 1.29 | 1.41 | 7 | |
| N0499 | 54.45 | 172.161 | 86.284 | 343.506 | 0.70 | 0.67 | 0.72 | 6 | |
| N0507 | 63.80 | 790.331 | 78.005 | 80.060 | 1.32 | 1.31 | 1.33 | 5 | |
| N0533 | 76.90 | 94.275 | 59.904 | 128.647 | 0.98 | 0.97 | 0.99 | 3 | |
| N0720 | 27.67 | 5.060 | 4.998 | 5.122 | 0.54 | 0.53 | 0.55 | 2 | |
| N0741 | 70.90 | 25.774 | 15.303 | 36.245 | 0.96 | 0.94 | 0.98 | 3 | |
| N1052 | 19.41 | 0.437 | 0.409 | 0.465 | 0.34 | 0.32 | 0.36 | 2 | |
| N1132 | 95.00 | 716.528 | 682.408 | 750.648 | 1.08 | 1.07 | 1.09 | 4 | |
| N1316 | 21.48 | 5.350 | 5.210 | 5.490 | 0.60 | 0.59 | 0.61 | 2 | |
| N1332 | 22.91 | 2.401 | 1.514 | 3.805 | 0.41 | 0.36 | 0.47 | 6 | |
| N1380 | 17.62 | 1.054 | 0.105 | 2.647 | 0.30 | 0.25 | 0.38 | 6 | |
| N1387 | 20.32 | 4.269 | 3.873 | 4.665 | 0.41 | 0.38 | 0.44 | 7 | a |
| N1395 | 24.10 | 2.437 | 1.537 | 3.863 | 0.65 | 0.60 | 0.69 | 6 | |
| N1399 | 19.95 | 33.942 | 17.011 | 67.724 | 1.21 | 1.18 | 1.24 | 6 | |
| N1400 | 26.42 | 0.899 | 0.820 | 0.978 | 0.55 | 0.51 | 0.59 | 7 | |
| N1404 | 20.99 | 16.983 | 12.987 | 20.980 | 0.58 | 0.57 | 0.59 | 3 | |
| N1407 | 28.84 | 10.027 | 7.019 | 13.036 | 0.87 | 0.86 | 0.88 | 3 | |
| N1550 | 51.10 | 1466.156 | 138.698 | 154.532 | 1.33 | 1.33 | 1.33 | 5 | |
| N1553 | 18.54 | 2.812 | 0.200 | 5.423 | 0.41 | 0.40 | 0.42 | 3 | |
| N1600 | 57.40 | 33.050 | 31.097 | 35.003 | 1.43 | 1.39 | 1.47 | 7 | |
| N1700 | 44.26 | 9.691 | 9.306 | 10.076 | 0.43 | 0.42 | 0.44 | 7 | |
| N2300 | 30.40 | 14.442 | 12.292 | 22.890 | 0.62 | 0.48 | 0.69 | 6 | |
| N2563 | 67.80 | 70.908 | 66.974 | 74.842 | 1.48 | 1.44 | 1.52 | 7 | |
| N3115 | 9.68 | 0.025 | 0.020 | 0.032 | 0.44 | 0.34 | 0.60 | 2 | |
| N3379 | 10.57 | 0.044 | 0.037 | 0.050 | 0.25 | 0.23 | 0.27 | 1 | |
| N3402 | 60.40 | 512.093 | 488.816 | 535.370 | 0.96 | 0.94 | 0.98 | 4 | |
| N3607 | 22.80 | 1.688 | 1.494 | 1.882 | 0.59 | 0.52 | 0.66 | 1 | |
| N3608 | 22.91 | 0.437 | 0.367 | 0.507 | 0.40 | 0.34 | 0.49 | 1 | |
| N3842 | 97.00 | 7.977 | 6.878 | 9.076 | 1.04 | 1.00 | 1.08 | 7 | a |
| N3923 | 22.91 | 4.410 | 4.346 | 4.474 | 0.45 | 0.44 | 0.46 | 2 | |
| N4104 | 120.00 | 470.995 | 468.941 | 473.049 | 1.52 | 1.47 | 1.57 | 8 | |
| N4125 | 23.88 | 3.180 | 3.126 | 3.235 | 0.41 | 0.40 | 0.42 | 2 | |
| N4261 | 31.62 | 7.213 | 7.098 | 7.327 | 0.76 | 0.75 | 0.77 | 1 | |
| N4278 | 16.07 | 0.260 | 0.246 | 0.274 | 0.30 | 0.29 | 0.31 | 1 | |
| N4291 | 26.18 | 8.021 | 7.660 | 20.148 | 0.59 | 0.52 | 0.65 | 6 | |
| N4325 | 110.00 | 1257.102 | 1188.887 | 1325.317 | 1.00 | 0.99 | 1.01 | 4 | |
| N4342 | 16.50 | 0.164 | 0.151 | 0.178 | 0.59 | 0.55 | 0.64 | 1 | |
| N4374 | 18.37 | 6.556 | 5.397 | 7.716 | 0.73 | 0.73 | 0.74 | 1 | |
| N4382 | 18.45 | 1.040 | 1.003 | 1.077 | 0.39 | 0.37 | 0.40 | 1 | |
| N4406 | 17.14 | 132.997 | 131.808 | 134.185 | 0.82 | 0.81 | 0.83 | 1 | |
| N4438 | 18.00 | 2.807 | 2.565 | 3.049 | 0.83 | 0.80 | 0.86 | 7 | |
| N4472 | 16.29 | 21.982 | 20.186 | 23.778 | 0.95 | 0.94 | 0.96 | 1 | |
| N4477 | 16.50 | 0.926 | 0.887 | 0.964 | 0.33 | 0.32 | 0.34 | 1 | |
| N4526 | 16.90 | 0.298 | 0.278 | 0.317 | 0.31 | 0.29 | 0.32 | 1 | |
| N4552 | 15.35 | 2.033 | 1.951 | 2.115 | 0.59 | 0.59 | 0.60 | 1 | |
| N4555 | 91.50 | 51.000 | 48.512 | 53.488 | 1.00 | 0.98 | 1.02 | 7 | |
| N4594 | 9.77 | 0.752 | 0.705 | 0.799 | 0.46 | 0.45 | 0.47 | 7 | |
| N4636 | 14.66 | 33.362 | 32.707 | 34.017 | 0.73 | 0.72 | 0.73 | 1 | |
| N4649 | 16.83 | 17.238 | 15.781 | 18.696 | 0.86 | 0.86 | 0.86 | 1 | |
| N4782 | 60.00 | 24.721 | 22.583 | 26.859 | 1.15 | 1.11 | 1.19 | 7 | |
| N5044 | 31.19 | 259.833 | 179.884 | 339.782 | 0.91 | 0.90 | 0.92 | 3 | |
| N5129 | 103.00 | 397.846 | 32.981 | 46.587 | 0.81 | 0.80 | 0.82 | 5 | |
| N5171 | 100.00 | 59.950 | 56.812 | 63.088 | 0.86 | 0.84 | 0.88 | 7 | |
| N5813 | 32.21 | 74.822 | 74.461 | 75.167 | 0.70 | 0.69 | 0.70 | 1 | |
| N5846 | 24.89 | 53.434 | 52.264 | 54.603 | 0.72 | 0.72 | 0.73 | 1 | |
| N5866 | 15.35 | 0.267 | 0.245 | 0.289 | 0.32 | 0.30 | 0.34 | 1 | |
| N6107 | 127.90 | 1512.000 | 1391.000 | 1633.000 | 1.86 | 1.69 | 2.03 | 8 | |
| N6338 | 123.03 | 2711.934 | 2291.479 | 3132.389 | 1.97 | 1.88 | 2.06 | 4 | |
| N6482 | 58.40 | 167.126 | 39.323 | 294.928 | 0.74 | 0.73 | 0.75 | 3 | |
| N6861 | 28.05 | 19.768 | 19.345 | 20.190 | 1.08 | 1.07 | 1.09 | 7 | |
| N6868 | 26.79 | 20.414 | 20.049 | 20.780 | 0.69 | 0.68 | 0.70 | 7 | |
| N7618 | 74.00 | 211.327 | 128.634 | 294.021 | 0.80 | 0.79 | 0.81 | 3 | |
| N7619 | 52.97 | 77.939 | 12.352 | 155.509 | 0.81 | 0.78 | 0.84 | 6 | |
| N7626 | 56.00 | 14.096 | 12.519 | 15.673 | 0.79 | 0.74 | 0.84 | 7 | a |

----------------------------------------------------------------------------------------------------

Note:
$L_{X,GAS}$: X-ray luminosity of the entire hot gas
Lxgas_lo: lower bound of $L_{X,GAS}$
Lxgas_up: upper bound of $L_{X,GAS}$
$T_{GAS}$: Tempeature of the hot gas (determined by fitting a single spectrum extracted from the entire hot gas)
Tgas_lo: lower bound of $T_{GAS}$
Tgas_up  upper bound of $T_{GAS}$

a: gas properties are measured within limited radii because of nearby sources
  N1387    r < 2 Re  (inside Fornax cluster)
  N3842    r < 1 Re  (embedded in a cluster)
  N7626    r < 5 Re  (inside N7619 group)

references
1. Kim, D.-W. and Fabbiano, G., 2015, ApJ, 812, 127
2. Boroson, B. Kim, D.-W. and Fabbiano, G., 2011, ApJ, 729, 12
3. Diehl, S. and Statler, T.S., 2008, ApJ, 687, 986
4. Lovisari, L. Reiprich, T.H. and Schellenberger, G. , 2015, A&A, 573A, 118
5. Eckmiller, H.J. Hudson, D.S. and Reiprich, T.H. , 2011, A&A, 535A, 105
6. O'Sullivan, E. Ponman, T.J. and Collins, R.S., 2003, MNRAS, 340, 1375
7. this work
8. O'Sullivan, E. Forbes, D.A. and Ponman, T.J., 2001, MNRAS, 328, 461

### 5.3 Radial Profiles of Hot Gas Temperature

The temperature profile in groups and clusters are often similar to each other (e.g., Vikhlinin et al. 2005; Sun et al. 2009). In a typical relaxed group or cluster, the temperature increases inwards from the virial radius ($R_{VIR}$) and peaks to $T_{MAX}$ at r ~ 0.1 $R_{VIR}$ (or ~100 kpc). Moving further inwards, the temperature decreases inside the cooling radius, depending on the cooling time (or the gas density). In ETGs, a similar trend is seen as the temperature often peaks in the inner part of the halo, albeit at a smaller radius (r ~ a few x 10 kpc). In this regard, the global temperature profiles of ETGs qualitatively follow those of groups and clusters. We note that the temperature peak may not be visible, if the outskirts of the hot ISM are too faint, fall out of the detector, or are embedded inside hotter ambient ICM/IGM.

In contrast to the overall temperature profile, the inner temperature behavior appears to be more complex (see also O'Sullivan et al. 2017; Lakhchaura et al. 2018). Inside roughly r < a few kpc (or Re), the temperature gradient can be negative (a hot core) in some galaxies and positive (a cool core) in other galaxies. This behavior at the central region is likely related to stellar and AGN feedback in addition to gravity which is dominant in a large scale, although the exact heating/cooling mechanisms are yet to be understood. Possibilities include gravitational heating from the central SMBH, a recent AGN outburst, or interaction with confined nuclear jets (see Pellegrini et al. 2012; Paggi et al. 2014). The temperature and density profiles and their universality or their deviation from the universal profiles will be addressed in Traynor et al. (2019 in prep.).

### 5.4 X-ray based mass profile

X-ray observations of the hot ISM can be used to measure the total mass within a given radius by balancing the gravitational force and the pressure gradient in the hot gas under the assumption of hydrostatic equilibrium (HE). While this method has been extensively used, the assumption of HE is seriously in question because of various spatial features often found in ETGs. Recent studies suggest that non-thermal pressure is not negligible, typically consisting of 10-30% of total pressure, but can be as large as >40% (de Plaa et al. 2012).

Using the Chandra Galaxy Atlas data products, Paggi et al. (2017) confirmed the presence of a nonthermal pressure component accounting for ~30% of the gas pressure in the central region of NGC 4649, likely linked to nuclear activity. In NGC 5846 where the X-ray gas morphology shows significant azimuthal asymmetries, especially in the NE direction, Paggi et al. (2017) found substantial departures from HE in this direction, consistent with bulk gas compression and decompression (due to sloshing); this effect disappears in the NW direction, where the emission is smooth and extended.

We will apply this method to the entire sample by removing the pie sectors where the hot gas shows azimuthal asymmetries, discontinuities, etc. to derive the mass profiles to the extent that the HE is valid (Kim et al. in prep).

### 5.5 Low Mass X-ray Binaries

Although not our primary goal, the X-ray properties of the LMXBs in each galaxy are readily available as byproducts, specifically their photometric and spectral information (hardness ratios and colors). Chandra detected LMXBs in ETGs have been extensively used to understand the stellar populations of ETGs via X-ray luminosity functions (XLFs) (e.g., Fig 4 in Kim & Fabbiano 2010) and their connections to GCs (Kundu et al. 2002; Sarazin et al. 2003), to merger histories (D'Abrusco et al. 2014), and to the age of stellar populations (Kim & Fabbiano 2010; Zhang et al. 2012). Vrtilek and Boroson (2013) have developed a new method using CCI (color-color-intensity) maps, that have proved to be particularly apt when used with the Chandra data in order to separate different types of X-ray points sources. We will apply a similar method to our ETG sample and further compare the sample of starburst galaxies (Islam et al. in prep).

### 5.6 Planned Works

We plan to add the followings in the next version.

1) New galaxies. We will keep adding new galaxies and will post the same type of data products on the CGA website.
2) Fe map. The metal abundance measurement is reliable only for high s/n data. We will add the Fe maps of a limited sample of hot gas rich systems.
3) Circum-nuclear region. We will apply the sub-pixel resolution algorithm to produce high resolution images in a special region of interest.
4) XMM-Newton data. We will further extend our program by adding XMM-Newton to allow us to trace the full extent of hot gas halos. With the higher effective area and larger field of view of XMM-Newton, we will be able to adequately investigate the faint diffuse emission in the galaxy outskirts which are critical to understand the interaction with the surrounding medium (e.g., by ram pressure stripping) and neighboring galaxies (sloshing, merging), as measuring the hot gas properties and mass profile on as large a scale as possible.

We have extracted archival data from the Chandra Data Archive, and the data analysis was supported by the CXC CIAO software and CALDB. We have used the NASA NED and ADS facilities. The computations in this paper were conducted on the Smithsonian High Performance Cluster (SI/HPC). This work was supported by the Chandra GO grants (AR4-15005X and AR5-16007X), by Smithsonian Competitive Grant Program for Science, by Smithsonian 2018 Scholarly Study Program and by NASA contract NAS8- 03060 (CXC).

## REFERENCES

Boroson, B, Kim, D.-W., Fabbiano 2011, ApJ, 729, 12
Cappellari, M. Emsellem, E. Krajnovi'c, D. et al. 2011, MNRAS, 413, 813
Cappellari M., Copin Y., 2003, MNRAS, 342, 345
Choi, E. Ostriker, J.P. Naab, T. et al. 2017, ApJ, 844, 31
Ciotti, L. Pellegrini, S. Negri, A. et al. 2017, ApJ, 835, 15
D'Abrusco, R., Fabbiano, G., Mineo, S., et al. 2014, ApJ, 783, 18
de Plaa, J. et al. 2012, A&A, 539, A34
de Vaucouleurs G., de Vaucouleurs A., Corwin H.G., Buta R.J., Paturel G., Fouque P. 1991 “Third Reference Catalogue of Bright Galaxies (RC3)” (Springer-Verlag: New York)
Dickey, J.M., and Lockman, F.J., 1990, ARAA, 28, 215
Diehl, S. & Statler, T. S. 2006, MNRAS, 368, 497
Diehl, S. & Statler, T. S. 2007, ApJ, 668, 150
Diehl, S. & Statler, T. S. 2008a, ApJ, 680, 897
Diehl, S. & Statler, T. S. 2008b, ApJ, 687, 986
Eckmiller, H.J. Hudson, D.S. and Reiprich, T.H. , 2011, A&A, 535A, 105
Goulding, A.D. Greene, J.E. Ma, C.-P. et al.  2016, ApJ, 826, 167
Grevesse, N. & Sauval, A.J. 1998, Space Science Reviews 85, 161
Humphrey, P. J., et al. 2013, MNRAS, 430, 1516
Kim, D.-W. and Pellegrini, S. 2012 “Hot ISM in Elliptical Galaxies” Astrophysics and Space Science Library Vol. 378 (Springer: Berlin)
Kim, D. W. & Fabbiano, G. 2010, ApJ, 721, 1523
Kim, D.-W. & Fabbiano, G. 2013, ApJ, 776, 116
Kim, D.-W. & Fabbiano, G. 2015, ApJ, 812, 127
Kim, D.-W., Anderson, C., Burke, D., et al. 2018, ApJ, 853, 129
Kundu, A., et al. 2002, ApJ, 574, L5
Lakhchaura, K., Werner, N., Sun, M., et al. 2018, MNRAS, 481, 4472
Lovisari, L. Reiprich, T.H. and Schellenberger, G. , 2015, A&A, 573A, 118
Markevitch, M., et al. 2003, ApJ, 583, 70
Negri, A. Posacki, S. Pellegrini, S. et al. 2014, MNRAS, 445, 1351
O'Sullivan, E. Forbes, D.A. and Ponman, T.J., 2001, MNRAS, 328, 461
O'Sullivan, E. Ponman, T.J. and Collins, R.S., 2003, MNRAS, 340, 1375
O'Sullivan, E. Vrtilek, J.M. Harris, D.E. et al. 2007, ApJ, 658, 299
O'Sullivan, E. David, L.P. and Vrtilek, J.M., 2014, MNRAS, 437, 730
O'Sullivan, E., Ponman, T.J., Kolokythas, K., et al. 2017, MNRAS, 472, 1482
Paggi, A. Fabbiano, G. Kim, D.-W. et al. 2014, ApJ, 787, 134
Paggi, A. Kim, D.-W. Anderson, C. et al. 2017, ApJ, 844, 5

Pellegrini, S. Wang, J. Fabbiano, G. et al. 2012, ApJ, 758, 94
Sanders, J. S. 2006, MNRAS, 371, 829
Sarazin, C. L., et al. 2003, ApJ, 595, 743
Sun, M. Voit, G.M. Donahue, M. et al. 2009, ApJ, 693, 1142
Tonry, J. L., Dressler, A., Blakeslee, J. P., et al. 2001, ApJ, 546, 681
Tully, R.B. Courtois, H.M. Dolphin, A.E. et al. 2013, AJ, 146, 86
Vikhlinin, A. Markevitch, M. Murray, S.S. et al. 2005, ApJ, 628, 655
Vikhlinin, A., 2006, ApJ, 640, 710
Voit, G.M. Ma, C.P. Greene, J. et al. 2017, arXiv170802189
Vrtilek, S. D. & Boroson, B. S. 2013, MNRAS, 428, 3693
Werner, N., Allen, S. W., and Simionescu, A. 2012, MNRAS, 425, 2731
Wood, R.A., Jones, C., Machacek, M.E., et al. 2017, ApJ, 847, 79
Zhang, Z., et al. 2012, A&A, 54

Table 2 Chandra Observations

| obsid (1) | gname (2) | mid (3) | obs_date (4) | OAA (5) | detector (6) | ccdid (7) | target (8) | eff_exp (9) | notes (10) |
|---|---|---|---|---|---|---|---|---|---|
| 02018 | I1262 | 90401 | 2001-08-23 | 0.2 | ACIS-S | 6,7 | 7 | 28.7 | |
| 06949 | I1262 | 90401 | 2006-04-17 | 0.0 | ACIS-I | 0,1,2,3 | 3 | 38.6 | |
| 07321 | I1262 | 90401 | 2006-04-19 | 0.0 | ACIS-I | 0,1,2,3 | 3 | 35.0 | |
| 07322 | I1262 | 90401 | 2006-04-22 | 0.0 | ACIS-I | 0,1,2,3 | 3 | 36.5 | |
| 02196 | I1459 | 02196 | 2001-08-12 | 0.0 | ACIS-S | 6,7 | 7 | 52.5 | a c |
| 10537 | I1860 | 10537 | 2009-09-12 | 0.0 | ACIS-S | 6,7 | 7 | 36.8 | |
| 03394 | I4296 | 03394 | 2001-12-15 | 0.1 | ACIS-S | 6,7 | 7 | 24.3 | a |
| 04053 | N0193 | 90201 | 2003-09-01 | 0.0 | ACIS-S | 6,7 | 7 | 16.6 | |
| 11389 | N0193 | 90201 | 2009-08-21 | 0.0 | ACIS-S | 6,7 | 7 | 90.1 | |
| 04156 | N0315 | 90201 | 2003-02-22 | 0.0 | ACIS-S | 6,7 | 7 | 51.2 | a c |
| 00855 | N0315 | 90201 | 2000-10-08 | 0.0 | ACIS-S | 7 | 7 | 4.7 | b c |
| 02147 | N0383 | 02147 | 2000-11-06 | 0.0 | ACIS-S | 6,7 | 7 | 42.9 | |
| 10536 | N0499 | 90401 | 2009-02-12 | 0.0 | ACIS-S | 6,7 | 7 | 18.4 | |
| 10865 | N0499 | 90401 | 2009-02-04 | 0.0 | ACIS-S | 6,7 | 7 | 4.4 | |
| 10866 | N0499 | 90401 | 2009-02-05 | 0.0 | ACIS-S | 6,7 | 7 | 7.8 | |
| 10867 | N0499 | 90401 | 2009-02-07 | 0.0 | ACIS-S | 6,7 | 7 | 6.8 | |
| 00317 | N0507 | 90201 | 2000-10-11 | 0.2 | ACIS-S | 6,7 | 7 | 17.7 | |
| 02882 | N0507 | 90201 | 2002-01-08 | 0.4 | ACIS-I | 0,1,2,3 | 3 | 42.6 | |
| 02880 | N0533 | 02880 | 2002-07-28 | 0.0 | ACIS-S | 6,7 | 7 | 35.3 | |
| 07062 | N0720 | 90401 | 2006-10-09 | 0.0 | ACIS-S | 6,7 | 7 | 20.2 | |
| 07372 | N0720 | 90401 | 2006-08-06 | 0.0 | ACIS-S | 6,7 | 7 | 46.3 | |
| 08448 | N0720 | 90401 | 2006-10-12 | 0.0 | ACIS-S | 6,7 | 7 | 7.5 | |
| 08449 | N0720 | 90401 | 2006-10-12 | 0.0 | ACIS-S | 6,7 | 7 | 18.4 | |
| 02223 | N0741 | 02223 | 2001-01-28 | 0.0 | ACIS-S | 6,7 | 7 | 29.1 | |
| 05910 | N1052 | 05910 | 2005-09-18 | 0.0 | ACIS-S | 6,7 | 7 | 57.2 | |
| 00801 | N1132 | 90201 | 1999-12-10 | 0.1 | ACIS-S | 6,7 | 7 | 11.7 | |
| 03576 | N1132 | 90201 | 2003-11-16 | 0.0 | ACIS-S | 6,7 | 7 | 26.6 | |
| 02022 | N1316 | 02022 | 2001-04-17 | 0.0 | ACIS-S | 6,7 | 7 | 23.6 | |
| 02915 | N1332 | 90201 | 2002-09-18 | 0.1 | ACIS-S | 6,7 | 7 | 4.6 | |
| 04372 | N1332 | 90201 | 2002-09-19 | 0.1 | ACIS-S | 6,7 | 7 | 49.1 | |
| 09526 | N1380 | 09526 | 2008-03-26 | 0.1 | ACIS-S | 6,7 | 7 | 37.0 | |
| 04168 | N1387 | 04168 | 2003-05-20 | 0.9 | ACIS-I | 0,1,2,3 | 3 | 45.4 | |
| 00799 | N1395 | 00799 | 1999-12-31 | 0.0 | ACIS-I | 0,1,2,3 | 3 | 16.4 | |
| 00239 | N1399 | 90602 | 2000-01-19 | 0.1 | ACIS-I | 0,1,2,3 | 3 | 3.6 | |
| 00319 | N1399 | 90602 | 2000-01-18 | 0.1 | ACIS-S | 6,7 | 7 | 51.9 | |

| | | | | | | | | |
|---|---|---|---|---|---|---|---|---|
| 04172 | N1399 | 90602 | 2003-05-26 | 1.5 | ACIS-I | 0,1,2,3 | 3 | 41.4 |
| 09530 | N1399 | 90602 | 2008-06-08 | 0.0 | ACIS-S | 6,7 | 7 | 56.5 |
| 14527 | N1399 | 90602 | 2013-07-01 | 0.0 | ACIS-S | 6,7 | 7 | 24.5 |
| 16639 | N1399 | 90602 | 2014-10-12 | 0.0 | ACIS-S | 6,7 | 7 | 28.9 |
| | | | | | | | | |
| 14033 | N1400 | 90202 | 2012-06-17 | 2.7 | ACIS-S | 7 | 7 | 49.7 |
| 07849 | N1400 | 90202 | 2007-07-11 | 0.0 | ACIS-S | 6,7 | 7 | 4.9 |
| | | | | | | | | |
| 02942 | N1404 | 91101 | 2003-02-13 | 0.1 | ACIS-S | 6,7 | 7 | 28.2 |
| 04174 | N1404 | 91101 | 2003-05-28 | 1.2 | ACIS-I | 0,1,2,3 | 3 | 43.6 |
| 09798 | N1404 | 91101 | 2007-12-24 | 1.2 | ACIS-S | 6,7 | 7 | 18.2 |
| 09799 | N1404 | 91101 | 2007-12-27 | 1.2 | ACIS-S | 6,7 | 7 | 16.4 |
| 16231 | N1404 | 91101 | 2014-10-20 | 1.9 | ACIS-S | 6,7 | 7 | 57.9 |
| 16232 | N1404 | 91101 | 2014-11-12 | 2.5 | ACIS-S | 6,7 | 7 | 67.6 |
| 16233 | N1404 | 91101 | 2014-11-09 | 2.5 | ACIS-S | 6,7 | 7 | 95.7 |
| 17540 | N1404 | 91101 | 2014-11-02 | 1.9 | ACIS-S | 6,7 | 7 | 28.1 |
| 17541 | N1404 | 91101 | 2014-10-23 | 1.9 | ACIS-S | 6,7 | 7 | 23.7 |
| 17548 | N1404 | 91101 | 2014-11-11 | 2.5 | ACIS-S | 6,7 | 7 | 47.5 |
| 17549 | N1404 | 91101 | 2015-03-28 | 2.7 | ACIS-S | 6,7 | 7 | 59.4 |
| | | | | | | | | |
| 00791 | N1407 | 00791 | 2000-08-16 | 0.0 | ACIS-S | 6,7 | 7 | 41.9 |
| | | | | | | | | |
| 03186 | N1550 | 90401 | 2002-01-08 | 0.9 | ACIS-I | 0,1,2,3 | 3 | 10.0 |
| 03187 | N1550 | 90401 | 2002-01-08 | 0.9 | ACIS-I | 0,1,2,3 | 1 | 9.6 |
| 05800 | N1550 | 90401 | 2005-10-22 | 4.0 | ACIS-S | 6,7 | 7 | 44.3 |
| 05801 | N1550 | 90401 | 2005-10-24 | 4.1 | ACIS-S | 6,7 | 7 | 42.9 |
| | | | | | | | | |
| 00783 | N1553 | 00783 | 2000-01-02 | 0.0 | ACIS-S | 6,7 | 7 | 16.6 |
| | | | | | | | | |
| 04283 | N1600 | 90201 | 2002-09-18 | 0.0 | ACIS-S | 6,7 | 7 | 21.9 |
| 04371 | N1600 | 90201 | 2002-09-20 | 0.0 | ACIS-S | 6,7 | 7 | 26.8 |
| | | | | | | | | |
| 02069 | N1700 | 02069 | 2000-11-03 | 0.1 | ACIS-S | 6,7 | 7 | 29.5 |
| | | | | | | | | |
| 04968 | N2300 | 90201 | 2004-06-23 | 6.6 | ACIS-S | 6,7 | 6 | 44.5 |
| 15648 | N2300 | 90201 | 2013-05-24 | 7.0 | ACIS-S | 6,7 | 6 | 21.2 |
| | | | | | | | | |
| 07925 | N2563 | 07925 | 2007-09-18 | 2.9 | ACIS-I | 0,1,2,3 | 1 | 48.1 |
| | | | | | | | | |
| 13820 | N3115 | 91101 | 2012-01-31 | 0.0 | ACIS-S | 6,7 | 7 | 179.1 |
| 02040 | N3115 | 91101 | 2001-06-14 | 0.1 | ACIS-S | 6,7 | 7 | 33.5 |
| 11268 | N3115 | 91101 | 2010-01-27 | 0.0 | ACIS-S | 6,7 | 7 | 40.1 |
| 12095 | N3115 | 91101 | 2010-01-29 | 0.0 | ACIS-S | 6,7 | 7 | 74.1 |
| 13817 | N3115 | 91101 | 2012-01-18 | 0.0 | ACIS-S | 6,7 | 7 | 166.3 |
| 13819 | N3115 | 91101 | 2012-01-26 | 0.0 | ACIS-S | 6,7 | 7 | 68.6 |
| 13821 | N3115 | 91101 | 2012-02-03 | 0.0 | ACIS-S | 6,7 | 7 | 151.1 |
| 13822 | N3115 | 91101 | 2012-01-21 | 0.0 | ACIS-S | 6,7 | 7 | 148.4 |
| 14383 | N3115 | 91101 | 2012-04-04 | 0.0 | ACIS-S | 6,7 | 7 | 117.4 |
| 14384 | N3115 | 91101 | 2012-04-06 | 0.0 | ACIS-S | 6,7 | 7 | 67.4 |
| 14419 | N3115 | 91101 | 2012-04-05 | 0.0 | ACIS-S | 6,7 | 7 | 42.5 |
| | | | | | | | | |
| 01587 | N3379 | 90501 | 2001-02-13 | 0.1 | ACIS-S | 6,7 | 7 | 30.5 |
| 07073 | N3379 | 90501 | 2006-01-23 | 0.0 | ACIS-S | 6,7 | 7 | 82.3 |
| 07074 | N3379 | 90501 | 2006-04-09 | 0.0 | ACIS-S | 6,7 | 7 | 68.3 |
| 07075 | N3379 | 90501 | 2006-07-03 | 0.0 | ACIS-S | 6,7 | 7 | 80.0 |
| 07076 | N3379 | 90501 | 2007-01-10 | 0.0 | ACIS-S | 6,7 | 7 | 64.1 |
| | | | | | | | | |
| 03243 | N3402 | 03243 | 2002-11-05 | 0.3 | ACIS-S | 6,7 | 7 | 28.0 |
| | | | | | | | | |
| 02073 | N3607 | 02073 | 2001-06-12 | 3.6 | ACIS-I | 0,1,2,3 | 2 | 38.5 |

| | | | | | | | | | |
|---|---|---|---|---|---|---|---|---|---|
| 02073 | N3608 | 02073 | 2001-06-12 | 2.3 | ACIS-I | 0,1,2,3 | 3 | 37.7 | |
| 04189 | N3842 | 04189 | 2003-01-24 | 1.8 | ACIS-S | 6,7 | 7 | 42.4 | |
| 01563 | N3923 | 90201 | 2001-06-14 | 0.1 | ACIS-S | 6,7 | 7 | 15.7 | |
| 09507 | N3923 | 90201 | 2008-04-11 | 0.1 | ACIS-S | 6,7 | 7 | 78.8 | |
| 06939 | N4104 | 06939 | 2006-02-16 | 0.7 | ACIS-S | 6,7 | 7 | 33.8 | |
| 02071 | N4125 | 02071 | 2001-09-09 | 0.1 | ACIS-S | 6,7 | 7 | 60.6 | |
| 09569 | N4261 | 90201 | 2008-02-12 | 0.1 | ACIS-S | 6,7 | 7 | 97.9 | |
| 00834 | N4261 | 90201 | 2000-05-06 | 0.0 | ACIS-S | 6,7 | 7 | 32.4 | a |
| 07077 | N4278 | 90901 | 2006-03-16 | 0.0 | ACIS-S | 7 | 7 | 104.4 | |
| 00398 | N4278 | 90901 | 2000-04-20 | 0.0 | ACIS-S | 7 | 7 | 1.2 | a |
| 04741 | N4278 | 90901 | 2005-02-03 | 0.0 | ACIS-S | 7 | 7 | 35.4 | |
| 07078 | N4278 | 90901 | 2006-07-25 | 0.0 | ACIS-S | 7 | 7 | 49.1 | |
| 07079 | N4278 | 90901 | 2006-10-24 | 0.0 | ACIS-S | 7 | 7 | 99.8 | |
| 07080 | N4278 | 90901 | 2007-04-20 | 0.0 | ACIS-S | 7 | 7 | 55.6 | |
| 07081 | N4278 | 90901 | 2007-02-20 | 0.0 | ACIS-S | 7 | 7 | 108.2 | |
| 11269 | N4278 | 90901 | 2010-03-15 | 0.0 | ACIS-S | 7 | 7 | 79.1 | |
| 12124 | N4278 | 90901 | 2010-03-20 | 0.0 | ACIS-S | 7 | 7 | 24.8 | |
| 11778 | N4291 | 11778 | 2010-12-11 | 0.0 | ACIS-S | 6,7 | 7 | 28.2 | |
| 03232 | N4325 | 03232 | 2003-02-04 | 0.0 | ACIS-S | 6,7 | 7 | 28.0 | |
| 04687 | N4342 | 90201 | 2005-02-11 | 0.0 | ACIS-S | 6,7 | 7 | 30.3 | |
| 12955 | N4342 | 90201 | 2011-02-17 | 0.0 | ACIS-S | 6,7 | 7 | 44.1 | |
| 05908 | N4374 | 90301 | 2005-05-01 | 0.0 | ACIS-S | 6,7 | 7 | 45.1 | |
| 06131 | N4374 | 90301 | 2005-11-07 | 0.0 | ACIS-S | 6,7 | 7 | 39.1 | |
| 00803 | N4374 | 90301 | 2000-05-19 | 0.0 | ACIS-S | 6,7 | 7 | 28.2 | |
| 02016 | N4382 | 02016 | 2001-05-29 | 0.1 | ACIS-S | 6,7 | 7 | 38.2 | |
| 00318 | N4406 | 90202 | 2000-04-07 | 0.2 | ACIS-S | 2,3,6,7 | 7 | 10.2 | |
| 00963 | N4406 | 90202 | 2000-04-07 | 0.2 | ACIS-S | 2,3,6,7 | 7 | 11.8 | d |
| 02883 | N4438 | 90201 | 2002-01-29 | 0.0 | ACIS-S | 6,7 | 7 | 21.2 | |
| 08042 | N4438 | 90201 | 2008-02-11 | 4.4 | ACIS-S | 6,7 | 7 | 3.8 | |
| 00321 | N4472 | 90501 | 2000-06-12 | 0.2 | ACIS-S | 6,7 | 7 | 27.8 | |
| 00322 | N4472 | 90501 | 2000-03-19 | 0.2 | ACIS-I | 0,1,2,3 | 3 | 10.4 | |
| 11274 | N4472 | 90501 | 2010-02-27 | 0.0 | ACIS-S | 6,7 | 7 | 37.6 | |
| 12888 | N4472 | 90501 | 2011-02-21 | 0.0 | ACIS-S | 6,7 | 7 | 155.7 | |
| 12889 | N4472 | 90501 | 2011-02-14 | 0.0 | ACIS-S | 6,7 | 7 | 131.0 | |
| 08066 | N4477 | 90401 | 2008-02-21 | 5.3 | ACIS-S | 6,7 | 6 | 4.8 | |
| 09527 | N4477 | 90401 | 2008-04-27 | 0.0 | ACIS-S | 6,7 | 7 | 36.4 | |
| 11736 | N4477 | 90401 | 2010-04-30 | 3.7 | ACIS-S | 6,7 | 6 | 56.3 | |
| 12209 | N4477 | 90401 | 2010-05-02 | 3.7 | ACIS-S | 6,7 | 6 | 19.9 | |
| 03925 | N4526 | 03925 | 2003-11-14 | 0.0 | ACIS-S | 6,7 | 7 | 36.1 | |
| 02072 | N4552 | 90401 | 2001-04-22 | 0.0 | ACIS-S | 6,7 | 7 | 52.1 | |
| 13985 | N4552 | 90401 | 2012-04-22 | 0.0 | ACIS-S | 6,7 | 7 | 49.4 | |
| 14358 | N4552 | 90401 | 2012-08-10 | 1.2 | ACIS-S | 6,7 | 7 | 48.9 | |
| 14359 | N4552 | 90401 | 2012-04-23 | 1.0 | ACIS-S | 6,7 | 7 | 46.6 | |

| | | | | | | | | |
|---|---|---|---|---|---|---|---|---|
| 02884 | N4555 | 02884 | 2003-02-04 | 0.0 | ACIS-S | 6,7 | 7 | 26.4 |
| 09532 | N4594 | 90301 | 2008-04-29 | 0.5 | ACIS-I | 0,1,2,3 | 3 | 83.6 |
| 09533 | N4594 | 90301 | 2008-12-02 | 0.0 | ACIS-I | 0,1,2,3 | 3 | 85.6 |
| 01586 | N4594 | 90301 | 2001-05-31 | 0.1 | ACIS-S | 6,7 | 7 | 17.5 |
| 00323 | N4636 | 90401 | 2000-01-26 | 0.0 | ACIS-S | 6,7 | 7 | 42.4 |
| 00324 | N4636 | 90401 | 1999-12-04 | 0.0 | ACIS-I | 0,1,2,3 | 3 | 5.5 |
| 03926 | N4636 | 90401 | 2003-02-14 | 0.0 | ACIS-I | 0,1,2,3 | 3 | 71.4 |
| 04415 | N4636 | 90401 | 2003-02-15 | 0.0 | ACIS-I | 0,1,2,3 | 3 | 71.8 |
| 14328 | N4649 | 90601 | 2011-08-12 | 0.1 | ACIS-S | 6,7 | 7 | 14.0 |
| 12975 | N4649 | 90601 | 2011-08-08 | 0.1 | ACIS-S | 6,7 | 7 | 83.1 |
| 12976 | N4649 | 90601 | 2011-02-24 | 0.1 | ACIS-S | 6,7 | 7 | 99.0 |
| 00785 | N4649 | 90601 | 2000-04-20 | 0.2 | ACIS-S | 6,7 | 7 | 23.0 |
| 08182 | N4649 | 90601 | 2007-01-30 | 0.1 | ACIS-S | 6,7 | 7 | 47.8 |
| 08507 | N4649 | 90601 | 2007-02-01 | 0.1 | ACIS-S | 6,7 | 7 | 17.3 |
| 03220 | N4782 | 03220 | 2002-06-16 | 0.3 | ACIS-S | 6,7 | 7 | 48.3 |
| 00798 | N5044 | 90201 | 2000-03-19 | 0.0 | ACIS-S | 6,7 | 7 | 20.2 |
| 09399 | N5044 | 90201 | 2008-03-07 | 0.0 | ACIS-S | 6,7 | 7 | 81.9 |
| 06944 | N5129 | 90201 | 2006-04-13 | 0.5 | ACIS-S | 6,7 | 7 | 19.9 |
| 07325 | N5129 | 90201 | 2006-05-14 | 0.5 | ACIS-S | 6,7 | 7 | 23.3 |
| 03216 | N5171 | 03216 | 2002-12-10 | 0.9 | ACIS-S | 6,7 | 7 | 34.4 |
| 05907 | N5813 | 90901 | 2005-04-02 | 0.0 | ACIS-S | 7 | 7 | 46.8 |
| 09517 | N5813 | 90901 | 2008-06-05 | 1.5 | ACIS-S | 7 | 7 | 97.5 |
| 12951 | N5813 | 90901 | 2011-03-28 | 2.5 | ACIS-S | 7 | 7 | 71.4 |
| 12952 | N5813 | 90901 | 2011-04-05 | 2.5 | ACIS-S | 7 | 7 | 141.8 |
| 12953 | N5813 | 90901 | 2011-04-07 | 2.5 | ACIS-S | 7 | 7 | 30.0 |
| 13246 | N5813 | 90901 | 2011-03-30 | 2.5 | ACIS-S | 7 | 7 | 40.4 |
| 13247 | N5813 | 90901 | 2011-03-31 | 2.5 | ACIS-S | 7 | 7 | 33.2 |
| 13253 | N5813 | 90901 | 2011-04-08 | 2.5 | ACIS-S | 7 | 7 | 114.9 |
| 13255 | N5813 | 90901 | 2011-04-10 | 2.5 | ACIS-S | 7 | 7 | 40.8 |
| 07923 | N5846 | 90201 | 2007-06-12 | 0.0 | ACIS-I | 0,1,2,3 | 3 | 86.2 |
| 00788 | N5846 | 90201 | 2000-05-24 | 0.1 | ACIS-S | 6,7 | 7 | 23.2 |
| 02879 | N5866 | 02879 | 2002-11-14 | 0.0 | ACIS-S | 6,7 | 7 | 30.7 |
| 08180 | N6107 | 08180 | 2007-09-29 | 0.1 | ACIS-S | 6,7 | 7 | 18.7 |
| 04194 | N6338 | 04194 | 2003-09-17 | 0.0 | ACIS-I | 0,1,2,3 | 3 | 46.6 |
| 03218 | N6482 | 03218 | 2002-05-20 | 0.0 | ACIS-S | 6,7 | 7 | 16.4 |
| 03190 | N6861 | 90201 | 2002-07-26 | 0.1 | ACIS-I | 0,1,2,3 | 1 | 18.9 |
| 11752 | N6861 | 90201 | 2009-08-13 | 0.0 | ACIS-I | 0,1,2,3 | 3 | 91.2 |
| 03191 | N6868 | 90201 | 2002-11-01 | 0.0 | ACIS-I | 0,1,2,3 | 2 | 22.4 |
| 11753 | N6868 | 90201 | 2009-08-19 | 0.0 | ACIS-I | 0,1,2,3 | 3 | 72.1 |
| 00802 | N7618 | 90301 | 1999-12-10 | 0.0 | ACIS-S | 6,7 | 7 | 10.5 |
| 07895 | N7618 | 90301 | 2007-09-08 | 0.0 | ACIS-S | 6,7 | 7 | 34.0 |
| 16014 | N7618 | 90301 | 2014-09-10 | 1.4 | ACIS-S | 6,7 | 7 | 30.7 |
| 03955 | N7619 | 90201 | 2003-09-24 | 3.6 | ACIS-S | 6,7 | 7 | 29.0 |
| 02074 | N7619 | 90201 | 2001-08-20 | 2.6 | ACIS-I | 0,1,2,3 | 2 | 26.0 |
| 02074 | N7626 | 02074 | 2001-08-20 | 5.8 | ACIS-I | 0,1,2,3 | 1 | 26.2 |

Column 1. Unique Chandra observation identification number
Column 2. Galaxy name (NGC or IC name)
Column 3. The merge id (mid in short) is an identification number when data from multiple obsids are used.
Column 4. observation date
Column 5. OAA (off-axis-angle) of the galaxy center in arcmin
Column 6. Detector (ACIS-I or ACIS-S)
Column 7. Chips used in CGA
Column 8. The chip where the center of the galaxy lies
Column 9. Effective exposure time (in ksec) of the target chip, after removing background flares
Column 10. Notes on individual obsids:

a. subarray (512 out of 1024 rows used)
b. subarray (256 out of 1024 rows used)
c. CCD readout streaks due to a strong AGN
d. no fid light in this observation (affecting aspect quality)

## Appendix A. Chandra Galaxy Atlas Data Products

The following downloadable data products are made available in two packages. The main package (package 1) consists of the high-level data products, e.g., point source removed and filled, exposure corrected images (jpg/png and FITS format) in multiple energy bands to best illustrate the diffuse hot gas distribution, derived T-maps (jpg/png and FITS format) in various adaptive binning methods to show the hot gas thermal structure. They can be used directly to get an overview, to obtain the necessary hot gas quantities and to compare them with data at other wavelengths with no additional X-ray data reduction.

For those who intend to perform their own analyses, e.g., to extract and fit X-ray spectra from user-specified regions, we provide all the necessary data in the supplementary package (package 2), e.g., event files per obsid and per ccd.

Those with **red** bold letters below are in the main package and the others are in the supplementary package. Those with faint **orange** letters are not in the package, due to the large number of files and their size. They can be made available upon requests.

Merged data (see section 3.1)

- **${gmv}_evt.fits** – a single event file containing the entire observations of a given galaxy. Here g, m and v in ${gmv} indicate galaxy name, merge id (mid) and version, respectively. ${gmv} looks like N1234_90201_v01.
- **${gmve}_img.fits** – a full-resolution image for a given energy band. 1 pixel corresponds to 0.492 arcsec. Here e in ${gmve} indicates an energy band (a short name in Table 3). ${gmve} looks like N1234_90201_v01_G.
- **${gmve}_exp.fits** – a matching exposure map for a given energy band in unit of $cm^2$ s (= effective area x exposure time). Due to the energy dependent effective area, the exposure map was made at the effective energy for each energy band given in Table 3.
- **${gmv}_frame.reg** – an ASCII region file of the full field-of-view. It contains one polygon per obsid in a physical pixel coordinate.
- **${gmv}_frame_fk5.reg** - same as the above but in fk5 (in sexagesimal).

Point source list (see section 3.2)

- **${gmve}_src.fits** – a source list (FITS table) for each energy band. They were detected by wavdetect on the merged image file.
- **${gmv}_src_psfsize.fits** – a list of sources in a FITS file which are to be removed. To avoid false sources to be excluded, we limit sources with net counts > 10 and significance > 3 $\sigma$. As described in section 3.2, the sources are determined in the C band detections by default, but when the diffuse emission is strong in the central region of gas-rich galaxies, the sources are determined in the H band detections, instead. The source sizes (used to remove them) are set by MARX PSF with variable encircled energy fractions depending on the source strength.
- **${gmv}_src_psfsize.dat** –same as above, but an ASCII list file with additional information about ra, dec, and net counts.

Event and image files after point sources removed (not filled)

- **${gmv}_psrem.evt** – an event file containing all data for all obsids and all ccds. All detected point sources were removed, but not filled.

- **${gmv}_C_psrem.img** – an image file containing data for all obsids and all ccds for C energy band. All detected point sources were removed, but not filled. This image file is the one used in the 2D adaptive binning (section 3.3).
- **${gmv}_${oc}_psrem_evt.fits** - a FITS event file for each obsid for each ccd. They are used for spectral extraction of each binned region. (section 3.4)
- **${gmv}_${oc}_psrem_bevt.fits** – corresponding background FITS event file for each obsid for each ccd. They were made appropriate for each obsid from the blank sky data (section 3.4).

Diffuse images after point sources removed and filled (see section 3.2)

- **${gmve}_diff_img.fits** – a diffuse image for a given energy band. This image was made after point sources were removed and holes were filled by surrounding pixels.
- **${gmve}_diff_flux.fits** – an exposure-corrected, flux diffuse image for a given energy band. It was made from a diffuse image divided by a corresponding exposure map.
- **${gmve}_sm_diff_img.fits** – a smoothed diffuse image for a given energy band by the 2D Gaussian smoothing with a 5σ kernel (7 pixels per σ) was applied.
- **${gmve}_sm_diff_flux.fits** – a smoothed, exposure-corrected flux diffuse image for a given energy band. The same smoothing was applied as in the above.

JPEG/PNG figures of the above images

- **${gmve}_img_${z}.jpg** – a ds9 jpg file for an X-ray image, ${gmve}_img. Here ${z} indicates a zoom factor (z1 = unzoomed, z05 = zoomed out by a factor of 1/2).
- **${gmve}_exp_${z}.jpg** – a ds9 jpg file for an exposure map, ${gmve}_exp.
- **${gmve}_flux_${z}.jpg** – a ds9 jpg file for a flux image, ${gmve}_flux.
- **${gmv}_rgb.png** - three color images to visualize the 2D spectral variation (including point sources). The energy bands used in rgb are 0.5-1.2, 1.2-2 and 2-7 keV, respectively.
- **${gm}_OX.png** – a ds9 jpg file to compare the X-ray (${gmve}.img where e=B band) and optical (DSS POSS2 Red) images. The $D_{25}$ ellipse (from RC3) and the fov (${gmv}_frame.reg) are overlaid.
- **${gmve}_diff_img_${z}.jpg** – a ds9 jpg file for a diffuse image, ${gmve}_diff.img.
- **${gmve}_diff_flux_${z}.jpg** – a ds9 jpg file for a diffuse flux image, ${gmve}_diff.flux.
- **${gmve}_sm_diff_img_${z}.jpg** – a ds9 jpg file for a smoothed diffuse image, ${gmve}_sm_diff.img.
- **${gmve}_sm_diff_flux_${z}.jpg** – a ds9 jpg file for a smoothed diffuse flux image, ${gmve}_sm_diff.flux.

Adaptively binned spatial regions and images (see section 3.3)

- **${gmvb}_Imap.fits** – a FITS image with pixel value = relative intensity in each bin. Here b in in the filename indicates one of four binning methods (AB, WB, CB, HB). ${gmvb} is like N1234_90201_v01_WB.
- **${gmvbn}.reg** – an ASCII region file for the n-th bin. Here n in in the file name is a 5-digit bin number. ${gmvbn} is like N1234_90201_v01_WB_00012.
- **${gmvb}_binno.fits** – a FITS image file with pixel value = bin number
- **${gmvb}_binrad.fits** – a FITS image file with pixel value = radius of bin (only for HB). Due to the nature of the HB method, the spectral extraction region is different from the bin

and the extraction circle is generally overlapping and bigger than the bin size given by the square grid.

Spectra, arf and rmf for each bin (see section 3.4)

- **${gmvbn}.pi** – A type I PHA file for the source spectrum for the n-th bin, extracted from individual event files for each obsid and for each ccd, then combined into a single file (see section 3.4)
- **${gmvbn}_bkg.pi** – same as the above but for the background spectrum. The same extraction region is used as in the source spectrum.
- **${gmvbn}_src.arf** – an ancilliary response file (ARF) for the n-th bin, made for each obsid and for each ccd, then combined into a single file.
- **${gmvbn}_src.rmf** – an response matrix file (RMF) for the n-th bin, made for each obsid and for each ccd, then combined into a single file.

Fitting results (see section 3.4)

- **${gmvb}_sum.dat**– an ASCII table containing (1) bin number, (2) galacto-centric distance in arcmin determined by photon weighted mean distance from the galaxy center, (3) area of the bin in pixels (one ACIS pixel is 0.492" x 0.492") (4-5) total and net counts, (6) reduced $\chi^2$, (7-9) best-fit T and its 1$\sigma$ lower and upper bounds, (10) T error in percent, (11-13) APEC normalization parameter divided by bin area and its 1$\sigma$ lower and upper bounds.
- **${gmvbns}.log** – an ASCII log file for SHERPA spectral fitting. Here, s in ${gmvbns}indicates spectral models (e.g., APEC, APECnPL = APEC + Power-law). ${gmvbns} looks like N1234_90201_v01_WB_00012_APECnPL.

Spectral maps (see section 3.4)

- **${gmvbsp}map.fits** - A FITS image with pixel value = spectral parameter. Here p in the file name indicate the spectral parameters used in the map (T=temperarure, N=norm of the APEC component / area of the bin, P=projected pseudo pressure and K=projected pseudo entropy). ${gmvbsp} is like N1234_90201_v01_WB_APECnPL_T. The T map is in keV unit and the other maps are in arbitrary units (see section 3.5).
- **${gmvbsp}map_lo.fits** – same as the above but with pixel value = lower limit of a given parameter
- **${gmvbsp}map_up.fits** - same as the above but with pixel value = upper limit of a given parameter
- **${gmvbsp}map_30pc.fits** – same as ${gmvbsp}.map, but the bins with a large error (> 30% in T) are masked out. Here ${gmvbrsp} looks like N1234_90201_v01_WB_sn20_APECnPL_T.
- **${gmvbs}_Cmap_30pc.fits** - A FITS image with pixel value = reduced chi2

PNG figures

- **${gmbr}_Imap.png** – a png file of the binned intensity map, ${gmvb}_Imap.fits.
- **${gmbrsp}map.png** – a png file of the masked (by 30% T error) spectral parameter maps, ${gmvbsp}map_30pc.fits.
- **${gmvbs}_Cmap.png** – a png file with pixel value = reduced chi2

## Appendix B. Notes on Individual Galaxies

Based on the 1D radial profiles and 2D spectral maps, we describe important features of individual galaxies. We also include the distinct features previously known by targeted studies. We use the terms, SB (surface brightness), T (temperature), EM (emission measure), P (projected or pseudo pressure) and K (projected or pseudo entropy). In describing temperature profiles, we call the case with a negative T gradient in the central region (roughly within a few kpc or ~0.5 $r_e$) a hot core and the case with a positive T gradient in the central region a cool core.

**IC 1262** (d=130.0 Mpc, 1' = 37.8 kpc, $r_e$= 7.7 kpc, $r_{25}$= 22.7 kpc)
It is a dominant galaxy in a small group. While the hot gas halo is roughly symmetric and smooth on a large scale (> 100 kpc), the hot ISM on a smaller scale exhibits rather complex substructures: a sharp discontinuity to the E and narrow arcs over 100 kpc long to the NS direction (Trinchieri et al. 2007). Also detected are cavities at 10-20 kpc to the N (Dong et al 2010). The hot gas in the enhanced surface brightness regions to the E from the center in r < 20kpc (~1 keV) and to the NS in r =20-100 kpc (~1.2 keV) is cooler than the surrounding gas (~1.6 keV). The gas thermal structures are also visible in the entropy map: the lowest entropy gas is extended to the E, the lower entropy gas is extended to the NS, surrounded by the higher entropy gas.

**IC 1459** (d= 29.2 Mpc, 1' = 8.5 kpc, $r_e$= 5.2 kpc, $r_{25}$= 22.3 kpc)
It is a dominant galaxy in a small group. A strong nuclear source ($L_X$ = 8 x $10^{40}$ erg $s^{-1}$) dominates the X-ray emission (Fabbiano et al. 2003). Because of the strong central source, there are (weak) ACIS readout streaks along the CCD column, visible in the smoothed diffuse image. Note that they are not excluded in this image, however, the streak regions were excluded in the spectral analysis (see section 4.3). The faint diffuse gas is extended to fill the $D_{25}$ ellipse. While the surface brightness is azimuthally symmetric, the 2D temperatures map shows slight asymmetry in that the gas along the minor axis (0.6-0.7 keV) is hotter than that along the major axis (0.4-0.5 keV).

**IC 1860** (d= 93.8 Mpc, 1' = 27.3 kpc, $r_e$= 8.4 kpc, $r_{25}$= 23.7 kpc)
It is the BCG (brightest cluster galaxy) of Abell S301. It is previously known to be sloshing (Gastadello et al 2013). The intensity and temperature maps show a narrow tail, extending to r = 1′ (or 27 kpc) toward the SE direction from the core. The gas in the tail is cooler (~1 kev) than that in the surrounding region (1.3 keV). The tail is also visible in the projected entropy map, but not obviously clear in the projected pressure map, indicating the pressure balance with the surrounding medium. Excluding the tail, the hot gas emission on a large scale (at or outside the $D_{25}$ ellipse) is more pronounced toward the SW and relatively weaker toward the NE direction.

**IC 4296** (d= 50.8 Mpc, 1' = 14.8 kpc, $r_e$= 11.9 kpc, $r_{25}$= 25.0 kpc)
It is the BCG in Abell 3565. A strong nuclear source ($L_X$ = 2.4 x $10^{41}$ erg $s^{-1}$) dominates the X-ray emission (Pellegrini et al. 2003; Humphrey & Buote 2006). The hot gas distribution is asymmetric, mainly extending to r = 1.6' or 20 kpc toward the SW direction in a fan-shape (PA=280-350°). The gas temperature in that region is hotter (~1.3 keV) than that in the central region (0.8 keV). The ACIS observation was done in a sub-array mode with 512 rows, but the entire D25 ellipse is included. A nearby spiral galaxy, IC 4299 which is at 6.2' to the S of IC 4296 is also detected just inside the fov.

**NGC 193** (d= 47.0 Mpc, 1' = 13.7 kpc, $r_e$= 4.4 kpc, $r_{25}$= 9.9 kpc)
It is a FR-I radio galaxy (Laing et al. 2011). The hot gas exhibits a well-structured shell at r = 1′ - 1.5′ (or 14-20 kpc). The radio lobes have inflated into a cocoon or a large cavity and the shell of shocked material around the cavity (Bogdan et al. 2014). The gas in the cavity is cooler (~0.7 keV) than that in the shell (~ 1 keV). It also has a distinct X-ray point source at the center.

**NGC 315** (d= 69.8 Mpc, 1' = 20.3 kpc, $r_e$= 12.5 kpc, $r_{25}$= 32.9 kpc)
It is a FR-I radio galaxy. The Chandra observations reveal a strong nuclear X-ray source ($L_X$ = 5 x $10^{41}$ erg $s^{-1}$) and also an X-ray jet ($\Gamma_{PH}$ ~ 2.2) extending to r = 1′ (or 20 kpc), coincident with a NW radio jet (Worrall et al. 2003, 2007). The strong nucleus causes (weak) CCD readout streaks along the CCD column, which are visible in the diffuse image. Note that they are not excluded in this image, however, the streak regions were excluded in the spectral analysis (see section 4.3).

**NGC 383** (d= 63.4 Mpc, 1' = 18.4 kpc, $r_e$= 6.3 kpc, $r_{25}$= 14.6 kpc)
It is a FR-I radio galaxy (3C 31). The Chandra observations reveal a strong nuclear X-ray source (several x $10^{41}$ erg $s^{-1}$) and also an X-ray jet extending to r = 10″ or 3 kpc, coincident with brighter, northern radio jet of 3C 31 (Hardcastle, et al. 2002). Because of its small scale, the northern jet is only seen in the unsmoothed image. A nearby galaxy, NGC 382, lying inside the D25 ellipse at ~0.6′ S (PA=200°) from NGC 383 is also detected with a similar gas temperature (~0.7 keV). A few nearby galaxies of the Arp 331 chain (see the table below) are also detected in the ACIS fov.

| name | RA (J2000) | DEC (J2000) | D(′) from N383 | PA(deg) from N383 |
|---|---|---|---|---|
| NGC 375 | 16.77466 | 32.34817 | 5.6′ | 226 |
| NGC 379 | 16.81537 | 32.52036 | 6.8′ | 343 |
| NGC 380 | 16.82330 | 32.48292 | 4.5′ | 340 |
| NGC 382 | 16.84946 | 32.40386 | 0.6′ | 200 |
| NGC 384 | 16.85460 | 32.29245 | 7.2′ | 180 |
| NGC 385 | 16.86352 | 32.31953 | 5.6′ | 175 |
| NGC 386 | 16.88039 | 32.36199 | 3.3′ | 156 |
| NGC 387 | 16.88775 | 32.39111 | 2.2′ | 127 |

**NGC 499** (d=54.45 Mpc, 1′=15.8 kpc, $r_e$=4.4 kpc, $r_{25}$ =12.8 kpc )
It is the 2nd brightest galaxy in the NGC 507 group (13.7′ to the NW, PA=334° from NGC 507). Given its high $L_{X,GAS}$ (> $10^{42}$ erg $s^{-1}$) and $T_{GAS}$ (~ 1 keV at the outskirts), it is likely a separate group which is currently merging with the NGC 507 group. NGC 499 and NGC 507 belong to one group in the 2MASS group catalog (Crook et al. 2007), but they are identified as two separate groups in Lyon group catalog (Garcia 1993). The temperature radial profile is a typical case with a hot core. In contrast to a cool core, the T gradient is negative in the inner region (r < 10 kpc) and positive in the outer region (r > 10 kpc). Although at the outskirts (r > 70 kpc) the temperature (as well as the surface brightness) goes up with increasing radius due to the hotter ICM of the NGC 507 group, the inner temperature profile is not affected by the ICM and therefore intrinsic to NGC 499. The 2D temperature map further shows that the gas distribution is azimuthally asymmetric. The inner hotter gas (~0.85 keV) extending to r = 20″ or 5 kpc is elongated along the major axis (PA = 70°) while the outer cooler gas (~0.65 keV) extending to r = 2′ or 30 kpc is elongated along the NE-SW direction (PA ~ 40°). There are two possible ghost cavities just outside the $D_{25}$ ellipse (at r =10-15 kpc), one to the N (PA ~ 20°) and another to the S (PA ~ 160°) from the galaxy center, seen in the diffuse image as well as in the projected pressure map. It is likely that all these features suggest the past, multiple AGN activities.

**NGC 507** (d=63.8 Mpc, 1′=18.6 kpc, $r_e$=12.9 kpc, $r_{25}$ =28.7 kpc )
It is the BCG in an optically rich group, merging with NGC 499 (at 13.7′ to the NW, PA=334). The T profile indicates the presence of a cool core and the SB profile shows the central cusp. The 2D maps illustrate rather complex substructures. Its core region has cavities, possibly related to the old bent radio lobes (Giacintucci et al 2011). Kraft et al. (2004) identified an abundance front. The hot gas is probably sloshing owing to interaction with NGC 499. NGC 508 at ~1.5′ to the N is also detected in the same observation.

**NGC 533** (d=76.9 Mpc, 1′= 22.4 kpc, $r_e$=16.2 kpc, $r_{25}$=42.5 kpc)
It is a relaxed group with a cool core (Eckmiller et al. 2011; Panagoulia et al. 2014). The T profile shows cooler gas (T ~ 0.8 keV) in the inner region (r < 5 kpc). The temperature steeply rises to ~1.3 keV at r = 16 kpc (close to $r_e$), then remains constant to r = 80 kpc (~5 $r_e$). The T map further indicates an asymmetric cool core. Inside r = 20 kpc, the hot gas is distributed along the major axis, generally following the optical light distribution (ellipticity e = 0.4). This is most clearly seen in the T and projected entropy maps, possibly indicating sloshing by a nearby galaxy. In the outer region (r = 20-60 kpc), the hot gas is more extended to the N and NE than toward the S. Note that the lower SB region just inside the $D_{25}$ ellipse to the S is affected by the node boundary which is not fully corrected even in the exposure corrected diffuse emission map. However, the lower SB region outside the $D_{25}$ ellipse toward the S is not affected. Shin et al. (2016) found two cavities at 1.5 kpc N and S of the nucleus. There is a possible ghost cavity at r~5 kpc from the center toward the NW.

**NGC 720** (d=27.7 Mpc, 1′ = 8.0 kpc, $r_e$=4.8 kpc, $r_{25}$=18.8 kpc)
It was extensively studied as a relaxed system where mass can be measured accurately (Buote et al. 2002 and Humphrey et al. 2011). Both T profile and T map show more or less constant T (~0.6 keV) throughout the galaxy (r = 2 - 20 kpc) with hints of increasing T toward the center and decreasing T at the outskirts. Unlike NGC 533, the gas distribution is considerably rounder than the optical figure (e = 0.5), as seen in the intensity map as well as the projected pressure and entropy maps.

**NGC 741** (d=70.9 Mpc, 1′ = 20.6 kpc, $r_e$=13.2 kpc, $r_{25}$=30.4 kpc)
It is a disturbed system, owing to a NAT radio galaxy (PKS 0153+05) and NGC 742 (at 0.8′ to the E) falling through the core. Jetha et al (2008) reported complex gas structures with X-ray filaments linking NGC 741 and NGC 742 and a possible ghost cavity to the W of NGC 741 (see also Schellenberger et al. 2017). Even if it is not relaxed, NGC 741 has a cool core (~0.7 keV) and hotter gas (1.2 keV) at the outskirts. NGC 742 has a nuclear X-ray source at the center which dominates its entire X-ray emission. Also detected in the ACIS field of view are two compact galaxies, ARK065 and ARK066, at the similar redshift with NGC 741 and 742.

| name | RA | DEC | D(′) | PA(deg) | other |
|---|---|---|---|---|---|
| | (J2000) | (J2000) | from N741 | from N741 | names |

| | | | | | | |
|---|---|---|---|---|---|---|
| NGC 742 | 29.10072 | 5.62668 | 0.8′ | 100 | | |
| ARK065 | 29.05004 | 5.58858 | 3.3′ | 223 | P007237 | |
| ARK066 | 29.07938 | 5.65208 | 1.5′ | 341 | P007250 | IC 1751 |

**NGC 1052** (d=19.4 Mpc, 1′ = 5.6 kpc, $r_e$=3.1 kpc, $r_{25}$=8.5 kpc)
It is a well-known LINER. A variable nuclear X-ray source dominates the entire X-ray emission (Hernández-García et al. (2013). Soft (0.4 keV) diffuse gas emission is seen inside a few kpc from the center. Asymmetric T maps (0.4 keV to the E and 0.7 keV to the W) seen in WB and HB need to be confirmed by deeper observations.

**NGC 1132** (d=95.0 Mpc, 1′ = 27.6 kpc, $r_e$=15.5 kpc, $r_{25}$=34.7 kpc)
It is a fossil group with an extended, luminous X-ray halo ($L_X$ ~ 7 x $10^{42}$ erg $s^{-1}$, Lovisari et al. 2015). The T profile shows a cool core (~0.8 keV). The azimuthally average temperature peaks at ~10 kpc (1.3 keV) and declines outwards to ~1 keV. In contrast to the expectation as a fossil system, the hot gas morphology indicates asymmetry, an edge to the E and extended emission to the W, possibly implying a rare case of a rejuvenated fossil group (e.g., von Benda- Beckmann et al. 2008). The detail observational results and implications are presented in a separate paper (Kim et al., 2018).

**NGC 1316** (d=21.5 Mpc, 1′ = 6.2 kpc, $r_e$=7.6 kpc, $r_{25}$=37.6 kpc)
It is a radio galaxy (Fornax A) with radio jets and extended lobes to the E-W direction (Ekers et al. 1983) and exhibits a number of signs for recent major mergers in 2-3 Gyr ago (Schweizer 1980). The hot gas morphology also indicates disturbed nature and cavities associated with the radio jets (Kim & Fabbiano 2003). Given the large optical luminosity and size of the stellar system, the amount ($L_{X,GAS}$ ~ a few x $10^{40}$ erg $s^{-1}$) and the extent (~10 kpc) of hot gas are very low, making its $L_X/L_K$ one of the lowest for among nearby ETGs. The azimuthally averaged temperature decreases from the center to r ~ 10 kpc, then increases outwards to 20 kpc. But the SB and temperature maps indicate the gas is not symmetric.

**NGC 1332** (d=22.9 Mpc, 1′ = 6.7 kpc, $r_e$=3.1 kpc, $r_{25}$=15.6 kpc)
It is an edge-on S0 galaxy. The temperature is almost constant at 0.6 keV inside the D25 ellipse, except that the hot gas may be slightly hotter (~0.7 keV) in the central region.

**NGC 1380** (d=17.6 Mpc, 1′ = 5.1 kpc, $r_e$=3.2 kpc, $r_{25}$=12.3 kpc)
It is an edge-on S0 galaxy in the Fornax cluster. It is located at 37.8′ to the NW from NGC 1399. The hot ISM (~0.3 keV) is detected but confined within a few kpc (~1 $r_e$). The hotter (1.2-1.5 keV) ICM in the Fornax cluster is also detected at r > 10 kpc.

**NGC 1387** (d=20.3 Mpc, 1′ = 5.9 kpc, $r_e$=3.5 kpc, $r_{25}$=8.3 kpc)
It is a barred S0 (SB0) galaxy in the Fornax cluster. It is located at 19′ to the W (PA=260) from NGC 1399. The hot ISM (~0.5 keV) is detected but confined within several kpc (or ~2 $r_e$). The hotter (1.2-1.5 keV) ICM in the Fornax cluster is also detected at r > 10 kpc. Note that the X-ray bright part is on the ACIS-I chip gap, although the exposure map appears to work properly.

**NGC 1395** (d=24.1 Mpc, 1′ = 7.0 kpc, $r_e$=5.4 kpc, $r_{25}$=20.6 kpc)

It is a large elliptical galaxy ($M_K$ = -25 mag), and the hot gas temperature is comparably high (0.8 - 0.9 keV). Given the short Chandra observations with significant background flares, the hot gas is limited roughly within the $D_{25}$ ellipse, and its temperature is more or less constant.

**NGC 1399** (d=20.0 Mpc, 1′ = 5.8 kpc, $r_e$=4.7 kpc, $r_{25}$=20.1 kpc)
It is at the center of the Fornax cluster and contains a large amount of extended hot halo. Although NGC 1316 (3.6° away in projection) is optically brighter by a factor of two (hence BGC), NGC 1399 is at the bottom of the potential well. On a galaxy scale (inside the $D_{25}$ ellipse), the intensity maps show two filaments to the north and one to the south. The radio jets are propagating between the two northern filaments and at the side along the southern filament (Paolillo et al. 2002 and Werner et al. 2012). The T maps show the cooler gas (0.8-1 keV) extending to the N-S direction along the filaments. The projected pressure map also shows the gaps where the radio jets are propagating. On a large cluster scale, the Chandra observations of the Fornax cluster (3x3 ACIS-I observations) reveal the asymmetric intracluster gas (Scharf et al. 2005). The hot halo is extended to r ~30′ (180 kpc) to the NE of NGC 1399. A few discontinuities, likely due to sloshing, are also detected by Su et al. (2017b).

**NGC 1400** (d=26.4 Mpc, 1′ = 7.7 kpc, $r_e$=2.9 kpc, $r_{25}$=8.8 kpc)
It is a member of the NGC 1407 / NGC 1400 merging group, 12′ away to the SW (PA=236°) from NGC 1407. Note that the galaxy (the NW side) is not fully covered by the deeper one of two observations. The hot ISM (0.5-0.6 keV) is confined inside the $D_{25}$ ellipse and surrounded by the hotter ambient gas (~1.2 keV) at r > 20 kpc. There is also a blob of hot gas at 3′-5′ away to the NE from NGC 1400 (Giacintucci et al. 2012, Su et al. 2014). Because the external hot gas has a similar temperature and abundance to that of NGC 1400, Su et al. (2014) suggested that it might have come out of NGC 1400 by ram pressure stripping.

**NGC 1404** (d=21.0 Mpc, 1′ = 6.1 kpc, $r_e$=2.7 kpc, $r_{25}$=10.1 kpc)
It is a member of the Fornax cluster, only 10′ (in projection 60 kpc) away to the SE (PA=152°) from NGC 1399. It is one of the most extensively studied ETGs due to a sharp discontinuity to the direction of NGC 1399 (Machacek et al 2005, Su et al. 2017a). The Chandra observations reveal a front at the NW edge and a tail to the SE, indicating that it is currently falling through the Fornax cluster. The temperature radial profile indicates a hot core as in the case of NGC 499. The temperature is 0.8 keV at the central region and decreases to 0.5 keV at r = 4 kpc (just outside $r_e$) and steeply rise at r=5-20 kpc to 1.3 keV. In the 2D spectral maps, the head-tail structure is clearly visible.

**NGC 1407** (d=28.8 Mpc, 1′ = 8.4 kpc, $r_e$=8.9 kpc, $r_{25}$=19.2 kpc)
It is the BCG in a small group. It has a cool core with an edge to the N (at r =7-8 kpc) and wings extending to the E-W, suggesting that the galaxy is moving to the N direction. The wings are inside a large scale, old diffuse radio structure, and they are bent likely as a consequence of motion to the N (Giacintucci et al, 2012). The temperature radial profile indicates another hot core case: T is ~ 1 keV in the center, decreases to ~0.8 keV at r = 1-3 kpc, then rises to ~1.3 keV at the outskirts (r > 20 kpc). The temperature maps further reveal the cooler E-W wings extending beyond the $D_{25}$ ellipse.

**NGC 1550** (d=51.1 Mpc, 1′ = 14.9 kpc, $r_e$=6.3 kpc, $r_{25}$=16.6 kpc)

It is a dominant galaxy in a group. This group is one of the most luminous local groups with $L_X \sim 10^{43}$ ergs $s^{-1}$ within 200 kpc (Sun et al. 2003). On a large scale, the hot gaseous halo is smooth and circularly symmetric as seen in the XMM-Newton observations (Kawaharada et al. 2009) but the central region is highly elongated. The 2D temperature map further suggests an asymmetric distribution of cooler gas (~1 keV) with a strong E-W elongation, being more pronounced to the W. A similar trend is also seen in the projected entropy map while it is not seen in the projected pressure map, indicating pressure balance between the cooler/low entropy and hotter/high entropy gas. The temperature radial profile shows that the temperature is constant at ~1 keV within $r < 5$ kpc and rises to ~1.5 keV at $r \sim 30$ kpc, then declines at the outskirts. Note that the CCD boundary of one of the two long ACIS-S observations falls at the southern end of the $D_{25}$ ellipse. While the boundary is clearly visible in the raw and binned images, it is properly treated in spectral fitting so that the spectral maps are smooth across the boundary.

**NGC 1553** (d=18.5 Mpc, 1′ = 5.4 kpc, $r_e$=5.1 kpc, $r_{25}$=12.0 kpc)
It is a face-on SB0 galaxy. A hard X-ray source ($L_X \sim 10^{40}$ erg $s^{-1}$) is present at the nucleus and the diffuse hot gas (~0.4 keV) roughly fills the $D_{25}$ ellipse. The gas emission is not smooth, the SB radial profile is rather flat and spiral arm like features starting from the center are seen along the minor axis (see also Blanton et al 2001). A deeper observation is necessary to confirm the distribution and its thermal structure of the hot gas.

**NGC 1600** (d=57.4 Mpc, 1′ = 16.7 kpc, $r_e$=13.5 kpc, $r_{25}$=20.5 kpc)
It is either a BCG in a loose group or an isolated elliptical galaxy surrounded by a number of satellite galaxies (Smith et al. 2008). Due to its low X-ray luminosity (a few x $10^{41}$ erg $s^{-1}$), it is not identified as a fossil group. It hosts a massive BH of 1.7 x $10^{10}$ $M_\odot$ (Thomas et al. 2016). The temperature profile is a typical one with a cool core. The cool core is at T ~ 0.8 keV and T reaches a peak (T ~ 1.5 keV) at a few $r_e$, then declines outward. Also detected in X-rays are NGC 1603 (2.5′ E of NGC 1600) and NGC 1601 (1.6′ N of NGC 1600). The hot gas in NGC 1603 shows a tail to the W, due to the ram pressure from the group halo as the galaxy is moving to the E (Sivakoff et al. 2004). There is no clear signature of sloshing in the main halo, likely because NGC 1603 is too small to perturb the hot halo of the main galaxy as $\Delta m_B = 2.7$.

**NGC 1700** (d=44.3 Mpc, 1′ = 12.9 kpc, $r_e$=3.86 kpc, $r_{25}$=21.4 kpc)
It is a giant ($M_K$=-25.5) elliptical galaxy. A hard X-ray source ($L_X$ ~ a few x $10^{40}$ erg $s^{-1}$) is present at the nucleus and the diffuse hot gas (~0.5 keV) roughly fills the $D_{25}$ ellipse. The hot gas distribution is significantly flattened likely due to the rotation, consistent with the stellar figure (E4) and kinematics (Statler & McNamara 2002). The temperature radial profile indicates a mild negative gradient (from 0.55 to 0.35 keV).

**NGC 2300** (d=30.4 Mpc, 1′ = 8.8 kpc, $r_e$=4.8 kpc, $r_{25}$=12.5 kpc)
It is the BCG in a group with neighboring stripped spiral galaxy NGC 2276 which is moving to the SW at ~850 km $s^{-1}$ and both galaxies are detected in X-rays (Rasmusen et al 2006). The temperature profile of NGC 2300 is a typical one with a cool core. The cool core is at ~0.6 keV and T reaches a peak ~1 keV at a few $r_e$, then declines outward. There may an edge to the E and NE and the diffuse gas is more extended to the SW. This may be due to sloshing by NGC 2276 (6′ away) after it has just passed the impact point.

**NGC 2563** (d=67.8 Mpc, 1′ = 19.7 kpc, $r_e$=6.4 kpc, $r_{25}$=20.6 kpc)
It is a dominant galaxy in a poor group with a typical cool core. The temperature is 0.8 keV near the center, peaks (T ~ 1.7 keV) at r=20 kpc (3 $r_e$), then declines outward. A nearby SB0 galaxy, NGC 2557 is also detected in X-rays. Rasmussen et al. (2012) studied individual group members with hot (X-rays) and cold (HI) gas to investigate the effect of ram pressure stripping and tidal interactions.

**NGC 3115** (d=9.7 Mpc, 1′ = 2.8 kpc, $r_e$=1.6 kpc, $r_{25}$=10.2 kpc)
An edge-on S0 galaxy with little hot gas. It is one of the gas poor ETGs with very deep Chandra observation, targeted to study the Bondi accretion (Wong et al. 2014; Lin et al. 2015a, b). The hot gas is detected within ~1 $r_e$.

**NGC 3379** (d=10.6 Mpc, 1′ = 3.1 kpc, $r_e$=2.4 kpc, $r_{25}$=8.3 kpc)
It is a typical old E galaxy with little hot gas. It is one of the gas poor ETGs with very deep Chandra observations, targeted to study a population of LMXBs (Brassington et al. 2008, 2010). The hot gas (T~0.3 keV) is detected within ~1 $r_e$ and its luminosity is one of the lowest ever measured from a hot phase of the ISM among genuine elliptical galaxies, as likely in the outflow phase (Trinchieri et al 2008).

**NGC 3402 = NGC 3411** (d=64.9 Mpc, 1 = 18.9 kpc, $r_e$=8.8 kpc, $r_{25}$=19.7 kpc)
It is the BCG in a small group, USGC S152. While the diffuse gas appears to be relaxed, the temperature map clearly indicates a shell-like structure at 20-40 kpc with cooler gas (~0.8 keV) surrounded by inner and outer hotter gas (~1 keV), as previously reported by O'Sullivan et al. (2007). The cooler gas is most obvious to the N (PA=-20 – 20°) and E (PA = 90-135°) and least to the SW (PA = 220-300°). This feature is not seen in the intensity, EM, projected pressure and projected entropy maps. It is not understood what caused the cooler shell. The possibilities include a previous AGN activity which could reheat the cool core and settling of material stripped from the halo of one of the other group member galaxies (O'Sullivan et al. 2007).

**NGC 3607** (d=22.8 Mpc, 1′ = 6.6 kpc, $r_e$=5.0 kpc, $r_{25}$=16.2 kpc)
It is a dominant E/S0 galaxy in a small group USGC U376 in the Leo cloud (Mazzei et al. 2014) with NGC 3608 (5.9′ N) and NGC 3605 (2.7′ SW). A hot core (~1 keV) may be present in the central region (r < 0.5 kpc) and then the gas temperature remains constant at 0.5 keV in r < $r_e$. The SB profile is also relatively flat (~$r^{-1}$) at r =1-10 kpc. Given the limited statistics of the Chandra data, it is not clear whether the hot gas indicates any sign of interactions. Both NGC 3607 and 3608 are known to be a LINER and the X-ray nuclear sources were studied by Flohic et al. (2006).

**NGC 3608** (d=22.9 Mpc, 1′ = 6.7 kpc, $r_e$=3.3 kpc, $r_{25}$=10.5 kpc)
It is an elliptical galaxy at 5.9′ (or ~40 kpc in projection) away from NGC 3607 and in the same fov of the Chandra observation of NGC 3607. As it is slightly smaller than NGC 3607 (0.8 mag less bright in K-band), its $L_X$ and $T_X$ are slightly smaller. The gas temperature is constant ~0.4 keV and the SB profile is relatively flat (~$r^{-1}$) at r < 2 $r_e$.

**NGC 3842** (d=97.0 Mpc, 1′ = 28.2 kpc, $r_e$=17.8 kpc, $r_{25}$=19.9 kpc)
It is the BCG of Abell 1367, but it is not at the center of the hot ICM, nor at the center of the cluster potential well. The diffuse hot gas at T ~ 1 keV is detected inside the $D_{25}$ ellipse. The hot ISM is embedded inside the hotter (5-6 keV) subcluster which is merging with the primary cluster of Abell 1367 centered at 20' SE of NGC 3842 (see Sun et al. 2005a, b). A few other galaxies in Abell 1367 subcluster (see the table below) are also detected in the same fov.

| name | RA (J2000) | DEC (J2000) | D(′) from N3842 | PA(deg) from N3842 | Notes |
|---|---|---|---|---|---|
| N3841 | 176.00896 | 19.97189 | 1.33 | 0.5 | E |
| N3837 | 175.98511 | 19.89458 | 3.57 | 202 | E |
| P169975 | 175.98833 | 19.95500 | 1.19 | 285 | S0 CGCG 097090 |
| U06697 | 175.95446 | 19.96844 | 3.26 | 290 | starburst CGCG 097087 |
| QSO | 175.98706 | 19.94705 | 1.23 | 263 | at z=0.35 |

**NGC 3923** (d=22.9 Mpc, 1′ = 6.7 kpc, re=5.8 kpc, $r_{25}$=19.6 kpc)
It is a young elliptical (E4) galaxy with a number of stellar shells (Bilek et al. 2016). The extended hot gas is detected inside the $D_{25}$ ellipse and is elongated along the major axis, but not as flat as the stellar system. In the inner region, the temperature decreases with increasing r, i.e., a hot core (0.7 keV at r=0.1 kpc). T reaches at the minimum (0.4 keV) at r=3kpc, then increases again to 0.6 keV at r=10-20 kpc. Kim & Fabbiano (2010) and Kim et al. (2012) investigated this galaxy among a sample of young elliptical galaxies, in terms of the X-ray binary luminosity function and the hot gas metallicity.

**NGC 4104** (d=120.0 Mpc, 1′ = 34.9 kpc, re=20.0 kpc, $r_{25}$=44.9 kpc)
It is a dominant galaxy in a small group. Given the shallow Chandra observation of this distant galaxy, its 2D spatial features are not clearly visible. Its radial temperature profile shows a cool core (~ 1 keV inside a few kpc) and a temperature peak (~1.5 keV) at r=20-40 kpc.

**NGC 4125** (d=23.9 Mpc, 1′ = 6.9 kpc, re=5.9 kpc, $r_{25}$=20.0 kpc)
It is a flattened elliptical (E6) galaxy. Similar to NGC 3923, the extended hot gas is detected inside the $D_{25}$ ellipse and is elongated along the major axis, but not as flat as the stellar system. It hosts a hot core, as in NGC 3923. The temperature peaks at 0.6 keV at r=0.1 kpc and decreases with increasing r, reaching a minimum (~0.3 keV) at the outer boundary where the gas temperature can be measured (~30 kpc). It also shows the characteristics of young ellipticals in their X-ray binary luminosity function (Kim & Fabbiano 2010; Zhang et al. 2012).

**NGC 4261** (d=31.6 Mpc, 1′ = 9.2 kpc, re=6.9 kpc, $r_{25}$=18.7 kpc)
It is a group dominant elliptical and also a FR-I radio galaxy (3C270). The bright AGN dominates the X-ray emission. The X-ray jets are detected at the position coincident with the radio jets and the compressed rims of the X-ray cavities are correlated with radio lobes (Zezas et al. 2005; Worrall et al. 2010; O'Sullivan et al 2011). The faint diffuse emission from the hot gas is detected inside the $D_{25}$ ellipse. The temperature is constant at ~0.7 keV in the inner region (r = 0.1 - 2 kpc), then abruptly increases to 1.3 keV at r = 5 kpc and remains high to the maximum radius (~30 kpc).

**NGC 4278** (d=16.1 Mpc, 1′ = 4.7 kpc, $r_e$=2.6 kpc, $r_{25}$=9.5 kpc)
It is one of the gas poor old elliptical galaxies. With deep Chandra observations (560 ksec), a population of LMXBs (Brassington et al. 2009; Fabbiano et al. 2010) and their connection to globular clusters (Kim et al. 2009; Fabbiano et al. 2010), and a central LINER activity in conjunction with optical and infrared data (Pellegrini et al. 2012) were extensively investigated. The hot gas is extended out to r ~ 5 kpc and the temperature is constant (~0.3 keV), but steeply increases to ~0.7 keV in the inner 0.3 kpc (see also Pellegrini et al 2012).

**NGC 4291** (d=26.2 Mpc, 1′ = 7.6 kpc, $r_e$=2.0 kpc, $r_{25}$=7.3 kpc)
Similar to NGC 4342, it is one of a few elliptical galaxies with unusually high BH-bulge mass ratios (see Bogdan et al. 2012a). The hot gas is extended roughly along the major axis, beyond the $D_{25}$ ellipse. It has a hot core with a negative T gradient to the minimum (0.4 keV) at r ~ 4 kpc (or 2 $r_e$), then a positive T gradient to ~0.8 keV at r ~ 10 kpc. The T map shows that the cooler gas may be extended more to the E than to the W.

**NGC 4325** (d=110.0 Mpc, 1′ = 32.0 kpc, $r_e$=10.5 kpc, $r_{25}$=15.3 kpc)
It is a dominant elliptical galaxy in a small group. The hot gas is extended (beyond the ACIS fov) and symmetric on a large scale (outside the $D_{25}$ ellipse), the core region is rather complex with a cool core and cavities (Russell et al. 2007; Lagana et al. 2015). The temperature is about 0.7 keV in the central region, increases to ~1.1 keV at r=35 kpc, then decreases outward. The T map further shows that the cooler gas inside the $D_{25}$ ellipse is elongated along the N-S, roughly following the major axis.

**NGC 4342** (d=16.5 Mpc, 1′ = 4.8 kpc, $r_e$=0.5 kpc, $r_{25}$=3.1 kpc)
It is one of a few elliptical galaxies with unusually high BH-bulge mass ratios (see Bogdan et al. 2012a). In contrast to a nearby massive elliptical NGC 4365, 20′ away (or 130 kpc in projection), it is optically faint but hot gas rich, hence associated with a large amount of dark matter (Bogdan et al. 2012b). Its $L_{X,GAS}/L_K$ is the highest among local ETGs, making it an extreme opposite to NGC 1316 with the lowest $L_{X,GAS}/L_K$. The SB map shows a head-tail structure with a discontinuity to the NE (just outside the $D_{25}$ ellipse) and a wide extended tail to the SW. The T map shows the extended tail is filled by cooler (0.6 keV) gas.

**NGC 4374** (d=18.4 Mpc, 1′ = 5.3 kpc, $r_e$=5.5 kpc, $r_{25}$=17.3 kpc)
It is one of the extensively studied elliptical galaxies in the Virgo cluster, also known as M84. It is at 17′ to the SW (PA=258) from NGC 4406 (M86) and at 89′ to the NW (PA=290) from the center of the Virgo cluster, NGC 4486 (M87). The hot gas exhibits many interesting features, including pronounced filaments and cavities in the central region (r < 10 kpc) which are associated with the radio jets (Finoguenov et al. 2008) and an extended tail to the SW from the galaxy center in r=10-30 kpc, likely due to the ram pressure (Randall et al. 2008). The T profile shows a hot core with a negative T gradient out to a T minimum (~0.6 keV) at r ~ 1 kpc, then a positive T gradient to a T maximum (~1.5 keV) at r~30 kpc. The T maps further show asymmetric, disturbed thermal structures, the cooler gas (~0.8 keV) filling the southern part of the $D_{25}$ ellipse and the hotter gas (~1.2 keV) filling the extended tail to the SW.

**NGC 4382** (d=18.5 Mpc, 1′ = 5.4 kpc, $r_e$=7.4 kpc, $r_{25}$=19.0 kpc)
It is a young S0 galaxy in the Virgo cluster, also known as M85. It contains a small amount of hot gas for its stellar luminosity (Sansom et al. 2006) and therefore it is often used to study LMXBs (Sivakoff et al. 2003). Within the limited statistics, the relatively cool (~0.4 keV) and smooth hot gas does not show a distinct feature.

**NGC 4406** (d=17.1 Mpc, 1 = 5.0 kpc, $r_e$=10.3 kpc, $r_{25}$=22.2 kpc)
It is one of the extensively studied elliptical galaxies in the Virgo cluster, also known as M86. It is at 17′ to the NE (PA=78) from NGC 4374 (M84) and at 75′ to the NW (PA=296) from the center of the Virgo cluster, NGC 4486 (M87). Its X-ray luminosity is the 2$^{nd}$ largest (M87 being the most luminous) among the Virgo galaxies. The X-ray emission from the extended plume to the NW from the galaxy center is as bright as that of the main body (e.g., Rangarajan et al. 1995). Based on its negative radial velocity (−250 km $s^{-1}$), it is moving supersonically in the Virgo cluster with few other galaxies in the group (e.g., NGC 4438) and the extended plume may be related to ram pressure stripping (e.g., Randall et al. 2008).

**NGC 4438** (d=18.0 Mpc, 1′ = 5.2 kpc, $r_e$=5.0 kpc, $r_{25}$=22.3 kpc)
It is an S0 galaxy in the Virgo cluster, possibly belongs to the M86 group. It is at 23′ to the E (PA=81) from NGC 4406 (M86) and at 58′ to the NW (PA=310) from the center of the Virgo cluster, NGC 4486 (M87). In addition to the hot gas in the central region, multiple filaments are visible to r~10 kpc to the W and SW, which may be due to the interaction with a nearby galaxy, NGC 4435, which is at 4.4′ to the NE (Machacek et al. 2004). Also detected are the extended H$\alpha$ filaments between M86 and NGC 4438, suggesting the interaction between these two galaxies (Kenney et al. 2008).

**NGC 4472** (d=16.3 Mpc, 1′ = 4.7 kpc, $r_e$=8.3 kpc, $r_{25}$=24.2 kpc)
It is one of the extensively studied hot gas rich elliptical galaxies in the Virgo cluster, at 4.4° to the S from the cluster center, also known as M49. Although it is not at the cluster center, it is brightest in the Virgo cluster. There are multiple cavities in the central region ($r < 10$ kpc), the contact discontinuity (a cold front) at ~20 kpc to the N and extended tails to the E as well as longer tails to the SW which are extended beyond the ACIS fov. These hot gas features are clearly indicating interactions with the radio jets and ICM in the Virgo cluster (e.g., see Biller et al. 2004; also see Kraft et al. 2011 for XMM-Newton data analysis.)

**NGC 4477** (d=16.5 Mpc, 1′ = 4.8 kpc, $r_e$=3.5 kpc, $r_{25}$=9.1 kpc)
It is a SB0 galaxy in the Virgo cluster at 75.6′ to the N (PA=351) from the center of the Virgo cluster, NGC 4486 (M87). It is also known as Seyfert 2 (Veron-Cetty & Veron 2006) and the X-ray nucleus is detected. The hot gas is relatively cold (0.3-0.4 keV) and confined within the $D_{25}$ ellipse. The SB map shows asymmetry, more extended to the N and W (than to the S and E) and also shows a distinct spiral like features. The Chandra observations were primarily obtained for a distant ($z \sim 1$) luminous cluster (Fassbender et al. 2011, Lerchster et al. 2011), XMMUJ1230+1339, which is at 3.7′ to the E from NGC 4477.

**NGC 4526** (d=16.9 Mpc, 1′ = 4.9 kpc, $r_e$=3.3 kpc, $r_{25}$=17.8 kpc)
It is an S0 galaxy in the Virgo cluster. It is at 66′ to the E (PA=106°) from NGC 4472 (M49) and at 4.8° to the S (PA=170°) from the center of the Virgo cluster, NGC 4486 (M87). A relatively strong X-ray point source is detected in the center, although it is not a known AGN. The hot gas is relatively weak and cold (0.3 keV) and confined inside 1 $r_e$.

**NGC 4552** (d=15.4 Mpc, 1′ = 4.5 kpc, $r_e$=3.1 kpc, $r_{25}$=11.4 kpc)

It is a S0 (listed as E0 in RC3) galaxy in the Virgo cluster, at 72′ to the E from the cluster center, also known as M89. The LINER nucleus source is detected in X-ray (Xu et al. 2005). In contrast to the relaxed old stellar system, the hot gas morphology shows an excellent example of the head-tail structure, likely caused by the relative motion inside the Virgo ICM, as it is falling into the cluster center (Machacek et al. 2006a, b). The cold front on the N has Kelvin–Helmholtz instability structures to the EW direction and the curved stripped tail is extended to the SE. There are also cavities in the core. See Machacek et al. (2006a) for the discussions on the gas stripping, Machacek et al. (2006b) on the nuclear outflow and Roediger et al. (2015) on theoretical modeling in terms of viscosity and KH instability. The T maps clearly show the extended tail filled with ~0.6 keV gas. The T profile shows a hot core, having a negative T gradient in the central region, a minimum (T ~ 0.4 keV) at 4-5 kpc, and then a positive gradient in the outer region.

**NGC 4555** (d=91.5 Mpc, 1′ = 26.6 kpc, $r_e$=13.2 kpc, $r_{25}$=25.4 kpc)
It is an isolated elliptical, but its relatively high gas temperature ($T_{GAS}$ ~ 1 keV) and luminosity ($L_X$ ~$10^{41.5}$ erg $s^{-1}$) may indicate a dominant galaxy in a very poor group with a massive dark halo (O'Sullivan & Ponman 2004). The 2D maps indicate that the hot gas is smooth and relaxed. The T map and profile suggest the presence of a cool core.

**NGC 4594** (d=9.8 Mpc, 1′ = 2.8 kpc, $r_e$=3.4 kpc, $r_{25}$=12.4 kpc)
It is a nearby edge-on S0 galaxy, also known as M104 and Sombrero. Its nucleus and X-ray binaries dominate the entire X-ray emission (Li et al. 2011). After excluding point sources, the temperature map suggests that low temperature (~0.5 keV) gas lies along the disk to the EW of the core. The gas to the perpendicular direction from the disk is slightly hotter (~0.7 keV).

**NGC 4636** (d=14.7 Mpc, 1′ = 4.3 kpc, $r_e$=6.7 kpc, $r_{25}$=12.8 kpc)
It is one of extensively studied hot gas rich elliptical galaxy in the Virgo cluster, at 10° to the S from the Virgo cluster center and at the northern end of the Virgo South Extension (centered around NGC 4697). The hot gas exhibits spiral-arm like features and cavities on a small scale (< 10 kpc), extension to the WSW on an intermediate scale (10-30 kpc) and another extension to the N on a large scale (> 50 kpc). The smaller scale features are related to the nuclear activities and radio jets and the larger scale features are likely sloshing due to the perturbation from nearby galaxies (O'Sullivan et al. 2005, Baldi et al. 2009). As seen in the SB maps, both T profile and map indicate complex thermal structures. The temperature of the hot gas is about 0.5 keV in the central region, increases to ~1 keV at ~15 kpc, then declines in the outer region. The T maps further show the asymmetric distribution of the inner cooler gas which is elongated to the N-S direction.

**NGC 4649** (d=16.8 Mpc, 1′ = 4.9 kpc, re=6.2 kpc, r25=18.1 kpc)
It is a giant elliptical galaxy in the Virgo cluster, at 3.3° to the E from the cluster center, also known as M60. It hosts a large amount of hot gas which had been considered as a prototype example of a smooth, relaxed hot halo. However, Chandra observations revealed the non-smooth, asymmetric features both on a small (< 3 kpc) scale related to the AGN activity (Paggi et al. 2014) and a large scale (20-30 kpc) related to the bulk motion (Wood et al. 2017). Also detected in the same fov is a nearby spiral galaxy, NGC 4647, located at 2.5′ form NGC 4649 in the NW direction. The temperature profile shows a negative gradient inside and a positive gradient outside with a minimum (T ~ 0.8 keV) at r ~ 1 kpc. A hot core in the center has been discussed in Pellegrini et al. (2012) and Paggi et al. (2014). The temperature map further indicates that the cooler gas (0.8-0.9 keV) extends preferentially to the NE and SW directions, the same directions where two extended wings are visible while the gas in the other directions (NW and SE) is hotter (1-1.2 keV). Woods et al. (2017) suggested that the two wings might be caused by the Kelvin-Helmholtz instability while the galaxy is infalling toward the center of the Virgo cluster. However, the cooler gas extended from the center may imply that the extended wings may also be related to the AGN outflows.

**NGC 4782** (d=60.0 Mpc, 1′ = 17.5 kpc, re=4.4 kpc, r25=15.5 kpc)
It is in a close pair VV201 with NGC 4783, aka Dumbbell galaxies. NGC 4782 is also a FR-I radio galaxy (3C278). The hot ISM (~0.5 keV) in both galaxies is embedded inside the hotter (~1.4 keV) ICM (see Machacek et al. 2007). The hot gas is highly disturbed. The hot gas in NGC 4782 exhibits a cavity and X-ray knots which are related to the radio jets and the hot gas of NGC 4783 exhibits a head-tail structure (a cold front and an extended tail) caused by the ram pressure as it is apparently moving to the E.

**NGC 5044** (d=31.2 Mpc, 1′ = 9.1 kpc, $r_e$=3.9 kpc, $r_{25}$=13.4 kpc)
It is a dominant galaxy in the X-ray brightest group in the sky. Its extended hot halo has been extensively studied from the early X-ray missions. The deep Chandra observations show the hot gas is sloshing with fronts visible in surface brightness and abundance, temperature maps with many small cavities (David et al. 2009, 2011, 2017). Also shown in XMM-Newton data is a large scale sloshing (O'Sullivan et al (2014)

**NGC 5129** (d=103.0 Mpc, 1′ = 30.0 kpc, re=14.3 kpc, $r_{25}$=25.4 kpc)
It is the dominant galaxy in a small group (Eckmiller et al. 2011, see also Bharadwaj et al. 2014). The hot gas has a cool core with a positive temperature gradient in the inner region to a peak (~1 keV) at r ~ 20 kpc and a negative gradient in the outer region to the fov limit ($r_{max}$~ 200 kpc).

**NGC 5171** (d=100.0 Mpc, 1′ = 29.1 kpc, $r_e$=12.4 kpc, $r_{25}$=15.9 kpc)
It is the dominant galaxy in a small group with multiple roughly-equal sized ellipticals. The hot ISM ($L_X$ ~ a few x $10^{40}$ erg $s^{-1}$) directly associated with NGC 5171 is confined within $r_e$ with T ~ 1 keV and is roughly symmetric inside $r_e$. Interestingly, there is a large amount (($L_X$ ~ a few x $10^{41}$ erg $s^{-1}$) of hotter gas (1.2-1.3 keV) filling gaps among group galaxies (see also Osmond et al. 2004), mostly to the N and the E from NGC 5171. Also detected in a single Chandra observation are three large galaxies (NGC 5179, NGC 5176 and NGC 5177) and two small galaxies (SDSS J132920.65+114424.1 and SDSS J132928.18+114625.2). The temperatures of their hot ISM are in the range of 0.3-0.6 keV (see also Jeltema et al. 2008).

| name | RA (J2000) | DEC (J2000) | D(′) from N5171 | PA(deg) from N5171 |
|---|---|---|---|---|
| N5176 | 202.35399 | 11.78148 | 2.9′ | 17 |
| N5177 | 202.35108 | 11.79703 | 3.8′ | 10 |
| N5179 | 202.37869 | 11.74583 | 2.4′ | 74 |
| SDSS J132920.65+114424.1 | | | 0.4′ | 326 |
| SDSS J132928.18+114625.2 | | | 2.8′ | 35 |

**NGC 5813** (d=32.2 Mpc, 1′ = 9.4 kpc, $r_e$=8.3 kpc, $r_{25}$=19.5 kpc)
It is the dominant galaxy in a small group. The hot gas morphology exhibits three sets of nested co-aligned cavities and shocks (Randall et al. 2011 and 2015). The temperature map shows cooler uplifted material along line of cavities.

**NGC 5846** (d=24.9 Mpc, 1′ = 7.2 kpc, $r_e$=7.2 kpc, $r_{25}$=14.7 kpc)
It is the dominant galaxy in a small group. The hot gas morphology exhibits small-scale cavities associated with radio jets in the central region and spiral-like tails and multiple cold fonts on a large scale which may be caused by sloshing due to a nearby galaxy NGC 5850 (Machacek et al. 2011, Gastadello et al. 2013, Paggi et al. 2017).

**N5866** (d=15.4 Mpc, 1′ = 4.5 kpc, $r_e$=2.8 kpc, $r_{25}$=10.4 kpc)
It is a nearby edge-on S0 galaxy, hosting a LINER nucleus. It could be M102 that has not been identified unambiguously. Li et al (2009) investigated the weak diffuse hot gas which is extended as far as 3.5 kpc away from the galactic plane. The faint spiral-like filament to the S lying outside the D25 ellipse may be interesting, but needs to be confirmed by deeper observations.

**NGC 6107** (d=127.9 Mpc, 1′ = 37.2 kpc, $r_e$=16.3 kpc, $r_{25}$=15.8 kpc)
It is the dominant galaxy in a small group. The hot ISM (~1 keV) inside the $D_{25}$ ellipse shows an elongated structure in the SE-NW direction (roughly along the minor axis), which is surrounded by the hotter (~1.5 keV) gas. The large scale ROSAT observation (Feretti et al 1995) showed that the hotter IGM extends to the entire regions connecting NGC 6107 and NGC 6109 (at 7.5′ to the NE from NGC 6107). It is also known as a radio galaxy, B2 1615+35, with a SE-NW radio extension (Feretti et al 1995, Condon et al. 2002), which is interestingly along the same direction with the inner hot gas feature.

**NGC 6338** (d=123.0 Mpc, 1′ = 35.8 kpc, $r_e$=17.1 kpc, $r_{25}$=27.1 kpc)
It is the dominant galaxy in a small group, possibly merging with PGC 59943 (or MCG +10-24-117) at 1.2 arcmin to the N. Both galaxies have stripped tails and multiple cavities (Pandge et el. 2012). The tail of PGC 59943 is stretched to the N, indicating that this galaxy is moving to the S. In this ACIS-I image, the chip gaps are visible in raw binned images and some spectral maps, mostly significant in CB (because a spatial bin was determined in a region with similar counts/area) but least in WB. The T profile and map show the presence of a cool core inside $r_e$.

**NGC 6482** (d=58.4 Mpc, 1′ = 17.0 kpc, $r_e$=6.3 kpc, $r_{25}$=16.9 kpc)
It is an isolated elliptical or a fossil group galaxy with a relatively relaxed hot gas morphology (Khoroshahi et al 2004; Buote 2017). In contrast to typical relaxed systems, the hot gas is hotter in the inner region than in the outer region with the temperature monotonically decreasing outward from 0.9 keV at the center to 0.5 keV at 50 kpc. The Suzaku data suggest a hint that the temperature may increase slightly (to 0.65 keV) at the outer region ~100 kpc (Buote 2017). All spectral maps show roughly circularly symmetric distribution, except in the inner 10 kpc region where the hotter gas is slightly elongated to the SE-NW direction (roughly along the minor axis direction), more pronounced to the SE.

**NGC 6861** (d=28.0 Mpc, 1′ = 8.2 kpc, $r_e$=3.1 kpc, $r_{25}$=11.5 kpc)

It is one of the two dominant galaxies (with NGC 6868 at 26′ to the E) in the Telescopium galaxy group. While NGC 6861 is slightly less luminous (by a factor of 1.4) than NGC 6868, but its velocity dispersion is higher (by a factor or 1.7). The hot gas morphology indicates the two galaxies (or two subgroups) are possibly merging. The hot gas in NGC 6861 has bifurcated tails trailing NGC 6861 at ~40 kpc to the W and NW, likely caused by the subgroup merger (Machacek et al. 2010).

**NGC 6868** (d=26.8 Mpc, 1′ = 7.8 kpc, $r_e$=3.9 kpc, $r_{25}$=13.8 kpc)
It is one of the two dominant galaxies in the Telescopium galaxy group, possibly merging with NGC 6861 at 26′ to the W. There is a cold font at ~23 kpc to the N, likely caused by sloshing due to the merger (Machacek et al. 2010). The T maps further show an asymmetry in the hot gas morphology with slightly cold gas (~0.6 keV) forming a shell-like feature which is more pronounced to the NE than to the S on a scale of the $D_{25}$ ellipse.

**NGC 7618** (d=74.0 Mpc, 1′ = 21.5 kpc, $r_e$=7.7 kpc, $r_{25}$=12.9 kpc)
It is a dominant galaxy in a small group, showing an excellent example of the pronounced spiral-like features caused by sloshing and the associated structures caused by turbulent instability. It is likely merging with UGC 12491 (PGC 71014), at 14′ to the NW from NGC 7618 (Kraft et al. 2006, Roediger et al. 2012). The T maps clearly show the curved, extended tail filled with 0.7-0.8 keV gas which is surrounded by ~1.2 keV hotter ambient gas. Note that the tail is more pronounced in the T map than in the SB map. The tail is clearly visible in the EM and projected entropy maps, but almost invisible in the projected pressure map, possibly indicating pressure balance with the ambient gas.

**N7619** (d=53.0 Mpc, 1′ = 15.4 kpc, $r_e$=8.8 kpc, $r_{25}$=19.4 kpc)
It is a dominant E galaxy in the Pegasus I group. It has a cold front to the NE and extended tails to the SW. The X-ray tail is metal-enriched suggesting that the hot gas is originated from the galaxy (Kim et al. 2008). A nearby galaxy, NGC 7626, at 7′ to the E is possibly interacting with NGC 7619.

**NGC 7626** (d=56.0 Mpc, 1 ′= 16.3 kpc, $r_e$=12.0 kpc, $r_{25}$=21.4 kpc)
It is an E galaxy in the Pegasus I group at 7′ to the E from the NGC 7619. The surface brightness and temperature maps suggest that it is possibly merging with NGC 7619 (Randall et al. 2009).

**Reference**

Baldi, A., Forman, W., Jones, C., et al. 2009, ApJ, 707, 1034
Bharadwaj, V. Reiprich, T.H. Schellenberger, G. et al. 2014, A&A, 572, A46
Bilek, M. Cuillandre, J.-C. Gwyn, S. et al. 2016, A&A, 588, 77
Biller, B. A., Jones, C., Forman, W. R., Kraft, R. & Ensslin, T. 2004, ApJ, 613, 238
Blanton, E.L., Sarazin, C.L., and Irwin, J.A., 2001, ApJ, 552, 106
Bogdan, A. Forman, W.R. Zhuravleva, I. et al. 2012a, ApJ, 753, 140
Bogdan, A. Forman, W.R. Kraft, R.P. et al. 2012b, ApJ, 755, 25
Bogdan, A., van Weeren, R.J., Kraft, R.P., et al. 2014, ApJ, 782, L19
Brassington, N. J. et al. 2009, ApJS, 181, 605
Brassington, N. J., Fabbiano, G., Kim, D. W., et al. 2008, ApJS, 179, 142
Brassington, N. J., Fabbiano, G., Blake, S., et al. 2010, ApJ, 725, 1805-1823
Buote, D. A., Jeltema, T. E., Canizares, C. R. & Garmire, G. P. 2002, ApJ, 577, 183
Buote, D.A., 2017, ApJ, 834, 164
Condon, J. J., Cotton, W. D., Broderick, J. J. 2002, AJ, 124, 675
Crook, A.C., Huchra, J.P., Martimbeau, N., et al. 2007, ApJ, 655, 790
David, L. P., Jones, C., Forman, W., et al. 2009, ApJ, 705, 624
David, L. P., O'Sullivan, E., Jones, C., et al. 2011, ApJ, 728, 162
David, L. P., Vrtilek, J., O'Sullivan, E., et al. 2017, ApJ, 842, 84
Dong, R., Rasmussen, J. & Mulchaey, J. S. 2010, ApJ, 712, 883
Eckmiller, H.J. Hudson, D.S. and Reiprich, T.H., 2011, A&A, 535, A105
Ekers, R. D., Gross, W. M., Wellington, K. J., et al. 1983, A&A, 127, 361
Fabbiano, G. Elvis, M. Markoff, S. et al. 2003, ApJ, 588, 175
Fabbiano, G. et al. 2010 ApJ, 725, 1824
Fassbender, R., B"ohringer, H., Santos, J.S., et al. 2011, A&A, 527A, 78
Feretti, L. Fanti, R. Parma, P. et al. 1995, A&A, 298, 699
Finoguenov, A., Ruszkowski, M., Jones, C., et al. 2008, ApJ, 686, 911
Flohic, H.M.L.G. Eracleous, M. Chartas, G. et al. 2006, ApJ, 647, 140
Garcia, A.M., 1993, A&AS, 100, 47
Gastaldello, F., Di Gesu, L., Ghizzardi, S., et al. 2013, ApJ, 770, 56
Giacintucci, S. O'Sullivan, E. Vrtilek, J. et al. 2011, ApJ, 732, 95
Giacintucci, S., O'Sullivan, E., Clarke, T. E., et al. 2012, ApJ, 755, 172
Hardcastle, M.J. Worrall, D.M. Birkinshaw, M. et al. 2002, MNRAS, 334, 182
Humphrey, P.J., and Buote, D.A., 2006, ApJ, 639, 136
Humphrey, P. J. et al. 2011, ApJ, 729, 53
Jeltema, T., Binder, B. &, Mulchaey, J. S. 2008, ApJ, 679, 1162
Jetha, N. N., Hardcastle, M. J., Babul, A., et al. 2008, MNRAS, 384, 1344
Kawaharada, M., Makishima, K., Kitaguchi, T., et al. 2009, ApJ, 691, 971
Kenney, J. D. P., Tal, T., Crowl, H. H., Feldmeier, J. & Jacoby, G. H. 2008, ApJ 687, L69
Khosroshahi, H.G. Jones, L.R. and Ponman, T.J., 2004, MNRAS, 349, 1240
Kim, D.-W., and Fabbiano, G. 2003, ApJ, 586, 826
Kim, D.-W., Kim, E., Fabbiano, G., et al. 2008, ApJ, 688, 931
Kim, D.-W., Fabbiano, G., Brassington, N.J., et al. 2009, ApJ, 703, 829
Kim, D.-W., and Fabbiano, G. 2010, ApJ, 721, 1523
Kim, D.-W., Fabbiano, G. and Pipino, A. 2012, ApJ, 715, 38

Kim, D.-W., Anderson, C., Burke, D., et al. 2018, ApJ, 853, 129
Kraft, R.P. Forman, W.R. Churazov, E. et al. 2004, ApJ, 601, 221
Kraft, R.P., Jones, C., Nulsen, P.E.J., et al. 2006, ApJ, 640, 762
Kraft, R. P., Forman, W. R., Jones, C., *et al.* 2011, *ApJ*, 727, 41
Lagana, T.F., Lovisari, L., Martins, L., et al. 2015, A&A, 573A, 66
Laing, R.A. Guidetti, D. Bridle, A.H. et al. 2011, MNRAS, 417, 2789
Lerchster, M., Seitz, S., Brimioulle, F., et al. 2011, MNRAS, 411, 2667
Li, J.-T. Wang, Q.D. Li, Z. et al. 2009, ApJ, 706, 693
Li, Z. Jones, C. Forman, W.R. et al. 2011, ApJ, 730, 84
Lin, D. Irwin, J.A. Wong, K.-W. et al. 2015a, ApJ, 808, 19
Lin, D. Irwin, J.A. Wong, K.-W. et al. 2015b, ApJ, 808, 20
Lovisari, L. Reiprich, T.H. and Schellenberger, G., 2015, A&A, 573A, 118
Machacek, M. E., Jones, C. & Forman, W. R. 2004, *ApJ*, 610, 183–200
Machacek, M. E., Dosaj, A., Forman, W., et al. 2005, ApJ, 621, 663
Machacek, M., Jones, C., Forman, W. R. & Nulsen, P. 2006, ApJ, 644, 155
Machacek, M., Nulsen, P. E. J., Jones, C. & Forman, W. R. 2006, ApJ, 648, 947
Machacek, M. E., Kraft, R. P., Jones, C. et al, 2007, ApJ, 664, 804.
Machacek, M. E., O'Sullivan, E. Randall, S.W. et al. 2010, ApJ, 711, 1316
Machacek, M. E., Jerius, D., Kraft, R., et al. 2011, ApJ, 743, 15
Mazzei, P., Marino, A., and Rampazzo, R., 2014, ApJ, 782, 53
O'Sullivan, E. and Ponman, T.J., 2004, MNRAS, 354, 935
O'Sullivan, E., Vrtilek, J. M. & Kempner, J. C. 2005, ApJ, 624, L77
O'Sullivan, E., Vrtilek, J.M. Harris, D.E. et al. 2007, ApJ, 658, 299
O'Sullivan, E., Worrall, D.M. Birkinshaw, M. et al. 2011, MNRAS, 416, 2916
O'Sullivan, E., David, L. P. & Vrtilek, J. M. 2014, MNRAS, 437, 730
Osmond, J. P. F., Ponman, T. J. & Finoguenov, A. 2004, MNRAS, 355, 11
Paggi, A., Fabbiano, G., Kim, D.-W. et al. 2014, ApJ, 787, 134
Paggi, A., Kim, D.-W., Anderson, A. et al. 2017, ApJ, 844, 5
Panagoulia, E.K. Fabian, A.C. and Sanders, J. S. 2014, MNRAS, 438, 2341
Pandge, M.B. Vagshette, N.D. David, L.P. et al. 2012, MNRAS, 421, 808
Paolillo, M., Fabbiano, G., Peres, G. & Kim, D.-W. 2002, ApJ, 565, 883
Pellegrini, S. Venturi, T. Comastri, A. e al. 2003, ApJ, 585, 677
Pellegrini, S., Wang, J., Fabbiano, G., et al. 2012, ApJ, 758, 94
Rangarajan, F. V. N., White, D. A., Ebeling, H. & Fabian, A. C. 1995, MNRAS, 277, 1047
Randall, S., Nulsen, P., Forman, W.R., et al. 2008, ApJ, 688, 208
Randall, S. W., Jones, C., Kraft, R., Forman, W. R. & O'Sullivan, E. 2009, ApJ, 696, 1431
Randall, S. W., Forman, W. R., Giacintucci, S., et al. 2011, ApJ, 726, 86
Randall, S. W., Nulsen, P. E. J., Jones, C., et al. 2015, ApJ, 805, 24
Rasmussen, J. Bai, X.-N. Mulchaey, J.S. et al. 2012, ApJ, 747, 31
Rasmussen, J. Ponman, T.J. and Mulchaey, J.S., 2006, MNRAS, 370, 453
Roediger, E., Kraft, R.P., Machacek, M.E. et al. 2012, ApJ, 754, 147
Roediger, E., Kraft, R. P., Nulsen, P. E. J., et al. 2015, ApJ, 806, 104
Russell, P.A., Ponman, T.J., and Sanderson, A.J.R. , 2007, MNRAS, 378, 1217
Sansom, A. E., O'Sullivan, E., Forbes, D. A., et al. 2006, *MNRAS*, 370, 1541
Scharf, C. A., Zurek, D. R. & Bureau, M. 2005, ApJ, 633, 154
Schellenberger, G., Vrtilek, J.M., David, L., et al. 2017, ApJ, 845, 84

Schweizer, F. 1980, ApJ, 237, 303
Shin, J., Woo, J.-H., and Mulchaey, J.S., 2016, ApJS, 227, 31
Sivakoff, G. R., Sarazin, C. L. and Irwin, J. A. 2003, *ApJ*, 599, 218
Sivakoff, G. R., Sarazin, C. L. and Carlin, J.L., 2004, ApJ, 617, 262 432, 530
Statler, T. S. & McNamara, B. R. 2002, ApJ, 581, 1032
Su, Y., Gu, L., White, R. & Irwin, J. 2014, ApJ, 786, 152
Su, Y. Kraft, R.P. Nulsen, P.E.J. et al. 2017a, ApJ, 835, 19
Su, Y., Nulsen, P.E.J., Kraft, R.P., et al. 2017b, ApJ, 851, 69
Sun, M. Forman, W. Vikhlinin, A. et al. 2003, ApJ, 598, 250
Sun, M. and Vikhlinin, A., 2005, ApJ, 621, 718
Sun, M. Vikhlinin, A. Forman, W. et al. 2005, ApJ, 619, 169
Thomas, J. Ma, C.-P. McConnell, N.J. et al. 2016, Nature, 532, 340
Trinchieri, G., Breitschwerdt, D., Pietsch, W., et al. 2007, A&A, 463, 153
Trinchieri, G., Pellegrini, S., Fabbiano, G., et al. 2008, ApJ, 688, 1000
Veron-Cetty, M.-P., and Veron, P., 2006, A&A, 455, 773
von Benda-Beckmann, A. M., D'Onghia, E., Gottlöber, S., et al. 2008, MNRAS, 386, 2345
Werner, N., Allen, S. W., and Simionescu, A. 2012, MNRAS, 425, 2731
Wong, K.-W. Irwin, J.A. Shcherbakov, R.V. et al. 2014, ApJ, 780, 9
Wood, R.A., Jones, C., Machacek, M.E., et al. 2017, ApJ, 847, 79
Worrall et al 2010, MNRAS 408, 701
Worrall, D.M. Birkinshaw, M. and Hardcastle, M.J., 2003, MNRAS, 343, L73
Worrall, D.M. Birkinshaw, M. Laing, R.A. et al. 2007, MNRAS, 380, 2
Xu, Y., Xu, H., Zhang, Z., et al. 2005, ApJ, 631, 809
Zezas, A., Birkinshaw, M., Worrall, D. M., Peters, A. & Fabbiano, G. 2005, ApJ, 627, 711
Zhang, Z., Gilfanov, M. & Bogdan, A. 2012, A&A, 546, A36